\documentclass[10pt,twoside]{lyxor_article}

\usepackage{amssymb}
\usepackage{amsfonts}
\usepackage[latin9]{inputenc}
\usepackage{amsmath}
\usepackage{amsbsy}
\usepackage{fancybox}
\usepackage{fancyhdr}
\usepackage{graphicx}
\usepackage{nicefrac}
\usepackage{longtable}
\usepackage{array}
\usepackage{eurosym}
\usepackage[hyphens]{url}
\usepackage{dsfont}
\usepackage{arydshln}
\usepackage{multirow}
\usepackage{multicol}
\usepackage{comment}
\usepackage{tikz}
\usepackage{cancel}
\usepackage{lipsum}
\usepackage{pdflscape}
\usepackage{algorithmic}
\usepackage{algorithm}
\usepackage{centernot}

\usetikzlibrary{positioning}
\usetikzlibrary{patterns,arrows,decorations.pathreplacing}

\usepackage{pifont}

\usepackage{color}
\definecolor{lyxor_light_blue}{RGB}{112,153,189}
\definecolor{lyxor_dark_blue}{RGB}{0,28,73}
\definecolor{lyxor_tan}{RGB}{195,187,175}
\definecolor{lyxor_cyan}{RGB}{0,164,167}
\definecolor{lyxor_pink}{RGB}{196,0,102}
\definecolor{lyxor_magenta}{RGB}{238,115,44}
\definecolor{lyxor_green}{RGB}{0,102,0}
\definecolor{lyxor_red}{RGB}{120,0,0}

\def\tableskip{\vskip 10pt plus 2pt minus 2pt\relax}
\def\figureskip{\vskip 10pt plus 2pt minus 2pt\relax}

\newlength{\figurewidth}
\newlength{\figureheight}
\newtheorem{theorem}{Theorem}
\newtheorem{remark}{Remark}
\newtheorem{example}{Example}

\def\limfunc#1{\mathop{\rm #1}}
\def\func#1{\mathop{\rm #1}\nolimits}

\newcommand{\TsIII}{\hspace{3pt}}

\newcommand{\TsV}{\hspace{5pt}}

\newcommand{\TsVIII}{\hspace{8pt}}

\newcommand{\TsX}{\hspace{10pt}}

\newcommand{\TsXII}{\hspace{12pt}}
\newcommand{\TsXIII}{\hspace{13pt}}

\DeclareFontFamily{OT1}{pzc}{}
\DeclareFontShape{OT1}{pzc}{m}{it}{<-> s * [1.3] pzcmi7t}{}
\DeclareMathAlphabet{\mathpzc}{OT1}{pzc}{m}{it}

\begin{document}

\setcounter{page}{1}

\title{\textbf{\color{lyxor_dark_blue}Risk Parity Portfolios with Skewness Risk:\\
An Application to Factor Investing and Alternative Risk Premia%
\footnote{We would like to thank Guillaume Lasserre and Guillaume
Weisang for their helpful comments. Correspondence should be
addressed to thierry.roncalli@lyxor.com.}}}

\author{
{\color{lyxor_dark_blue} Benjamin Bruder} \\
Quantitative Research \\
Lyxor Asset Management, Paris \\
\texttt{benjamin.bruder@lyxor.com} \and
{\color{lyxor_dark_blue} Nazar Kostyuchyk} \\
Quantitative Research \\
Lyxor Asset Management, Paris \\
\texttt{nazar.kostyuchyk@lyxor.com} \and
{\color{lyxor_dark_blue} Thierry Roncalli} \\
Quantitative Research \\
Lyxor Asset Management, Paris \\
\texttt{thierry.roncalli@lyxor.com} }

\date{\color{lyxor_light_blue}September 2016}

\maketitle

\begin{abstract}
This article develops a model that takes into account skewness
risk in risk parity portfolios. In this framework, asset returns are
viewed as stochastic processes with jumps or random variables
generated by a Gaussian mixture distribution. This dual
representation allows us to show that skewness and jump risks are
equivalent. As the mixture representation is simple, we obtain
analytical formulas for computing asset risk contributions of a
given portfolio. Therefore, we define risk budgeting
portfolios and derive existence and uniqueness conditions. We then
apply our model to the equity/bond/volatility asset mix policy. When
assets exhibit jump risks like the short volatility strategy, we
show that skewness-based risk parity portfolios produce better
allocation than volatility-based risk parity portfolios. Finally, we
illustrate how this model is suitable to manage the skewness risk of
long-only equity factor portfolios and to allocate between
alternative risk premia.
\end{abstract}

\medskip


\section{Introduction}

Over the last few years, three concepts have rapidly emerged in
asset management and definitively changed allocation policies of
large institutional investors and pension funds around the world. These
three concepts are risk parity, factor investing and
alternative risk premia. They constitute the key components of risk-based
investing. Risk parity is an asset allocation model, which focuses
on risk diversification (Roncalli, 2013). The primary idea of factor
investing is to capture risk factors, which are rewarded in the
universe of equities (Ang, 2014), whereas the concept of alternative
risk premia is an extension of the factor investing approach, and
concerns all the asset classes (Hamdan \textsl{et al.}, 2016). These
three methods share the same philosophy. For institutional
investors, the main issue is the diversification level of
their allocation, that is the exposure to systematic risk factors,
because idiosyncratic risks and specific bets generally disappear in
large portfolios. This explains that most of factor investing
portfolios and alternative risk premia portfolios are managed with a
risk parity approach.\bigskip

However, risk parity portfolios are generally designed to manage a
long-only portfolio of traditional assets (equities and bonds). For
instance, the ERC portfolio of Maillard \textsl{et al.} (2010)
consists in allocating the same amount of risk across the assets of
the portfolio. The simplicity and robustness of this portfolio
have attracted many professionals. Nevertheless, one may think that
it is not adapted when assets present a high jump risk. In this
case, the volatility is certainly not the best metric to assess the
risk of the asset. A typical example is the volatility carry risk
premium, which has a very low volatility but a high skewness. Other
examples are the value strategy or the long/short
winners-minus-losers strategy in equities.\bigskip

Portfolio allocation with skewness is an extensive field of research
in finance (Kraus and Litzenberger, 1976; Harvey and Siddique, 2000;
Patton, 2004; Jurczenko and Maillet, 2006; Jondeau and Rockinger,
2006; Martellini and Ziemann, 2010; Xiong and Idzorek, 2011). Most
of these studies are based on the maximization of a utility
function that takes into account high-order moments. Whereas these
approaches are appealing from a theoretical point of view, they are
not actually implemented by professionals because of their complexity
and lack of robustness. Another approach consists in using the
Cornish-Fisher value-at-risk. Again, this approach is not used in
practice because it requires the computation of co-skewness
statistics, which are difficult to estimate.\bigskip

In this paper, we propose an approach as an extension of the
original framework of risk parity portfolios. By noting that
skewness risk is related to jump risk, we consider a
regime-switching model with two states: a normal state and a state
where the jumps occur. It follows that asset returns follow a
Gaussian mixture model, which is easy to manipulate. Our model is
related to the econometric literature on Markov switching models
(Hamilton, 1989), without the ambitious nature of forecasting. The
primary objective of the model is to measure the total risk of a
portfolio, including both volatility and skewness risks. For that,
we use the expected shortfall, a coherent and convex risk
measure (thus) satisfying the Euler decomposition. We can
then calculate the risk contribution of assets and define risk
budgeting portfolios. We apply this framework to the equity/bond
asset mix policy when we introduce a short volatility strategy. We
then compare the results with those obtained with a traditional
risk parity model. In particular, we illustrate the skewness
aggregation dilemma found by Hamdan \textsl{et al.} (2016) and
discuss whether the skewness risk can be hedged like the volatility
risk.\bigskip

This article is organized as follows. In section two, we describe
the skewness model of asset returns, and analyze the relationship
between jump and skewness risks. In section three, we provide
analytical formulas of the expected shortfall and the corresponding
risk contributions. We also define risk budgeting portfolios and
study the existence/uniqueness problem. The application to the
equity/bond/volatility asset mix policy is presented in section
four. Section five deals with the allocation of risk factors and
alternative risk premia. Finally, section six offers our concluding
remarks.

\section{A skewness model of asset returns}

In this section, we develop a simple model for modeling
skewness. We consider a standard Lévy process and
approximate its density by a mixture of two Gaussian
distributions. The mixture representation presents some appealing
properties for computational purposes, whereas the jump-diffusion
representation helps in understanding the dynamics of the model. Moreover, this
framework highlights the relationship between skewness risk and jump
risk.

\subsection{A model of asset returns}

\subsubsection{The jump-diffusion representation}

We consider $n$ risky assets represented by the vector of prices $%
S_{t}=\left( S_{1,t},\ldots ,S_{n,t}\right) $. We note $B_{t}$ the
price of the risk-free asset. We assume that asset prices follow a
diffusion, which is governed by a standard Lévy process:
\begin{equation*}
\left\{
\begin{array}{l}
\mathrm{d}B_{t}=rB_{t}\,\mathrm{d}t \\
\mathrm{d}S_{t}=\limfunc{diag}\left( S_{t}\right) \,\mathrm{d}L_{t} \\
\mathrm{d}L_{t}=\mu \,\mathrm{d}t+\Sigma ^{1/2}\,\mathrm{d}W_{t}+\mathrm{d}%
Z_{t}%
\end{array}%
\right.
\end{equation*}%
where $r$ is the interest rate, $\mu $ and $\Sigma $ are the vector
of expected returns and the covariance matrix, $W_{t}$ is a
$n$-dimensional standard Brownian motion and $Z_{t}$ is a pure
$n$-dimensional jump process. We also assume that the components of
$W_{t}$ and $Z_{t}$ are mutually independent.\bigskip

In order to keep the model simple, the jump process $Z_{t}$ is set
to a compound Poisson process with a finite number of jumps:
\begin{equation*}
Z_{t}=\sum_{i=1}^{N_{t}}Z_{i}
\end{equation*}%
where $N_{t}$ is a scalar Poisson process with constant intensity
parameter $\lambda >0$, and $Z_{1},\ldots ,Z_{N_{t}}$ are vectors of
\textit{i.i.d.} random jump amplitudes with law $\nu \left(
\mathrm{d}z\right) $. Using standard assumptions, we have $\nu
\left( \mathrm{d}z\right) =\lambda f\left( z\right) \mathrm{d}z$
where $f\left( z\right) $ is the probability
density function of the multivariate Gaussian distribution $\mathcal{N}%
\left( \tilde{\mu},\tilde{\Sigma}\right) $, $\tilde{\mu}$ is the
expected value of jump amplitudes and $\tilde{\Sigma}$ is the
associated  covariance matrix. In Appendix
\ref{appendix:characteristic-function}, we show that the
characteristic function of asset returns $R_{t}=\left(
R_{1,t},\ldots ,R_{n,t}\right) $ for the holding period
$\mathrm{d}t$ may be approximated with the following expression:
\begin{equation}
\mathbb{E}\left[ e^{-iu.R_{t}}\right] \approx \left( 1-\lambda \,\mathrm{d}%
t\right) \cdot e^{\left( iu^{\top }\mu -\frac{1}{2}u^{\top }\Sigma
u\right) \,\mathrm{d}t}+\left( \lambda \,\mathrm{d}t\right) \cdot
e^{iu^{\top }\left( \mu \,\mathrm{d}t+\tilde{\mu}\right)
-\frac{1}{2}u^{\top }\left( \Sigma
\,\mathrm{d}t+\tilde{\Sigma}\right) u}  \label{eq:cf1}
\end{equation}

\subsubsection{The Gaussian mixture representation}

For the sake of simplicity, we make the following assumptions: jumps
occur simultaneously for all assets; portfolio's rebalancing is
done at frequency $\mathrm{d}t$; between two rebalancing dates, the
assets can jump with probability $\lambda \,\mathrm{d}t$. Let
$R_{t}=\left( R_{1,t},\ldots ,R_{n,t}\right) $ be the asset returns
for the holding period $\mathrm{d}t$. Using the previous
assumptions, we consider a Gaussian mixture model with two regimes
to define $R_{t}$:

\begin{itemize}
\item The continuous component, which has probability $\left( 1-\lambda
\,\mathrm{d}t\right) $ to occur, is driven by the Gaussian distribution $%
\mathcal{N}\left( \mu \,\mathrm{d}t,\Sigma \,\mathrm{d}t\right) $;

\item The jump component, which has probability $\lambda \,\mathrm{d}t$
to occur, is driven by the Gaussian distribution $\mathcal{N}\left( \tilde{\mu%
},\tilde{\Sigma}\right) $.
\end{itemize}

\noindent Therefore, asset returns have the following multivariate density function:%
\begin{eqnarray*}
f\left( y\right)  &=&\frac{1-\lambda \,\mathrm{d}t}{\left( 2\pi
\right)
^{n/2}\left\vert \Sigma \,\mathrm{d}t\right\vert ^{1/2}}\exp \left( -\frac{1%
}{2\,}\left( y-\mu \,\mathrm{d}t\right) ^{\top }\left( \Sigma \,\mathrm{d}%
t\right) ^{-1}\left( y-\mu \,\mathrm{d}t\right) \right) + \\
&&\frac{\lambda \,\mathrm{d}t}{\left( 2\pi \right) ^{n/2}\left\vert \Sigma \,%
\mathrm{d}t+\tilde{\Sigma}\right\vert ^{1/2}}\exp \left(
-\frac{1}{2}\left( y-\left( \mu \,\mathrm{d}t+\tilde{\mu}\right)
\right) ^{\top }\left( \Sigma \,\mathrm{d}t+\tilde{\Sigma}\right)
^{-1}\left( y-\left( \mu \,\mathrm{d}t+\tilde{\mu}\right) \right)
\right)
\end{eqnarray*}%
It follows that the characteristic function of $R_{t}$ is equal to:%
\begin{equation}
\mathbb{E}\left[ e^{-iu.R_{t}}\right] =\left( 1-\lambda
\,\mathrm{d}t\right)
\cdot e^{\left( iu^{\top }\mu -\frac{1}{2}u^{\top }\Sigma u\right) \,\mathrm{%
d}t}+\left( \lambda \,\mathrm{d}t\right) \cdot e_{{}}^{iu^{\top
}\left( \mu \,\mathrm{d}t+\tilde{\mu}\right) -\frac{1}{2}u^{\top
}\left( \Sigma \, \mathrm{d}t+\tilde{\Sigma}\right) u }
\label{eq:cf2}
\end{equation}%
We notice that the characteristic functions (\ref{eq:cf1}) and (\ref{eq:cf2}%
) coincide. This justifies the use of a Gaussian mixture
distribution for modeling asset returns.

\begin{remark}
Without loss of generality, we set $\mathrm{d}t$ equal to one in the
sequel. This simplifies the notations without changing the results.
If one prefers to consider another holding period, it suffices to
scale the parameters $\mu$, $\Sigma$ and $\lambda$ by a factor
$\mathrm{d}t$.
\end{remark}

\subsection{Distribution function of the portfolio's return}

Let $x=\left( x_{1},\ldots ,x_{n}\right) $ be the vector of weights
in the
portfolio. We assume that the portfolio is fully invested meaning that $%
\sum_{i=1}^{n}x_{i}=1$. We note $V_{t}$ the value of the portfolio
$x$ at time $t$. Let $n_{i}=x_{i}V_{t}/S_{i,t}$ the number of shares
invested in
asset $i$. We have:%
\begin{align*}
\mathrm{d}V_{t}& =\sum_{i=1}^{n}n_{i}\,\mathrm{d}S_{i,t} \\
& =V_{t}x^{\top }\left( \limfunc{diag}S_{t}\right) ^{-1}\,\mathrm{d}S_{t} \\
& =V_{t}x^{\top }\,\mathrm{d}L_{t}
\end{align*}%
It follows that:%
\begin{equation*}
\frac{\mathrm{d}V_{t}}{V_{t}}=x^{\top }\,\mathrm{d}L_{t}
\end{equation*}
We deduce that the return of the portfolio $R\left( x\right)
=\sum_{i=1}^{n}x_{i}R_{i}=x^{\top }R$ is then a mixture of two
normal random variables:
\begin{equation*}
R\left( x\right) = Y = B_{1}\cdot Y_{1}+B_{2}\cdot Y_{2}
\end{equation*}%
where $B_{1}$ and $B_{2}=1-B_{1}$ are Bernoulli random variables
with probability $\pi _{1}=1-\lambda$ and $\pi _{2}=\lambda$. The
first regime corresponds to the normal case: $Y_{1}\sim
\mathcal{N}\left( \mu _{1}\left( x\right) ,\sigma _{1}^{2}\left(
x\right) \right) $ where $\mu _{1}\left( x\right) =x^{\top }\mu $
and $\sigma _{1}^{2}\left( x\right) =x^{\top }\Sigma x$. The second
regime incorporates the jumps: $Y_{2}\sim \mathcal{N}\left( \mu
_{2}\left( x\right) ,\sigma _{2}^{2}\left( x\right) \right) $ where
$\mu _{2}\left( x\right) =x^{\top }\left( \mu +\tilde{\mu}\right) $
and $\sigma _{2}^{2}\left( x\right) =x^{\top }\left( \Sigma
+\tilde{\Sigma}\right) x$. It follows that the portfolio's return
$R\left( x\right) $ has the following density function:
\begin{eqnarray}
f\left( y\right)  &=&\pi_1 f_{1}\left( y\right) + \pi_2
f_{2}\left( y\right)   \notag \\
&=&\left( 1-\lambda \right) \frac{1}{\sigma _{1}\left( x\right)
}\phi \left( \frac{y-\mu _{1}\left( x\right) }{\sigma _{1}\left( x\right) }%
\right) + \lambda \frac{1}{\sigma _{2}\left( x\right) }\phi \left(
\frac{y-\mu _{2}\left( x\right) }{\sigma _{2}\left( x\right)
}\right) \label{eq:jump1}
\end{eqnarray}%
\bigskip

\begin{figure}[tbph]
\centering
\caption{Probability density function of $R\left(x\right)$ when $\lambda = 0.20$}
\label{fig:density1}
\figureskip
\includegraphics[width = \figurewidth, height = \figureheight]{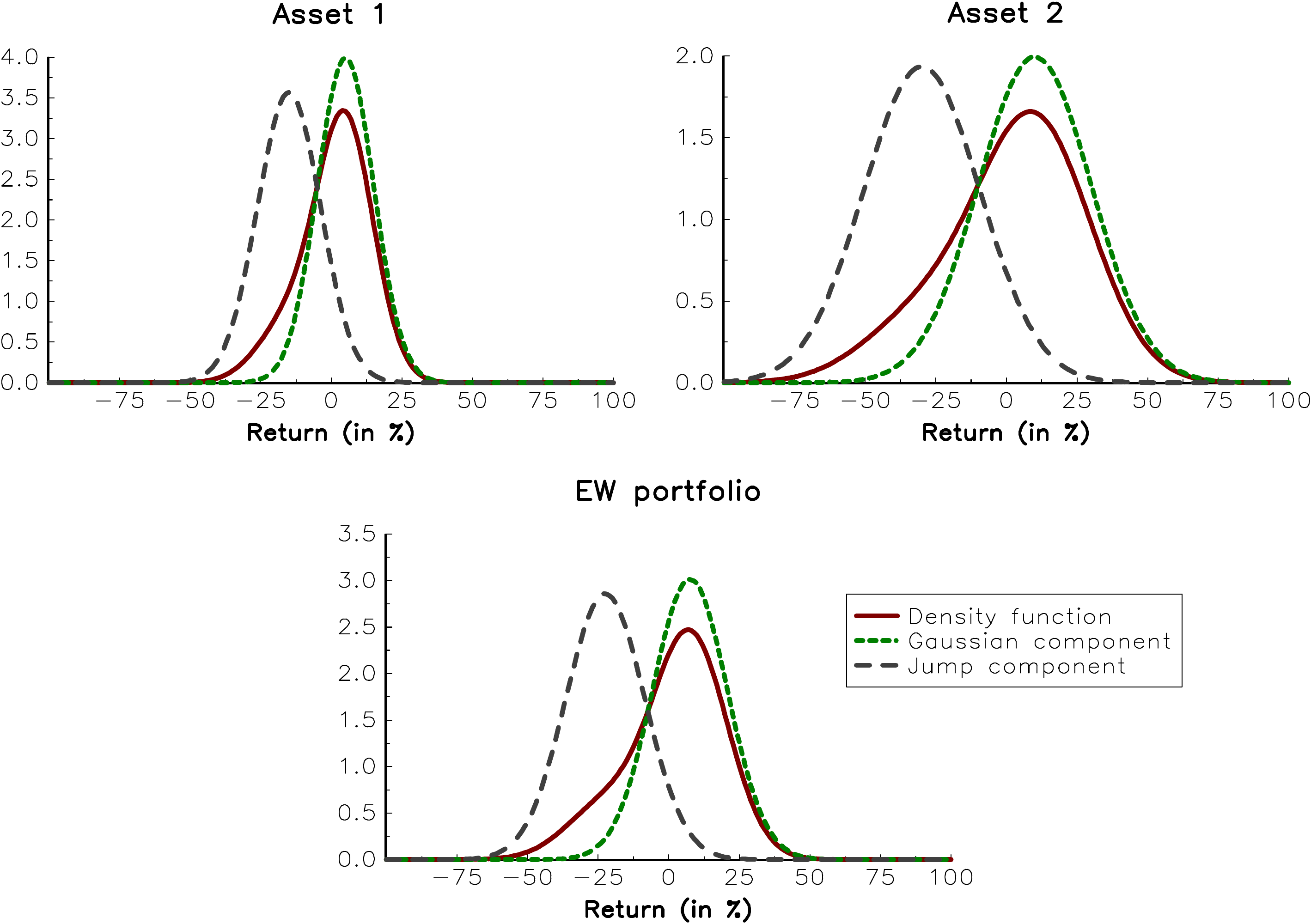}
\end{figure}

\begin{figure}[tbph]
\centering
\caption{Probability density function of $R\left(x\right)$ for several values of $\lambda$}
\label{fig:density2}
\figureskip
\includegraphics[width = \figurewidth, height = \figureheight]{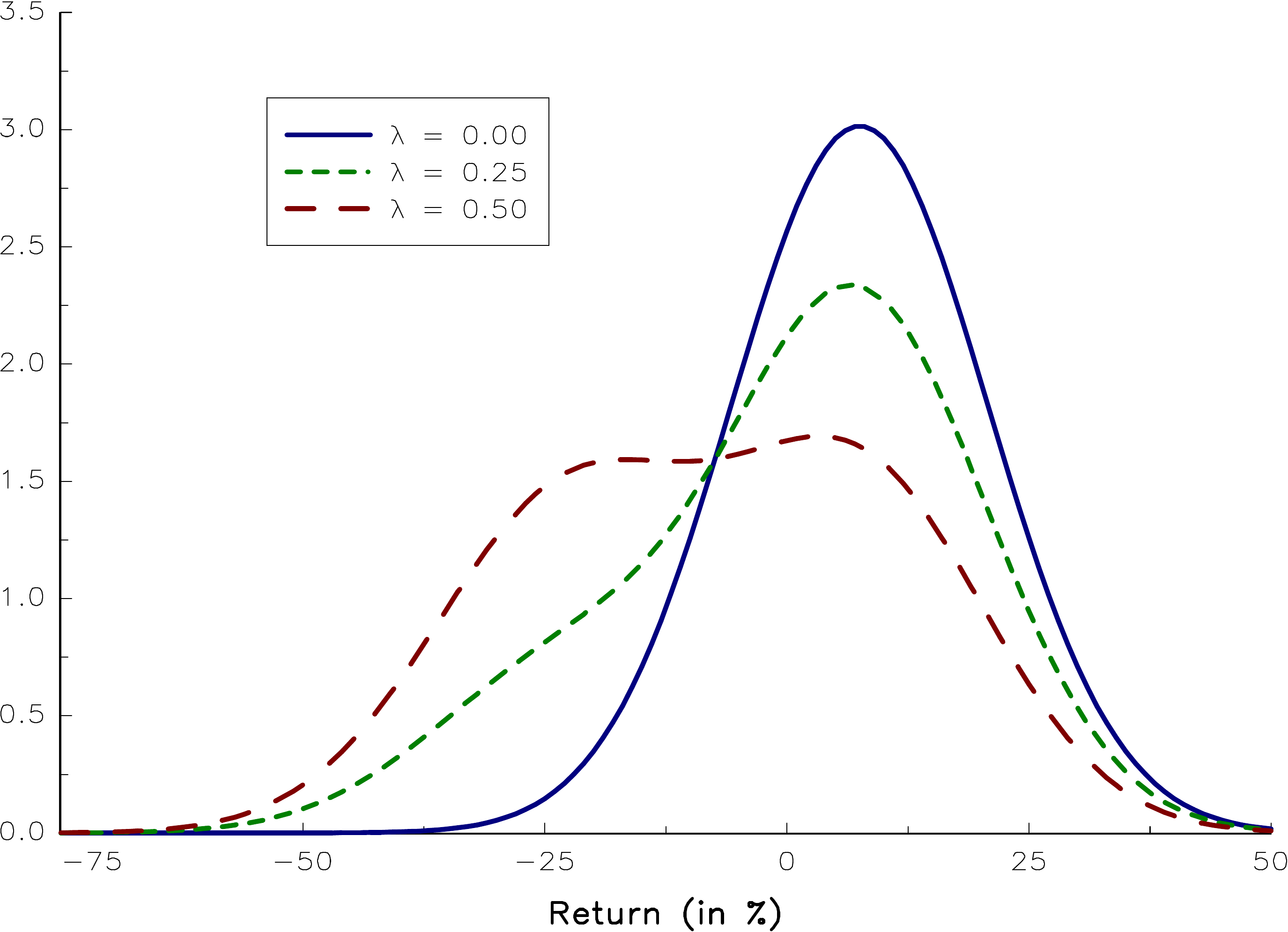}
\end{figure}

We consider a universe of two assets with the following
characteristics: $\mu_1 = 5\%$, $\sigma_1 = 10\%$, $\tilde{\mu}_1 =
-20\%$, $\tilde{\sigma}_1 = 5\%$, $\mu_2 = 10\%$, $\sigma_2 = 20\%$,
$\tilde{\mu}_2 = -40\%$ and $\tilde{\sigma}_2 = 5\%$. The
cross-correlations are set to $\rho = 50\%$ and $\tilde{\rho} =
60\%$. When the jump frequency $\lambda$ is equal to $20\%$, the
probability density function of asset returns and portfolio's
returns is given in Figure \ref{fig:density1}. The shape of the
distribution function depends highly on the parameter $\lambda$.
Indeed, if this parameter is equal to $0$ or $1$, the portfolio's
return is Gaussian. For $\lambda \in \left] 0,1\right[$, the
distribution function exhibits skewness and kurtosis as shown in
Figure \ref{fig:density2}.

\subsection{Relationship between jump risk and skewness risk}

Using Equation (\ref{eq:appendix-skewness1}) found in Appendix
\ref{appendix:skewness}, we deduce the skewness of the
portfolio's return $R\left( x\right) $ is equal to:
\begin{equation*}
\gamma _{1} =\frac{\left( \lambda -\lambda ^{2}\right) \left( \left(
1-2\lambda \right) \left( x^{\top }\tilde{\mu}\right) ^{3}+3\left(
x^{\top }\tilde{\mu}\right) \left( x^{\top }\tilde{\Sigma}x\right) \right) }{%
\left( x^{\top }\Sigma x+\lambda x^{\top }\tilde{\Sigma}x+\left( \lambda
-\lambda ^{2}\right) \left( x^{\top }\tilde{\mu}\right) ^{2}\right)^{\nicefrac{3}{2}}}
\end{equation*}%
It follows that the model exhibits skewness, except in some limit
cases. Indeed, we have:
\begin{equation*}
\gamma _{1} =0\Leftrightarrow x^{\top }\tilde{\mu}=0\text{ or
}\lambda =0\text{ or }\lambda =1
\end{equation*}%
In the other cases, $\gamma _{1}$ may be positive or
negative depending on the value of the parameters:%
\begin{equation*}
\limfunc{sgn}\gamma _{1} =\limfunc{sgn}\left( 1-2\lambda
\right) \left( x^{\top }\tilde{\mu}\right) ^{3}+3\left( x^{\top }\tilde{\mu}%
\right) \left( x^{\top }\tilde{\Sigma}x\right)
\end{equation*}%
If the expected value $x^{\top }\tilde{\mu}$ of the portfolio's jump
is positive, the skewness is then positive. If $x^{\top
}\tilde{\mu}<0$, the skewness is generally negative, except if the
frequency $\lambda $ is high and the variance $x^{\top
}\tilde{\Sigma}x$ of the portfolio's jump is low.\bigskip

\begin{figure}[tbph]
\centering
\caption{Skewness coefficient of the portfolio's return $R\left(x\right)$}
\label{fig:skewness3}
\figureskip
\includegraphics[width = \figurewidth, height = \figureheight]{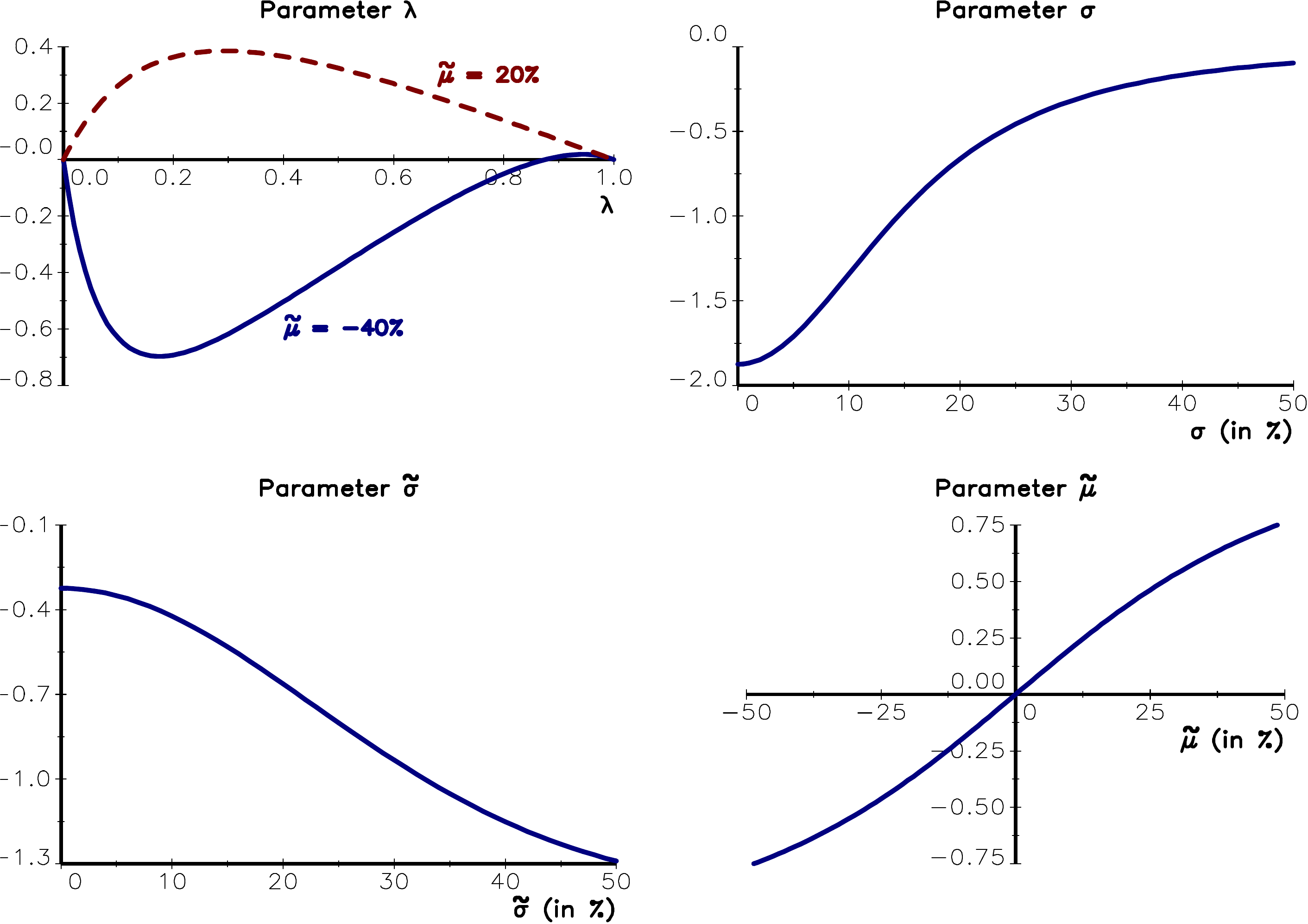}
\end{figure}

Let us consider a portfolio with only one asset\footnote{%
We have $x=1$.}. The parameters are equal to\footnote{%
We notice that the skewness does not depend on the parameter $\mu
$.} $\sigma =20\%$, $\tilde{\mu}=-40\%$, $\tilde{\sigma}=20\%$ and
$\lambda =25\%$. In this case, we obtain $\gamma _{1} = -0.663$. In
Figure \ref{fig:skewness3}, we show the evolution of the skewness
coefficient with respect to a given parameter while the other
parameters remain constant.
Skewness coefficient increases with the normal volatility\footnote{%
When the skewness is positive, it decreases with the normal
volatility.} (top-right panel). That's because all else being equal
if the normal volatility increases, jumps get more and more
indistinct from normal movements of returns. We observe that the
jump's volatility generally increases the magnitude of the skewness
(bottom-left panel). The impact of the frequency $\lambda $ is more
difficult to analyze, but we generally obtain a bell curve or an
inverted bell curve (top-left panel).\bigskip

These results are in line with those found by Hamdan \textsl{et al.}
(2016) on the skewness risk of alternative risk premia. These authors
observed that the skewness is maximum in absolute value when the
portfolio's volatility is low. We now understand the mechanism
behind this stylized fact. When the volatility is high, jumps have a
moderate impact on the probability distribution of asset returns,
because they may appear as the realization of a high volatility
regime. When the volatility is low, jumps substantially change the
probability distribution and are not compatible with the normal
dynamics of asset returns. In particular, we have:%
\begin{equation*}
\lim_{\sigma _{1}\left( x\right) \rightarrow 0}\gamma _{1}=\Delta
_{\mu }\left( x\right) \cdot \frac{\left( \lambda -\lambda
^{2}\right) \left( \left( 1-2\lambda \right) \Delta _{\mu }\left(
x\right) ^{2}+3\Delta _{\sigma }\left( x\right) ^{2}\right) }{\left(
\left( \lambda -\lambda ^{2}\right) \Delta _{\mu }\left( x\right)
^{2}+\lambda \Delta _{\sigma }\left( x\right) ^{2}\right)
^{\nicefrac{3}{2}}}
\end{equation*}%
where $\Delta _{\mu }\left( x\right) =\mu _{2}\left( x\right) -\mu
_{1}\left( x\right) =x^{\top }\tilde{\mu}$\ and $\Delta _{\sigma
}\left( x\right) =\sqrt{\sigma _{2}^{2}\left( x\right) -\sigma
_{1}^{2}\left( x\right) }=\sqrt{x^{\top }\tilde{\Sigma}x}$ are the
portfolio's expected return and volatility due to the jump
component. In this case, the (negative) skewness risk is an
increasing function of $\Delta _{\mu }\left( x\right) $ and $\Delta
_{\sigma }\left( x\right) $.

\subsection{Estimation of the parameters}

Our Gaussian mixture model contains 5 parameters:%
\begin{equation*}
\theta =\left( \lambda ,\mu ,\Sigma
,\tilde{\mu},\tilde{\Sigma}\right)
\end{equation*}%
A first idea to estimate $\theta $ is to use the method of maximum
likelihood. We consider a sample $\left\{ R_{t},t=1,\ldots
,T\right\} $ of asset returns, where $R_{t}=\left( R_{1,t},\ldots
,R_{n,t}\right) $ is the vector of asset returns observed at time
$t$ and $R_{i,t}$ is the return of the asset $i$ for the same
period. The log-likelihood function is then equal
to:%
\begin{equation*}
\ell \left( \theta \right) =\sum_{t=1}^{T}\ln f\left( R_{t}\right)
\end{equation*}%
where $f\left( y\right) $ is the multivariate probability density function%
\footnote{%
Recall that $\pi _{1}=1-\lambda $, $\pi _{2}=\lambda $ and $\phi
_{n}\left( y;\mu ,\Sigma \right) $ is the probability density
function of the Gaussian distribution $\mathcal{N}\left( \mu ,\Sigma
\right) $.}:
\begin{equation*}
f\left( y\right) =\pi _{1}\phi _{n}\left( y;\mu ,\Sigma \right) +\pi
_{2}\phi _{n}\left( y;\mu +\tilde{\mu},\Sigma +\tilde{\Sigma}\right)
\end{equation*}%
However, maximizing the log-likelihood function is a difficult task
using standard numerical optimization procedure. A better approach
is to consider the EM algorithm of Dempster \textsl{et al.} (1977),
which is used extensively for such problems (Xu and Jordan,
1996).\bigskip

Following Redner and Walker (1984), we introduce the notations: $\mu
_{1}=\mu $, $\Sigma _{1}=\Sigma $, $\mu _{2}=\mu +\tilde{\mu}$,
$\Sigma _{2}=\Sigma +\tilde{\Sigma}$. The log-likelihood function
becomes:
\begin{equation*}
\ell \left( \theta \right) =\sum_{t=1}^{T}\ln \sum_{j=1}^{2}\pi
_{j}\phi _{n}\left( R_{t};\mu _{j},\Sigma _{j}\right)
\end{equation*}%
Let $\pi _{j,t}$ be the posterior probability of the regime $j$ at
time $t$. We
have:%
\begin{eqnarray*}
\pi _{j,t} &=&\Pr \left\{ B_{j}=1\mid R_{t}\right\}  \\
&=&\frac{\pi _{j}\phi _{n}\left( R_{t};\mu _{j},\Sigma _{j}\right) }{%
\sum_{s=1}^{2}\pi _{s}\phi _{n}\left( R_{t};\mu _{s},\Sigma
_{s}\right) }
\end{eqnarray*}%
In Appendix \ref{appendix:em-algorithm}, we derive the following EM
algorithm:

\begin{enumerate}
\item The E-step consists in updating the posterior distribution of $B_{j}$
for all the observations:%
\begin{equation*}
\pi _{j,t}^{\left( k\right) }=\frac{\pi _{j}^{\left( k\right) }\phi
_{n}\left( R_{t};\mu _{j}^{\left( k\right) },\Sigma _{j}^{\left(
k\right) }\right) }{\sum_{s=1}^{2}\pi _{s}^{\left( k\right) }\phi
_{n}\left( R_{t};\mu _{s}^{\left( k\right) },\Sigma _{s}^{\left(
k\right) }\right) }
\end{equation*}

\item The M-step consists in updating the estimators $\hat{\pi}_{j}$, $\hat{%
\mu}_{j}$ and $\hat{\Sigma}_{j}$:%
\begin{eqnarray*}
\pi _{j}^{\left( k+1\right) } &=&\frac{\sum_{t=1}^{T}\pi
_{j,t}^{\left(
k\right) }}{T} \\
\mu _{j}^{\left( k+1\right) } &=&\frac{\sum_{t=1}^{T}\pi
_{j,t}^{\left(
k\right) }R_{t}}{\sum_{t=1}^{T}\pi _{j,t}^{\left( k\right) }} \\
\Sigma _{j}^{\left( k+1\right) } &=&\frac{\sum_{t=1}^{T}\pi
_{j,t}^{\left( k\right) }\left( R_{t}-\mu _{j}^{\left( k+1\right)
}\right) \left( R_{t}-\mu _{j}^{\left( k+1\right) }\right) ^{\top
}}{\sum_{t=1}^{T}\pi _{j,t}^{\left( k\right) }}
\end{eqnarray*}
\end{enumerate}

\noindent Starting from initial values $\pi _{j}^{\left( 0\right)
}$, $\mu _{j}^{\left( 0\right) }$ and $\Sigma _{j}^{\left( 0\right)
}$, the EM algorithm generally converges and we have
$\hat{\pi}_{j}=\pi _{j}^{\left(
\infty \right) }$, $\hat{\mu}_{j}=\pi _{j}^{\left( \infty \right) }$ and $%
\hat{\Sigma}_{j}=\Sigma _{j}^{\left( \infty \right) }$.

\section{Risk parity portfolios with jumps}

As shown in the previous section, jumps occur rarely and their
impacts mainly affect the tail of the distribution. It is thus natural
to change the volatility risk measure in order to perform
portfolio allocation. As value-at-risk is not a convex risk
measure, expected shortfall is a better alternative.

\subsection{The expected shortfall risk measure}

The expected shortfall of Portfolio $x$ is defined by:%
\begin{equation*}
\limfunc{ES}\nolimits_{\alpha }\left( x\right) =\frac{1}{1-\alpha }%
\int_{\alpha }^{1}\limfunc{VaR}\nolimits_{u}\left( x\right) \,\mathrm{d}u
\end{equation*}%
Acerbi and Tasche (2002) interpreted the expected shortfall as the average
of the value-at-risk at level $\alpha $ and higher. We also notice that it
is equal to the expected loss given that the loss is beyond the
value-at-risk:
\begin{equation*}
\limfunc{ES}\nolimits_{\alpha }\left( x\right) =\mathbb{E}\left[ L\left(
x\right) \mid L\left( x\right) \geq \limfunc{VaR}\nolimits_{\alpha }\left(
x\right) \right]
\end{equation*}%
where $L\left( x\right) =-R\left( x\right) $ is the portfolio's
loss. Using results in Appendix \ref{appendix:expected-shortfall},
we show that the expression of the expected shortfall when the
portfolio's return $R\left( x\right) $ follows the distribution
function (\ref{eq:jump1}) is equal to:
\begin{equation}
\func{ES}_{\alpha }\left( x\right) =\left( 1-\lambda\right) \cdot
\varphi \left( \limfunc{VaR}\nolimits_{\alpha }\left( x\right) ,\mu
_{1}\left( x\right) ,\sigma _{1}\left( x\right) \right) + \lambda
\cdot \varphi \left( \limfunc{VaR}\nolimits_{\alpha }\left( x\right)
,\mu _{2}\left( x\right) ,\sigma _{2}\left( x\right) \right)
\label{eq:es1}
\end{equation}%
where the function $\varphi \left( a,b,c\right) $ is defined by:%
\begin{equation*}
\varphi \left( a,b,c\right) =\frac{c}{1-\alpha }\phi \left( \frac{a+b}{c}%
\right) -\frac{b}{1-\alpha }\Phi \left( -\frac{a+b}{c}\right)
\end{equation*}%
Here, the value-at-risk $\limfunc{VaR}\nolimits_{\alpha }\left( x\right) $
is found by solving the following equation:%
\begin{equation*}
\left( 1-\lambda \right) \cdot \Phi \left( \frac{\limfunc{VaR}%
\nolimits_{\alpha }\left( x\right) +\mu _{1}\left( x\right) }{\sigma
_{1}\left( x\right) }\right) + \lambda \cdot \Phi \left(
\frac{\limfunc{VaR}\nolimits_{\alpha }\left( x\right) +\mu
_{2}\left( x\right) }{\sigma _{2}\left( x\right) }\right) =\alpha
\end{equation*}

\subsection{Analytical expression of risk contributions}

Using results in Appendix \ref{appendix:marginal-expected-shortfall}, we
show that the vector of risk contributions is equal to :
\begin{eqnarray}
\mathcal{RC}\left( x\right)  &=&\frac{\varpi _{1}\left( x\right) }{1-\alpha }%
\left( x\circ \delta _{1}\left( x\right) \right) +\frac{\varpi _{2}\left(
x\right) }{1-\alpha }\left( x\circ \delta _{2}\left( x\right) \right) - \notag \\
&&\frac{1}{1-\alpha }\left( \left( 1-\lambda \right) \left( x\circ
\mu \right) \Phi \left( -h_{1}\left( x\right) \right) + \lambda
 \left( x\circ \left( \mu +\tilde{\mu}\right)
\right) \Phi \left( -h_{2}\left( x\right) \right) \right) \label{eq:rc1}
\end{eqnarray}%
where $\circ$ is the Hadamard product,
\begin{eqnarray*}
\delta _{1}\left( x\right)  &=&\left( 1+\frac{h_{1}\left( x\right) }{\sigma
_{1}\left( x\right) }\limfunc{VaR}\nolimits_{\alpha }\left( x\right) \right)
\Sigma x- \\
&&\limfunc{VaR}\nolimits_{\alpha }\left( x\right) \frac{\varpi _{1}\left(
x\right) \frac{h_{1}\left( x\right) }{\sigma _{1}\left( x\right) }\Sigma
x+\varpi _{2}\left( x\right) \left( \frac{h_{2}\left( x\right) }{\sigma
_{2}\left( x\right) }\left( \Sigma +\tilde{\Sigma}\right) x-\tilde{\mu}%
\right) }{\varpi _{1}\left( x\right) +\varpi _{2}\left( x\right) }
\end{eqnarray*}%
and:%
\begin{eqnarray*}
\delta _{2}\left( x\right)  &=&\left( 1+\frac{h_{2}\left( x\right) }{\sigma
_{2}\left( x\right) }\limfunc{VaR}\nolimits_{\alpha }\left( x\right) \right)
\left( \Sigma +\tilde{\Sigma}\right) x- \\
&&\limfunc{VaR}\nolimits_{\alpha }\left( x\right) \frac{\varpi _{1}\left(
x\right) \left( \frac{h_{1}\left( x\right) }{\sigma _{1}\left( x\right) }%
\Sigma x+\tilde{\mu}\right) +\varpi _{2}\left( x\right) \frac{h_{2}\left(
x\right) }{\sigma _{2}\left( x\right) }\left( \Sigma +\tilde{\Sigma}\right) x%
}{\varpi _{1}\left( x\right) +\varpi _{2}\left( x\right) }
\end{eqnarray*}%
The other notations are $h_{i}\left( x\right) =\sigma _{i}\left(
x\right) ^{-1}\left( \limfunc{VaR}\nolimits_{\alpha }\left( x\right)
+\mu _{i}\left( x\right) \right) $, $\varpi _{i}\left( x\right) =\pi
_{i}\sigma _{i}\left( x\right) ^{-1}\phi \left( h_{i}\left( x\right)
\right) $ where $\pi _{1}=1-\lambda$ and $\pi _{2}=\lambda$.

\begin{example}
\label{example1} We consider three assets, whose expected returns
are equal to $10\%$, $15\%$ and $20\%$. Their volatilities are equal
to $20\%$, $25\%$ and $30\%$ while the
correlation matrix of asset returns is provided by the following matrix:%
\begin{equation*}
\rho =\left(
\begin{array}{ccc}
1.00 &      &       \\
0.50 & 1.00 &       \\
0.20 & 0.40 & 1.00  \\
\end{array}%
\right)
\end{equation*}
For the jumps, we assume that $\tilde{\mu}_i = -10\%$,
$\tilde{\sigma}_i = 20\%$ and $\tilde{\rho}_{i,j} = 50\%$.
Moreover, the intensity $\lambda$ of jumps is equal to $0.25$, meaning
that we observe a jump every four years on average.
\end{example}

We calculate the one-year expected shortfall at the $95\%$
confidence level. Results (expressed in $\%$) are reported in Tables
\ref{tab:example1-1} and \ref{tab:example1-2}. The first table
corresponds to the traditional Gaussian expected shortfall when we
do not take into account the jumps. In this case, the risk is equal
to $28.46\%$ of the portfolio's value. By considering jumps, the
expected shortfall increases and is equal to $37.22\%$. The
introduction of jumps has also modified the risk decomposition. For
instance, the risk contribution of Asset 1 has increased whereas
this of Asset 3 has been reduced.

\begin{table}[tbph]
\centering
\caption{Expected shortfall decomposition (without jumps)}
\label{tab:example1-1}
\tableskip
\begin{tabular}{|c|cccc|}
\hline
Asset & $x_i$ & $\mathcal{MR}_i$ & $\mathcal{RC}_i$ & $\mathcal{RC}_i^{\star}$ \\ \hline
$1$ & $20.00$ & ${\TsV}8.90$ & ${\TsV}1.78$ & ${\TsV}6.26$ \\
$2$ & $20.00$ &      $18.22$ & ${\TsV}3.64$ &      $12.80$ \\
$3$ & $60.00$ &      $38.40$ &      $23.04$ &      $80.94$ \\ \hline
\multicolumn{3}{|l}{$\func{ES}_{\alpha }\left( x\right)$} & $28.46$ & \\ \hline
\end{tabular}
\end{table}

\begin{table}[tbph]
\centering
\caption{Expected shortfall decomposition (with jumps)}
\label{tab:example1-2}
\tableskip
\begin{tabular}{|c|cccc|}
\hline
Asset & $x_i$ & $\mathcal{MR}_i$ & $\mathcal{RC}_i$ & $\mathcal{RC}_i^{\star}$ \\ \hline
$1$ & $20.00$ & $20.39$ & ${\TsV}4.08$ & $10.96$ \\
$2$ & $20.00$ & $27.31$ & ${\TsV}5.46$ & $14.67$ \\
$3$ & $60.00$ & $46.13$ &      $27.68$ & $74.37$ \\ \hline
\multicolumn{3}{|l}{$\func{ES}_{\alpha }\left( x\right)$} & $37.22$ & \\ \hline
\end{tabular}
\end{table}

\begin{figure}[tbph]
\centering
\caption{Evolution of risk contributions (in \%) with respect to the intensity $\lambda$}
\label{fig:example2}
\figureskip
\includegraphics[width = \figurewidth, height = \figureheight]{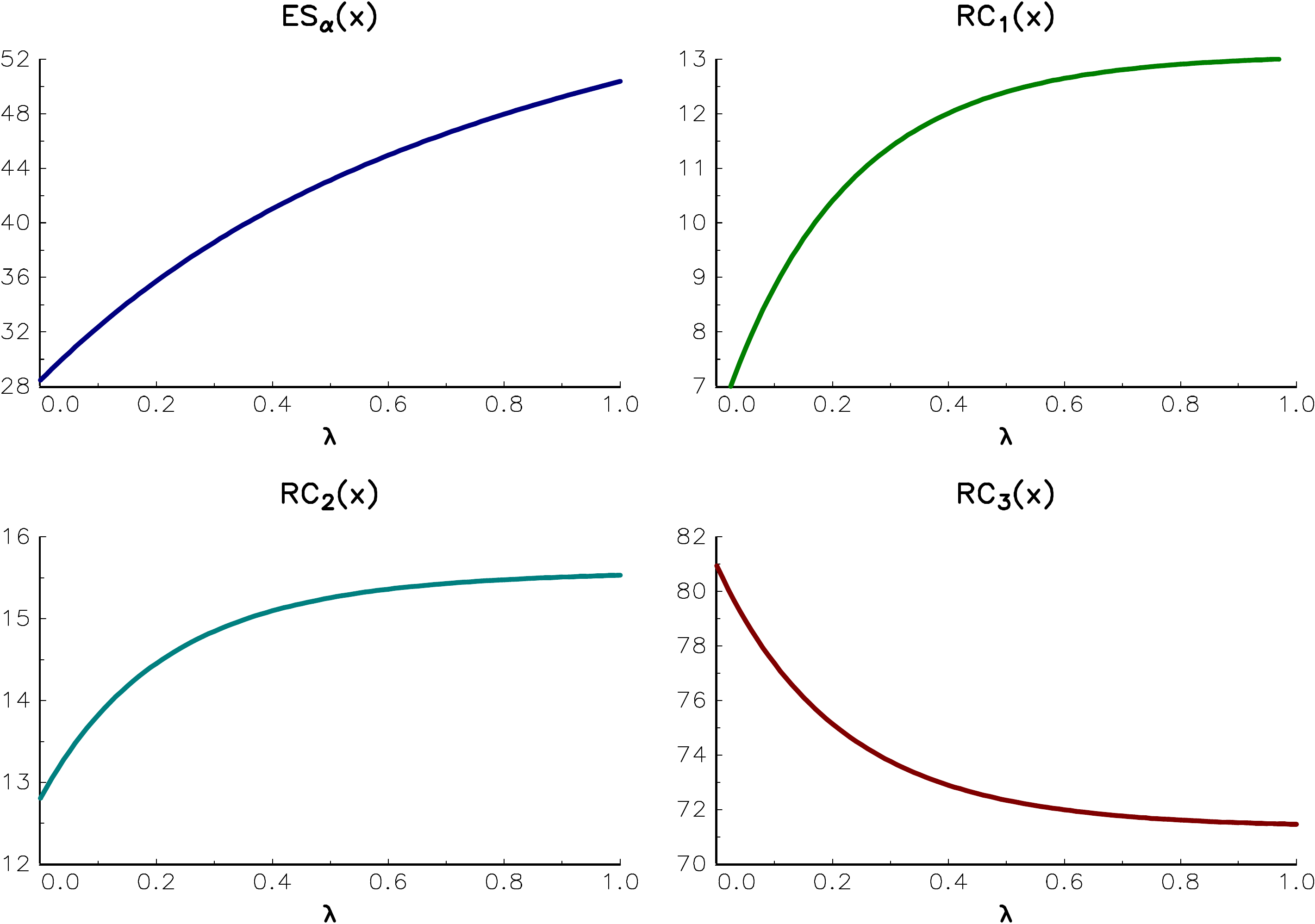}
\end{figure}

\subsection{Risk budgeting portfolios}

Roncalli (2013) defines the RB portfolio using the following non-linear
system:%
\begin{equation}
\left\{
\begin{array}{l}
\mathcal{RC}_{i}\left( x\right) =b_{i}\mathcal{R}\left( x\right) \\
b_{i}>0 \\
x_{i}\geq 0 \\
\sum_{i=1}^{n}b_{i}=1 \\
\sum_{i=1}^{n}x_{i}=1%
\end{array}%
\right.  \label{eq:rb1}
\end{equation}%
where $b_{i}$ is the risk budget of asset $i$ expressed in relative terms.
The constraint $b_{i}>0$ implies that no risk budget can be set to
zero. This restriction is necessary in order to ensure that the RB portfolio
is unique (Roncalli, 2013).\bigskip

The program (\ref{eq:rb1}) is valid for any coherent and convex risk
measure. Therefore, we can use the expected shortfall, whose
expression is given in Equation (\ref{eq:es1}) and risk contributions are defined
by Formula (\ref{eq:rc1}).

\subsubsection{Existence and uniqueness of the RB portfolio}
\label{section:uniqueness}

In order to ensure the existence of the RB portfolio, we have to impose the
following restriction:%
\begin{equation*}
\mathcal{R}\left( x\right) =\limfunc{ES}\nolimits_{\alpha }\left( x\right)
\geq 0
\end{equation*}%
where $\mathcal{R}\left( x\right) $ is the risk measure and corresponds to
the expected shortfall in our case. Indeed, a coherent convex risk measure
satisfies the homogeneity property $\mathcal{R}\left( \delta x\right)
=\delta \mathcal{R}\left( x\right) $ where $\delta $ is a positive scalar.
Suppose that there is a portfolio $x\in \left[ 0,1\right] ^{n}$ such that $%
\mathcal{R}\left( x\right) <0$. We can then leverage the portfolio by a
scaling factor $\delta >1$, and we obtain $\mathcal{R}\left( \delta
x\right) <\mathcal{R}\left( x\right) <0$. It follows that$%
\lim\limits_{\delta\rightarrow \infty }\mathcal{R}\left( \delta
x\right) =-\infty $. This is why it is necessary for the risk
measure to always be positive. As Acerbi and Tasche (2002) proved
that the expected shortfall is a coherent convex risk measure,
$\limfunc{ES}\nolimits_{\alpha }\left( x\right) \geq 0$ must be
satisfied to ensure the existence of the RB portfolio.\bigskip

In Appendix \ref{appendix:uniqueness}, we study the existence
problem. Let $\alpha^{-}$ be the root of the equation below:
\begin{equation*}
\frac{1-\lambda }{1-\alpha ^{-}}\phi \left( \Phi ^{-1}\left( \frac{%
\alpha ^{-}-\lambda }{1-\lambda }\right) \right) +\lambda \Phi
^{-1}\left( \frac{\alpha ^{-}-\lambda }{1-\lambda }\right) = \left(
1+\lambda \right) \func{SR}\nolimits_{1}^{+}
\end{equation*}%
where $\func{SR}\nolimits_{1}^{+}$ is the maximum Sharpe ratio under
the first regime. We obtain two cases that depend on the parameter
$\alpha^{-}$. In particular, we obtain the following theorem:
\begin{theorem}
If $\alpha \geq \max \left( \alpha ^{-},\lambda\right) $, the RB
portfolio exists and is unique. It is the solution of the following
optimization program:
\begin{eqnarray}
x^{\star }\left( c\right)  &=&\arg \min \limfunc{ES}\nolimits_{\alpha
}\left( x\right)   \label{eq:rb2} \\
&\text{u.c.}&\left\{
\begin{array}{l}
\sum_{i=1}^{n}b_{i}\ln x_{i}\geq c \\
1^{\top }x=1 \\
c\geq \mathbf{0}%
\end{array}%
\right.   \notag
\end{eqnarray}%
where $c$ is a constant to be determined.
\end{theorem}

\begin{figure}[tbph]
\centering
\caption{Relationship between $\lambda$, $\func{SR}\nolimits_{1}^{+}$ and $\alpha^{-}$}
\label{fig:exist2}
\figureskip
\includegraphics[width = \figurewidth, height = \figureheight]{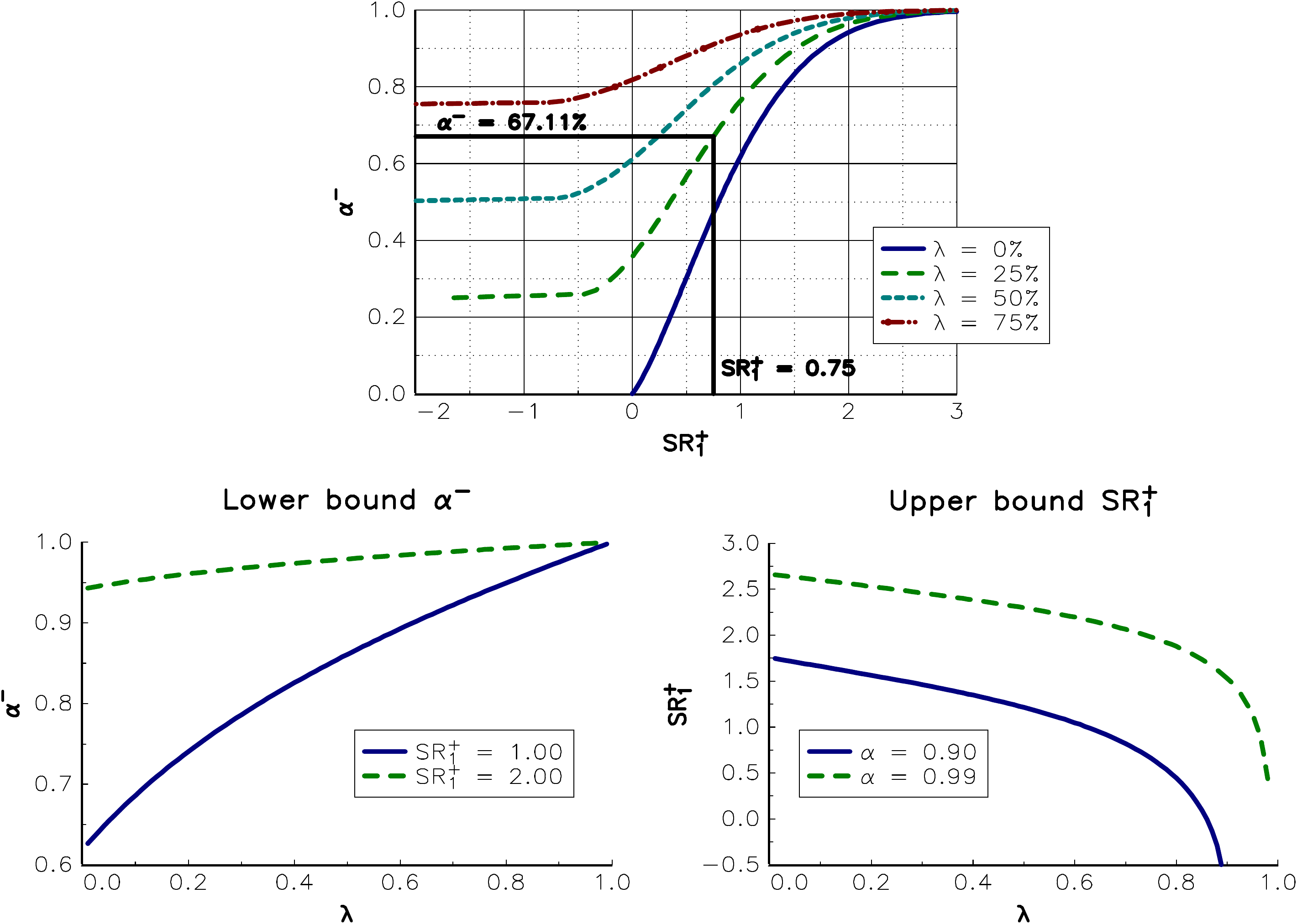}
\end{figure}

This theorem states that the RB portfolio may not exist when $\alpha
\leq \max \left( \alpha ^{-},\lambda\right) $. In general, $\alpha$
is equal to $99\%$ and $\lambda$ is lower than $25\%$. This case
corresponds to a jump every four years. Therefore, the stronger
constraint is $\alpha \geq \alpha ^{-}$. In Figure \ref{fig:exist2},
we report the relationship between $\alpha^{-}$ and the parameters
$\lambda$ and $\func{SR}\nolimits_{1}^{+}$. We notice that
$\alpha^{-}$ is an increasing function of the maximum Sharpe ratio.
For instance, if $\lambda$ is equal to $25\%$, the lower bound
$\alpha^{-}$ is equal to $67.11\%$ when the maximum Sharpe ratio is
equal to $75\%$. In this case, the existence condition is equivalent
to:
\begin{equation*}
\alpha \geq 67.11\%
\end{equation*}%
or:
\begin{equation*}
\limfunc{SR}\nolimits_{1}^{+}\leq 0.75
\end{equation*}%
The worst scenario occurs when the Sharpe ratio of the tangency
portfolio is high. In this case, the expected shortfall may be
negative, meaning that there is no risk, except if we consider a
high confidence level. For example, when we consider a
trend-following strategy, the maximum Sharpe ratio may be high at
some periods due to important trends. However, in most cases, the
vector of expected returns $\mu$ is equal to zero, because the
portfolio manager has no views on the future performance of the
assets. Therefore, his objective is to build a diversified
portfolio. In this case, the maximum Sharpe ratio is equal to zero,
implying that there is always a unique solution to the optimization
program\footnote{Because we assume that $\alpha \gg \lambda$.}.

\subsubsection{Some examples}

To find the numerical solution of the RB portfolio, we use the
following optimization program (Roncalli, 2013; Spinu, 2013):
\begin{eqnarray*}
y^{\star } &=&\arg \min \limfunc{ES}\nolimits_{\alpha }\left( y\right)
-\sum_{i=1}^{n}b_{i}\ln y_{i} \\
&\text{u.c.}&y\geq \mathbf{0}
\end{eqnarray*}%
The RB portfolio corresponds then to the following weights:
\begin{equation*}
x_{i}^{\star }=\frac{y_{i}^{\star }}{\sum_{j=1}^{n}y_{j}^{\star }}
\end{equation*}

\begin{example}
\label{example2} We consider three assets, whose expected returns
are equal to $3\%$, $8\%$ and $12\%$. Their volatilities are equal
to $8\%$, $20\%$ and $30\%$ while the
correlation matrix of asset returns is provided by the following matrix:%
\begin{equation*}
\rho =\left(
\begin{array}{ccc}
1.00 &      &       \\
0.50 & 1.00 &       \\
0.20 & 0.40 & 1.00  \\
\end{array}%
\right)
\end{equation*}
For the jumps, we have $\tilde{\mu}_1 = -15\%$, $\tilde{\mu}_2 =
-40\%$, $\tilde{\mu}_3 = 0\%$, $\tilde{\sigma}_1 = 15\%$,
$\tilde{\sigma}_2 = 20\%$ and $\tilde{\sigma}_3 = 10\%$. We also
assume that the correlation matrix between the jumps is equal to:
\begin{equation*}
\tilde{\rho} =\left(
\begin{array}{ccc}
1.00 &      &       \\
0.50 & 1.00 &       \\
0.00 & 0.00 & 1.00  \\
\end{array}%
\right)
\end{equation*}
Moreover, the intensity $\lambda$ of jumps is equal to $0.25$.
\end{example}

We consider the equal risk contribution (ERC) portfolio where the
risk budgets are the same for all the assets. In Table
\ref{tab:example2-1}, we report the weights of the ERC portfolio
when the risk measure is the portfolio's volatility (Maillard
\textsl{et al.}, 2010). In this case, the weight of Asset 1 is equal
to $60.94\%$ whereas the weight of Asset 3 is equal to $16.87\%$.
The difference between the two weights can be explained by the
difference in terms of volatility between the three assets. If we
use a Gaussian expected shortfall (Roncalli, 2015), the allocation
does not change significantly (Table \ref{tab:example2-2}). This is
normal as the Sharpe ratio of the three assets is similar. However,
when we introduce jumps, we note a big impact on the allocation (Table
\ref{tab:example2-3}). Indeed, Asset 3 is highly volatile, but the
risk of jumps is limited. This is not the case for the two other
assets. In particular, the skewness risk of Asset 1 is high because
its volatility is low, but the expected value of a jump is
approximatively five times the expected return in the normal regime.
As a result, the allocation is equal to $\left(44.70\%, 19.87\%,
35.42\%\right)$.

\begin{table}[tbph]
\centering
\caption{ERC portfolio (volatility risk measure)}
\label{tab:example2-1}
\tableskip
\begin{tabular}{|c|cccc|}
\hline
Asset & $x_i$ & $\mathcal{MR}_i$ & $\mathcal{RC}_i$ & $\mathcal{RC}_i^{\star}$ \\ \hline
$1$ & $60.94$ & ${\TsV}5.96$ & $3.63$ & $33.33$ \\
$2$ & $22.20$ &      $16.35$ & $3.63$ & $33.33$ \\
$3$ & $16.87$ &      $21.52$ & $3.63$ & $33.33$ \\ \hline
\multicolumn{3}{|l}{$\sigma\left( x\right)$} & $10.89$ & \\ \hline
\end{tabular}
\end{table}

\begin{table}[tbph]
\centering
\caption{ERC portfolio ($95\%$ Gaussian expected shortfall)}
\label{tab:example2-2}
\tableskip
\begin{tabular}{|c|cccc|}
\hline
Asset & $x_i$ & $\mathcal{MR}_i$ & $\mathcal{RC}_i$ & $\mathcal{RC}_i^{\star}$ \\ \hline
$1$ & $60.85$ & ${\TsV}9.24$ & $5.62$ & $33.33$ \\
$2$ & $21.96$ &      $25.60$ & $5.62$ & $33.33$ \\
$3$ & $17.19$ &      $32.72$ & $5.62$ & $33.33$ \\ \hline
\multicolumn{3}{|l}{$\func{ES}_{\alpha }\left( x\right)$} & $16.87$ & \\ \hline
\end{tabular}
\end{table}

\begin{table}[tbph]
\centering
\caption{ERC portfolio ($95\%$ expected shortfall with jumps)}
\label{tab:example2-3}
\tableskip
\begin{tabular}{|c|cccc|}
\hline
Asset & $x_i$ & $\mathcal{MR}_i$ & $\mathcal{RC}_i$ & $\mathcal{RC}_i^{\star}$ \\ \hline
$1$ & $44.70$ & $24.70$ & $11.04$ & $33.33$ \\
$2$ & $19.87$ & $55.56$ & $11.04$ & $33.33$ \\
$3$ & $35.42$ & $31.17$ & $11.04$ & $33.33$ \\ \hline
\multicolumn{3}{|l}{$\func{ES}_{\alpha }\left( x\right)$} & $33.12$ & \\ \hline
\end{tabular}
\end{table}

\begin{remark}
Using the previous example, we find that $\func{SR}\nolimits_{1}^{+}
= 0.52$. It follows that $\alpha^{-}$ is equal to $57.4\%$.
\end{remark}

\section{Illustration with the equity/bond/volatility asset mix policy}

\begin{figure}[tbh]
\centering
\caption{Cumulative performance of US bonds, US equities and US short volatility}
\label{fig:carry0}
\figureskip
\includegraphics[width = \figurewidth, height = \figureheight]{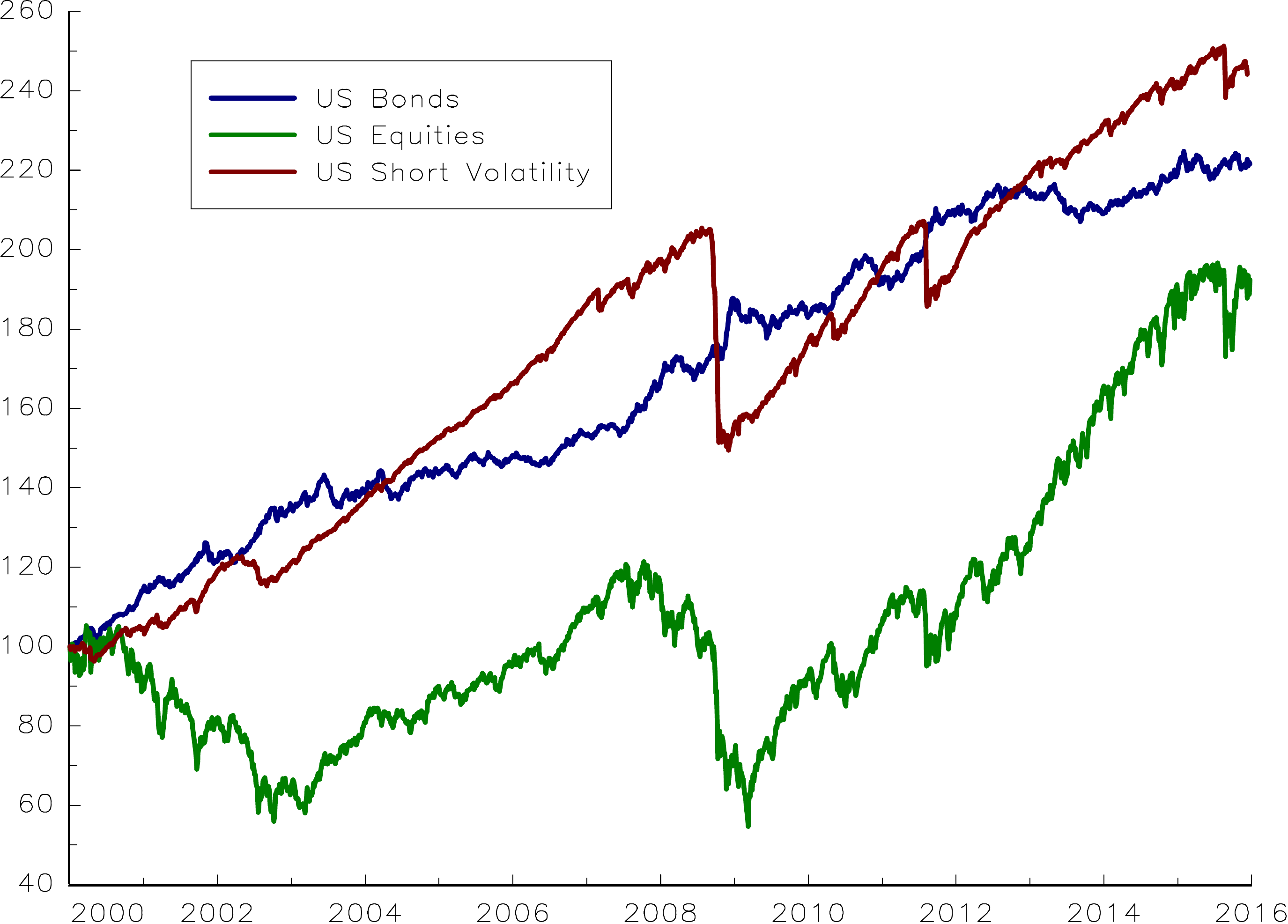}
\end{figure}

We consider the traditional equity-bond asset mix policy. Nowadays,
a lot of investors prefer to use a risk parity portfolio instead of
a constant-mix portfolio in order to obtain a risk-balanced
allocation between the two assets. While bonds and equities are not
Gaussian, it is however accepted that the volatility measure is a
good approximation of the risks taken by the investor. Nevertheless,
if we introduce a short volatility (or volatility carry) exposure,
this assumption is not
valid. For instance, we report the cumulative performance%
\footnote{We use the Barclays US Government Bond Index, the S\&P 500
TR Index and the generic equities/volatility/carry/US index
calculated by Hamdan \textsl{et al.} (2016).} of bonds, equities and
the short volatility strategy in Figure \ref{fig:carry0} for the
study period January 2000--December 2015. We notice that the short
volatility strategy presents a high jump risk compared to bonds and
equities. This is confirmed by the extreme returns and statistics of
skewness coefficients calculated with different frequencies (see
Tables \ref{tab:carry1-jumps} and \ref{tab:carry1-skewness}). Indeed, we
observe a high asymmetry between best and worst returns for the
short volatility strategy. Moreover, the skewness of daily returns
is about $-7$ for the volatility carry index, whereas it is close to
zero for the bond and equity indices.

\begin{table}[tbph]
\centering
\caption{Best returns, worst returns and maximum drawdown}
\label{tab:carry1-jumps}
\tableskip
\begin{tabular}{|c|ccccc|}
\hline
Asset    & Daily           & Weekly           & Monthly          & Annually         & Maximum         \\ \hline
         & \multicolumn{5}{c|}{Best returns (in \%)}                                                  \\
Bonds    & ${\TsVIII}1.89$ &  ${\TsXII}3.06$  &   ${\TsXII}7.10$ & ${\TsVIII}15.75$ &                 \\
Equities & ${\TsIII}11.58$ & ${\TsVIII}19.21$ & ${\TsVIII}22.64$ & ${\TsVIII}75.32$ &                 \\
Carry    & ${\TsVIII}2.27$ &  ${\TsXII}3.13$  &   ${\TsXII}4.39$ & ${\TsVIII}16.99$ &                 \\ \hline
         & \multicolumn{5}{c|}{Worst returns (in \%)}                                                 \\
Bonds    &  $-1.67$        & ${\TsV}$$-2.81$  & ${\TsV}$$-4.40$  & ${\TsV}$$-3.41$  & ${\TsV}$$-6.03$ \\
Equities &  $-9.03$        &        $-18.29$  &        $-29.67$  &        $-49.69$  &        $-55.25$ \\
Carry    &  $-6.82$        &        $-11.04$  &        $-23.43$  &        $-23.37$  &        $-27.30$ \\ \hline
\end{tabular}
\end{table}

\begin{table}[tbh]
\centering
\caption{Skewness coefficients}
\label{tab:carry1-skewness}
\tableskip
\begin{tabular}{|c|cccc|}
\hline
Asset    & Daily           & Weekly  & Monthly         & Annually       \\ \hline
Bonds    &        $ -0.12$ & $-0.17$ & ${\TsVIII}0.07$ & ${\TsVIII}0.22$ \\
Equities & ${\TsVIII}0.01$ & $-0.44$ &         $-0.81$ &         $-0.57$ \\
Carry    &         $-7.24$ & $-5.77$ &         $-6.32$ &         $-2.23$ \\
\hline
\end{tabular}
\end{table}

\subsection{Estimation of the mixture model}

\begin{table}[b]
\centering
\caption{Estimation of the Gaussian model (daily model)}
\label{tab:carry1-1}
\tableskip
\begin{tabular}{|c|ccccc|}
\hline Asset    & $\mu_i$ & $\sigma_i$ & \multicolumn{3}{c|}{$\rho_{i,j}$} \\ \hline
Bonds    & $5.54$ & ${\TsV}4.37$ & ${\TsIII}100.00$ &      $      $ & $      $ \\
Equities & $6.25$ &      $20.39$ &         $-35.06$ &      $100.00$ & $      $ \\
Carry    & $5.96$ & ${\TsV}5.62$ &         $-22.28$ & ${\TsV}58.07$ & $100.00$ \\
\hline
\end{tabular}
\end{table}

In Table \ref{tab:carry1-1}, we report the maximum likelihood results of the
Gaussian model:
\begin{equation}
f\left( y\right) = \phi _{3}\left( y;\mu \,\mathrm{d}%
t,\Sigma \,\mathrm{d}t\right) \label{eq:em-pdf1}
\end{equation}%
The parameters $\mu$ and $\Sigma$ (expressed in \%) are estimated
using daily returns. In Appendix \ref{appendix:additional-results}
on Page \pageref{tab:carry1-2}, we also give the estimated values of
$\mu$ and $\Sigma$ when we consider weekly, monthly and annually
overlapping returns. On average, the volatility of US bonds and
volatility carry is about $5\%$, whereas the volatility of US
equities is $4$ times larger and approximately equal to $20\%$. US
bonds have a negative correlation with US equites and volatility
carry, which are both positively correlated. We notice that the
frequency to compute the returns has an influence on the parameters,
but the impact is relatively low, except for the
cross-correlations.\bigskip

\begin{figure}[tbph]
\centering
\caption{Estimated EM parameters (bonds)}
\label{fig:carry6-2-1}
\figureskip
\includegraphics[width = \figurewidth, height = \figureheight]{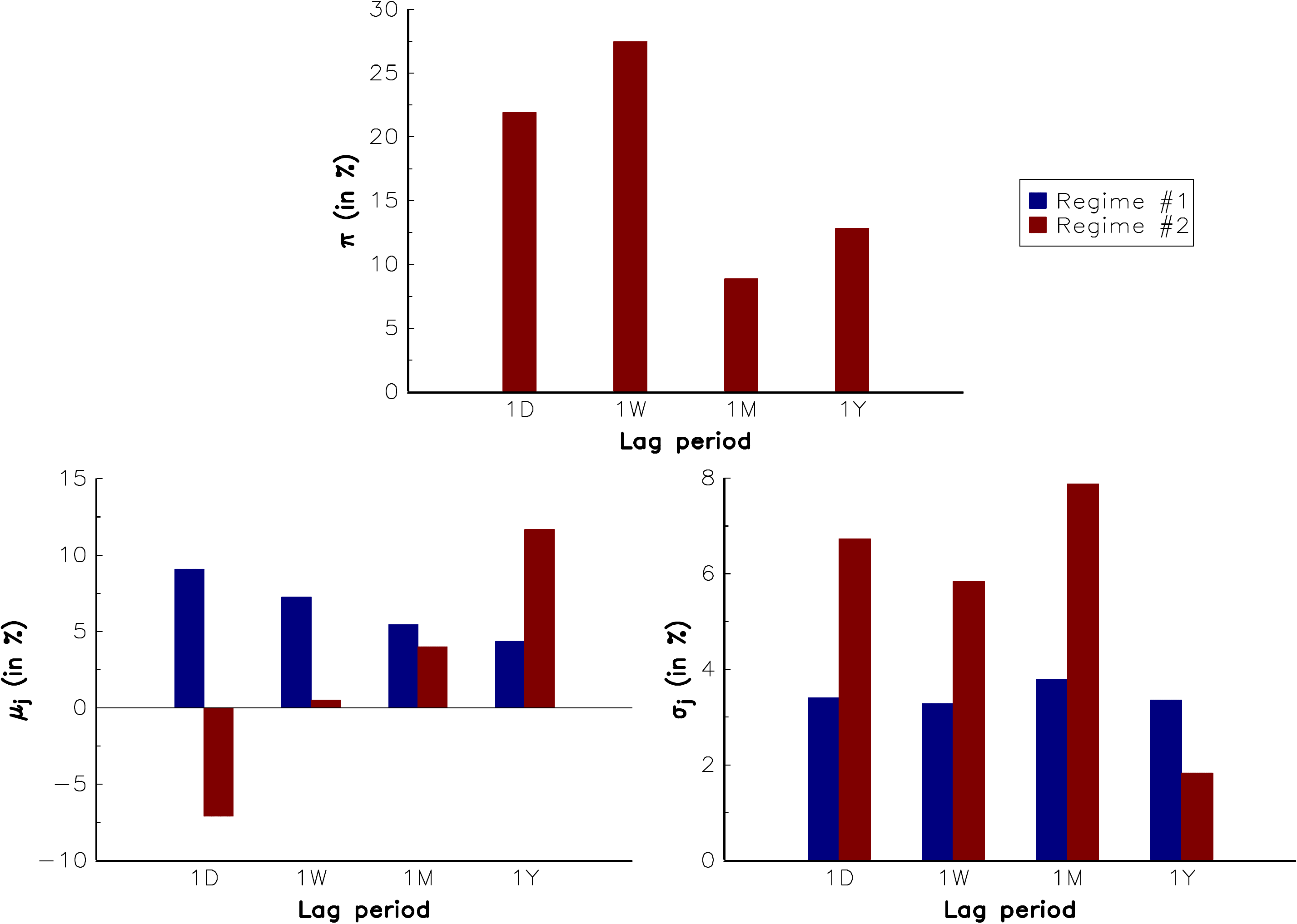}
\end{figure}

\begin{figure}[tbph]
\centering
\caption{Estimated EM parameters (equities)}
\label{fig:carry6-2-2}
\figureskip
\includegraphics[width = \figurewidth, height = \figureheight]{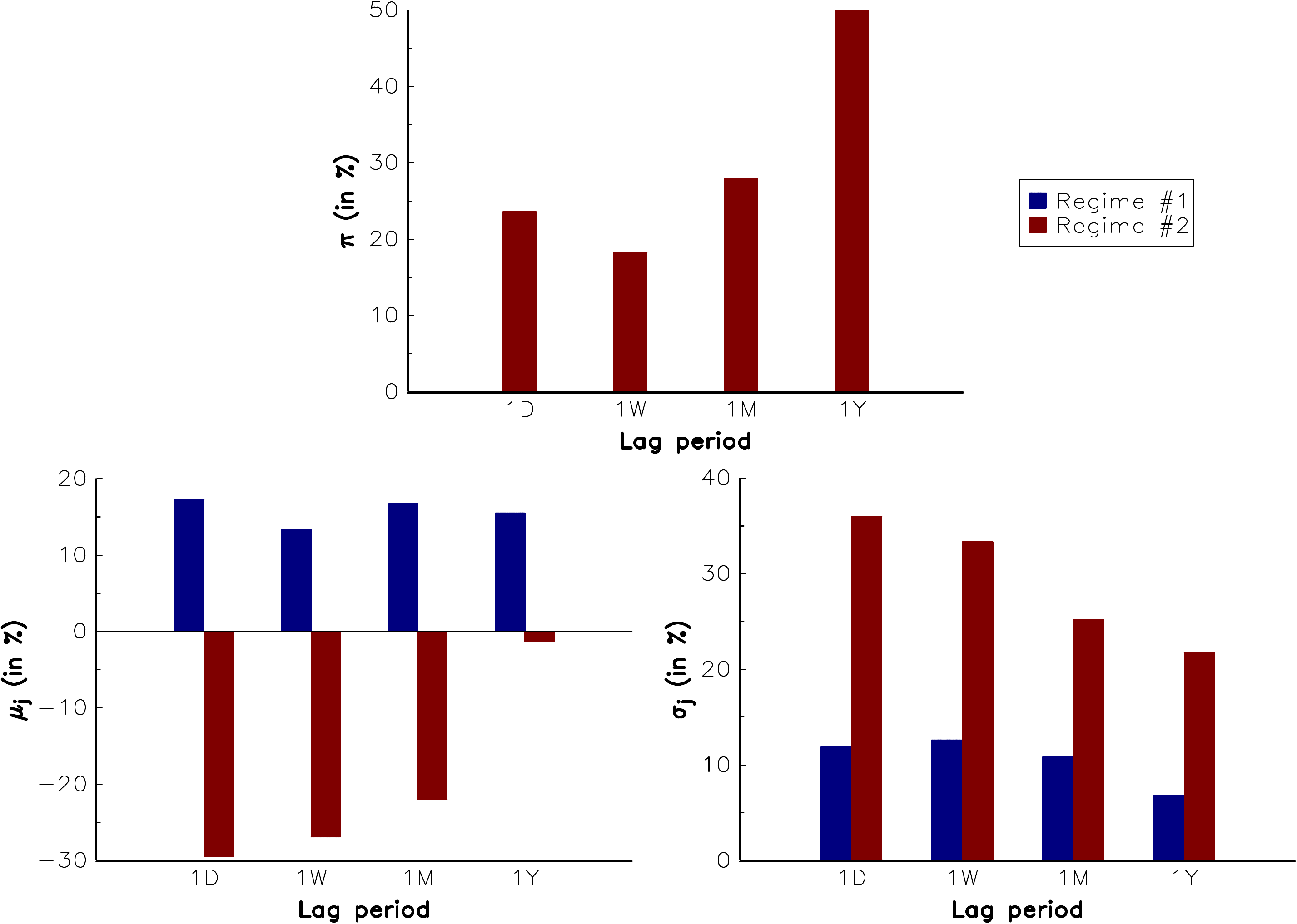}
\end{figure}

\begin{figure}[tbph]
\centering
\caption{Estimated EM parameters (carry)}
\label{fig:carry6-2-3}
\figureskip
\includegraphics[width = \figurewidth, height = \figureheight]{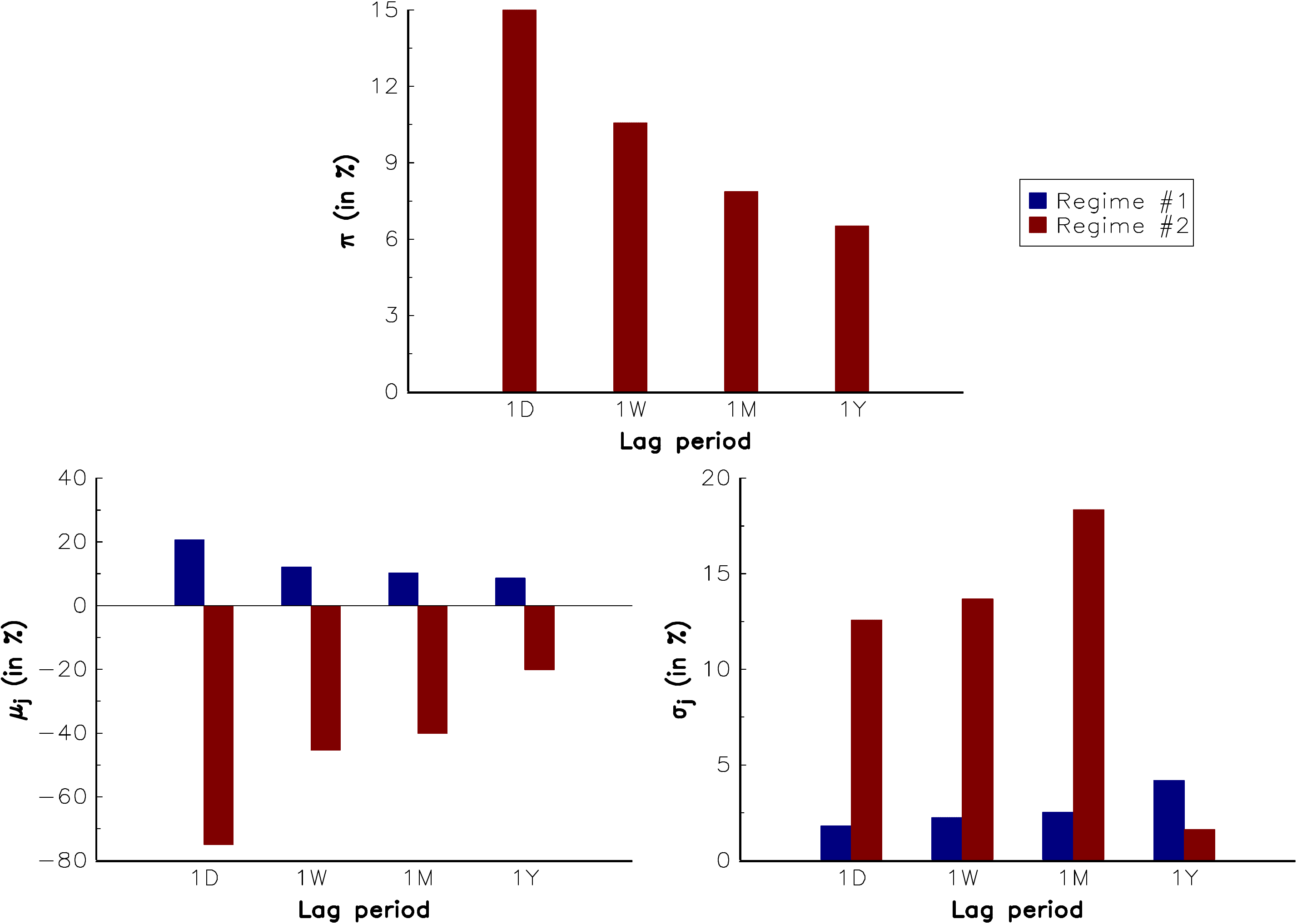}
\end{figure}

We now estimate the Gaussian mixture model with the following
parametrization of the probability density function:
\begin{equation}
f\left( y\right) = \left(1 - \pi\right)\phi _{1}\left(
y;\mu_{1}\,\mathrm{d}t,\sigma _{1}^{2}\,\mathrm{d}t\right) +\pi\phi
_{1}\left( y;\mu _{2}\,\mathrm{d}t,\sigma
_{2}^{2}\,\mathrm{d}t\right) \label{eq:em-pdf2}
\end{equation}%
We only consider the one-dimensional case in order to illustrate
some issues related to the choice of parametrization and frequency.
In Figures \ref{fig:carry6-2-1}, \ref{fig:carry6-2-2} and
\ref{fig:carry6-2-3}, we report the estimated values of the
parameters $\pi$, $\mu _{1}$, $\mu _{2}$, $\sigma _{1}$ and $\sigma
_{2}$. The frequency $\pi$ is large for bonds and
equities. These results suggest that a two-volatility regime model
is better to describe the returns of equities and bonds than a model
with jumps. The first regime is a low volatility environment
associated with positive returns, whereas the second regime is a
high volatility environment associated with negative returns. In the
case of the carry risk premium, the frequency $\pi$ is lower. The
large discrepancy between $\mu _{1}$ and $\mu _{2}$, but also
between $\sigma _{1}$ and $\sigma _{2}$ justifies that a model with
jumps is more relevant than a model with two volatility regimes.
Contrary to the standard Gaussian model, the choice of the period
for calculating the returns is an important factor. We notice that
using one-year returns is not appropriate, because it may produce
incoherent results. For instance, the second regime for bonds
corresponds to an environment of high returns with low volatility.
For the carry risk premium, the volatility is also higher for the
first regime than for the second regime. Similarly
the case of daily returns is an issue. Indeed, the period
may be too short to observe a complete drawdown risk. This is why we
prefer to consider a weekly or monthly period to estimate the jump
component.

\begin{figure}[tbph]
\centering
\caption{Density function of the jump component (carry)}
\label{fig:carry7-4}
\figureskip
\includegraphics[width = \figurewidth, height = \figureheight]{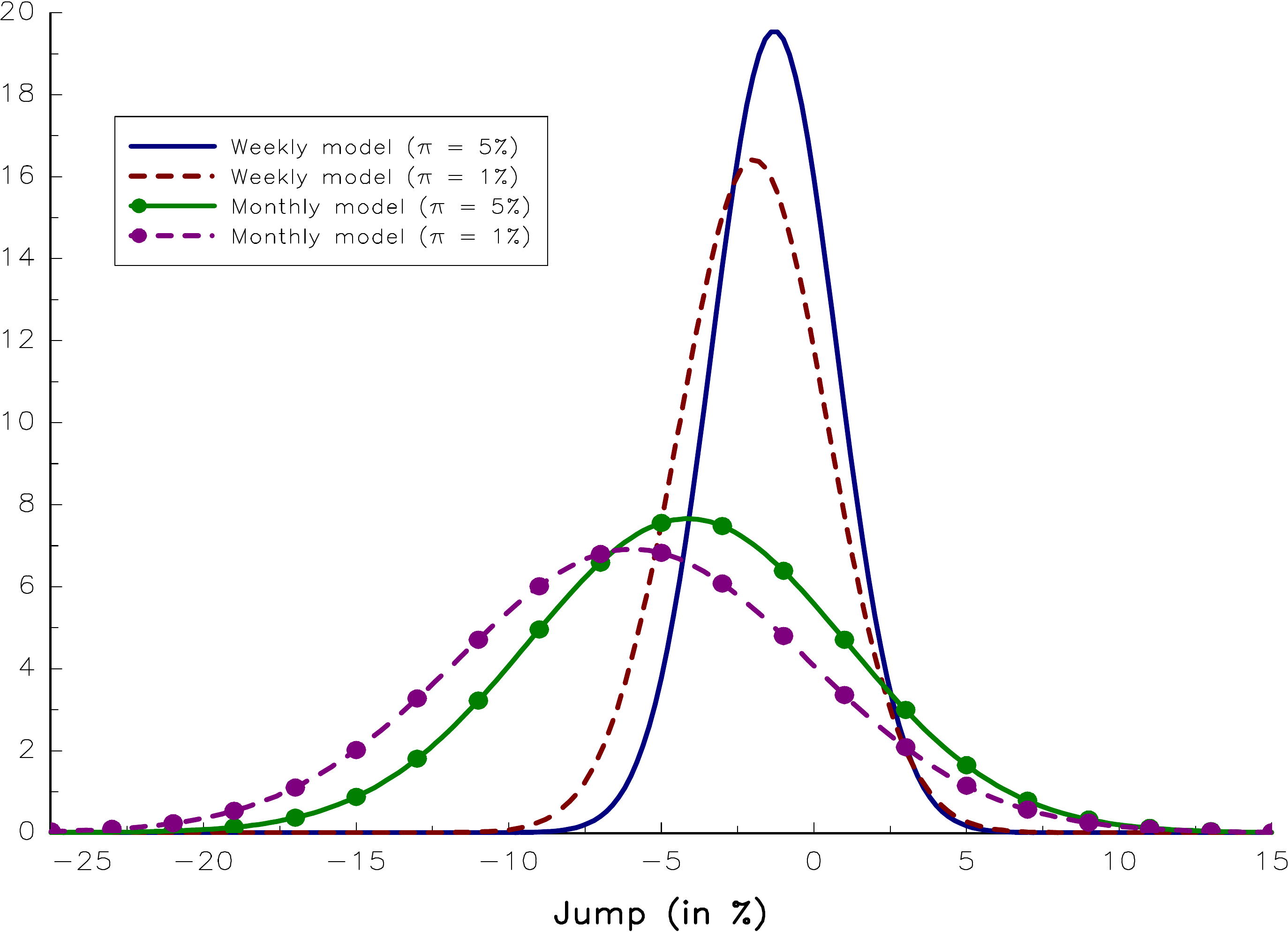}
\end{figure}

Let us consider the multivariate case, where the probability density
function is specified as follows:%
\begin{equation}
f\left( y\right) =\left( 1-\pi \right) \phi _{3}\left( y;\mu \,\mathrm{d}%
t,\Sigma \,\mathrm{d}t\right) +\pi \phi _{3}\left( y;\mu \,\mathrm{d}t+%
\tilde{\mu},\Sigma \,\mathrm{d}t+\tilde{\Sigma}\right)
\label{eq:em-pdf3}
\end{equation}%
This parametrization ensures that the estimated model is a Gaussian
mixture model with a jump component, and not a two-volatility regime
model. The reason is that $\Sigma$ is a one-year covariance matrix,
whereas $\tilde{\Sigma}$ is a covariance matrix, whose period
corresponds to the frequency of calculated returns. Moreover, we
assume that $\pi$ is given in order to control the jump frequency.
Results are provided in Tables \ref{tab:carry7-1-1} --
\ref{tab:carry7-2-2} on Page \pageref{tab:carry7-1-1}. We confirm
that the expected value of the jump $\tilde{\mu}_i$ is positive for
bonds, whereas it is negative for equities and carry. Moreover,
the absolute value of $\tilde{\mu}_i$ increases when
the jump frequency $\pi$ decreases. For the carry risk premium, we
report the probability density function of the jump component
in Figure \ref{fig:carry7-4}. Regarding the dependence of jumps, we
notice that the correlations $\tilde{\rho}_{i,j}$ tend to be higher
when the jump probability is lower.

\begin{remark}
With the parametrization (\ref{eq:em-pdf3}), it is not possible to
compare results for different values of the frequency $\mathrm{d}t$.
Indeed, fixing $\pi$ at $5\%$ is equivalent to observing a jump
every $20$ periods on average. The return period is then
approximately equal to five months in the case of weekly returns and
one year and eight months in the case of monthly returns. Conversely, if
fixing the return period is preferable, it is better to consider the
parametrization $\pi = \lambda\,\mathrm{d}t$, where $\lambda$ is the
annually jump frequency:
\begin{equation}
f\left( y\right) =\left( 1-\lambda\,\mathrm{d}t \right) \phi _{3}\left( y;\mu \,\mathrm{d}%
t,\Sigma \,\mathrm{d}t\right) + \left(\lambda\,\mathrm{d}t\right) \phi _{3}\left( y;\mu \,\mathrm{d}t+%
\tilde{\mu},\Sigma \,\mathrm{d}t+\tilde{\Sigma}\right)
\label{eq:em-pdf4}
\end{equation}%
In this case, the return period (expressed in years) is equal to
$1/\lambda$. For example, if $\lambda$ is equal to $10\%$, the
return period of jumps is equal to ten years and we obtain results
that are given in Tables \ref{tab:carry7-3-1} and
\ref{tab:carry7-3-2} on Page \pageref{tab:carry7-3-1}.
\end{remark}

\begin{figure}[tbph]
\centering
\caption{Expected weekly drawdown (in \%) of the carry risk premium}
\label{fig:carry8-1-1}
\figureskip
\includegraphics[width = \figurewidth, height = \figureheight]{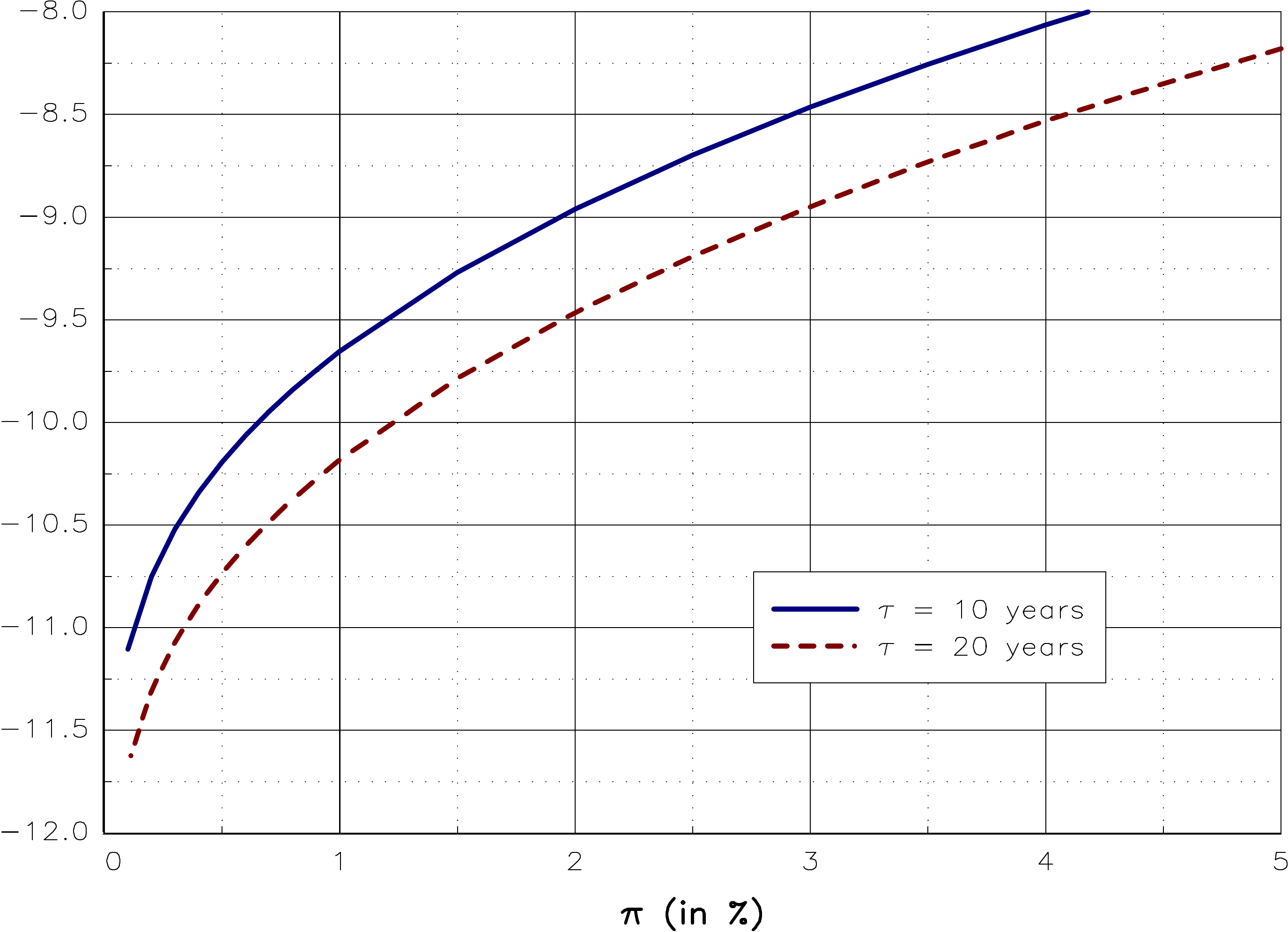}
\end{figure}

In order to deduce the final model, we calculate the expected
drawdown associated to a given return period $\tau $:
\begin{equation*}
\mathbb{E}\left[ \limfunc{DD}\nolimits_{i}\left( \tau \right) \right] =%
\tilde{\mu}_{i}+\Phi ^{-1}\left( \frac{\mathrm{d}t}{\tau }\right) \tilde{%
\sigma}_{i}
\end{equation*}%
As the carry strategy constitutes our most skewed asset, it is the only one used in calibrating the parameter $\pi$.
Results are reported
in Figure \ref{fig:carry8-1-1} for weekly returns.
We observe that the expected drawdown is about $-10.5\%$
for $\pi =0.5\%$ when the return period is between 10 and 20 years.
This expected drawdown is then close to the worst weekly return
observed during the period January 2000 -- December 2015 (see Table
\ref{tab:carry1-jumps} on Page \pageref{tab:carry1-jumps}). We also introduce
the constraint that bonds have no jump component,
because the previous estimations suggest that the expected jump is positive for this asset class.
Moreover, it is not statistically significant.

\begin{table}[tbph]
\centering
\caption{Estimation of the constrained mixture model when $\pi = 0.5\%$ (weekly model)}
\label{tab:carry8-2-1}
\tableskip
\begin{tabular}{|c|ccccc|}
\hline Asset    & $\mu_i$ & $\sigma_i$ & \multicolumn{3}{c|}{$\rho_{i,j}$} \\ \hdashline
Bonds    & ${\TsVIII}5.38$ & ${\TsV}4.17$ & ${\TsIII}100.00$ &      $      $ & $      $ \\
Equities & ${\TsVIII}7.89$ &      $15.64$ &         $-36.80$ &      $100.00$ & $      $ \\
Carry    & ${\TsIII}10.10$ & ${\TsV}2.91$ &         $-25.17$ & ${\TsV}57.43$ & $100.00$ \\ \hline
Asset    & $\tilde{\mu}_i$ & $\tilde{\sigma}_i$ & \multicolumn{3}{c|}{$\tilde{\rho}_{i,j}$} \\ \hdashline
Bonds    & ${\TsVIII}0.00$ & ${\TsV}0.00$ &      $100.00$   &      $      $ & $      $ \\
Equities &         $-1.20$ & ${\TsV}6.76$ &  ${\TsX}0.00$   &      $100.00$ & $      $ \\
Carry    &         $-2.23$ & ${\TsV}2.57$ &  ${\TsX}0.00$   & ${\TsV}60.45$ & $100.00$ \\
\hline
\end{tabular}
\end{table}

\begin{figure}[tbph]
\centering
\caption{Probability density function of asset returns in the normal regime (weekly model)}
\label{fig:carry8-3}
\figureskip
\includegraphics[width = \figurewidth, height = \figureheight]{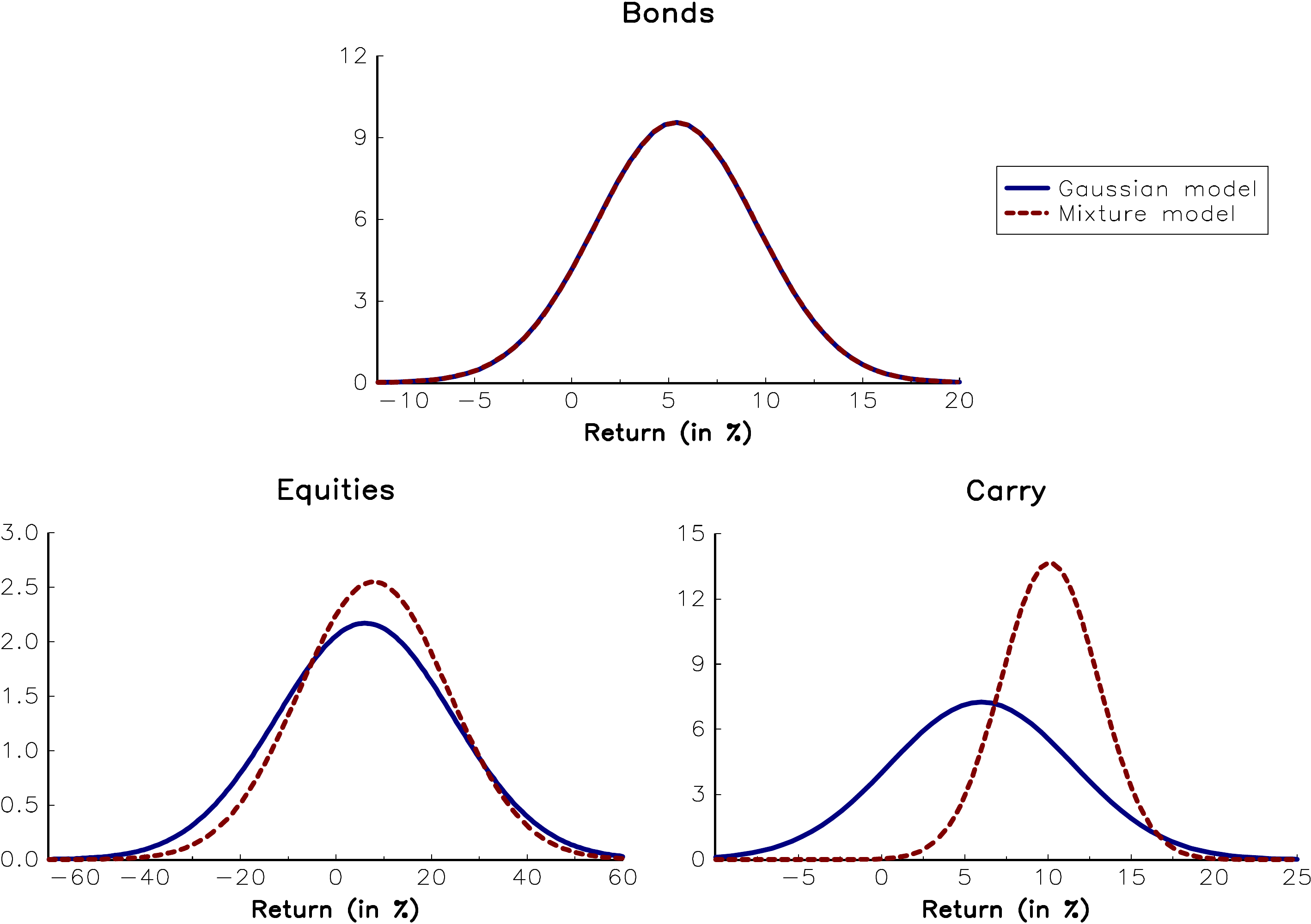}
\end{figure}

Using the method of maximum likelihood%
\footnote{The EM algorithm can no longer be applied due to the
constraints.},
we finally obtain the results%
\footnote{For monthly returns, results are given in Table
\ref{tab:carry8-2-2} on Page \pageref{tab:carry8-2-2}. We have
selected $\pi = 1.0\%$, because the expected drawdown is about
$-22\%$, which is close to the worst monthly return of $-23.43\%$
(see Figure \ref{fig:carry8-1-2}).} given in Table
\ref{tab:carry8-2-1}. In the case of bonds, we retrieve the results
of the Gaussian model (see Table \ref{tab:carry1-2} on Page
\pageref{tab:carry1-2}). The normal expected return and volatility
are equal to $5.38\%$ and $4.17\%$. For equities and the short
volatility strategy, the introduction of the jump component
increases the expected return and decreases volatility. Indeed,
$\mu_i$ is respectively equal to $7.89\%$ and $10.10\%$, whereas it
was equal to $6.09\%$ and $6.00\%$ in the Gaussian case.
The impact on volatility is particularly pronounced for
the carry risk premium since it is divided by a factor of $1.9$
($2.91\%$ versus $5.50\%$). The cross-correlations in the normal
regime are close to the values obtained in the Gaussian model. In
order to compare the two models, we report the probability density
function of asset returns in the normal regime in Figure
\ref{fig:carry8-3}. We confirm that the jump model impacts mainly
the modeling of the carry risk premium%
\footnote{As expected, the jump component is important for the short
volatility strategy and we obtain $\tilde{\mu}_3=-2.23\%$. For
equities, we observe a lower expected jump, but a large dispersion
of weekly jumps.}.

\subsection{Comparing in-sample ERC portfolios}

We now calculate the ERC portfolio using the $99\%$ expected
shortfall risk measure. For that, we use the previous estimates. By
construction, we introduce a backtesting bias because the portfolio
allocation is based on the full sample of historical data. However,
we assume that the expected returns $\mu_i$ are equal to zero and
not to the historical estimates in order to limit the in-sample
bias. In the case of the Gaussian model, we obtain the traditional
ERC portfolio based on the volatility risk measure (Maillard
\textsl{et al.}, 2010). In the case of the weekly model, the
allocation is the following: $59.17\%$ in bonds, $10.43\%$ in
equities and $30.40\%$ in the short volatility strategy. If we
consider the jump model and consider only the normal regime, the
allocation becomes: $46.46\%$ in bonds, $9.18\%$ in equities and
$44.36\%$ in the short volatility strategy. Therefore, the carry
risk premium represents a large part of the portfolio. If we take
into account the skewness risk, the allocation is lower and is equal
to $36.93\%$. This difference in terms of allocation is even
larger when we consider the monthly model. In this case, the
allocation for the carry risk premium is $45.58\%$ without jumps and
$25.53\%$ with jumps.

\begin{table}[tbph]
\centering
\caption{Weights (in \%) of the ERC portfolio}
\label{tab:carry9-1}
\tableskip
\begin{tabular}{|c|c:cc|c:cc|}
\hline
\multirow{3}{*}{Asset}
         & \multicolumn{3}{c|}{Weekly model} & \multicolumn{3}{c|}{Monthly model}   \\
         & & \multicolumn{2}{c|}{Jump model} & & \multicolumn{2}{c|}{Jump model}    \\
         & Gaussian  & Normal       & Mixture & Gaussian  & Normal       & Mixture  \\ \hline
Bonds    & $59.17$   &      $46.46$ & $52.71$ & $61.36$   &      $44.36$ & $62.22$  \\
Equities & $10.43$   & ${\TsV}9.18$ & $10.36$ & $12.09$   &      $10.06$ & $12.25$  \\
carry    & $30.40$   &      $44.36$ & $36.93$ & $26.55$   &      $45.58$ & $25.53$  \\ \hline
\end{tabular}
\end{table}

\begin{figure}[tbph]
\centering
\caption{One-year rolling volatility (in \%) of the carry risk premium (weekly model)}
\label{fig:carry9-3}
\figureskip
\includegraphics[width = \figurewidth, height = \figureheight]{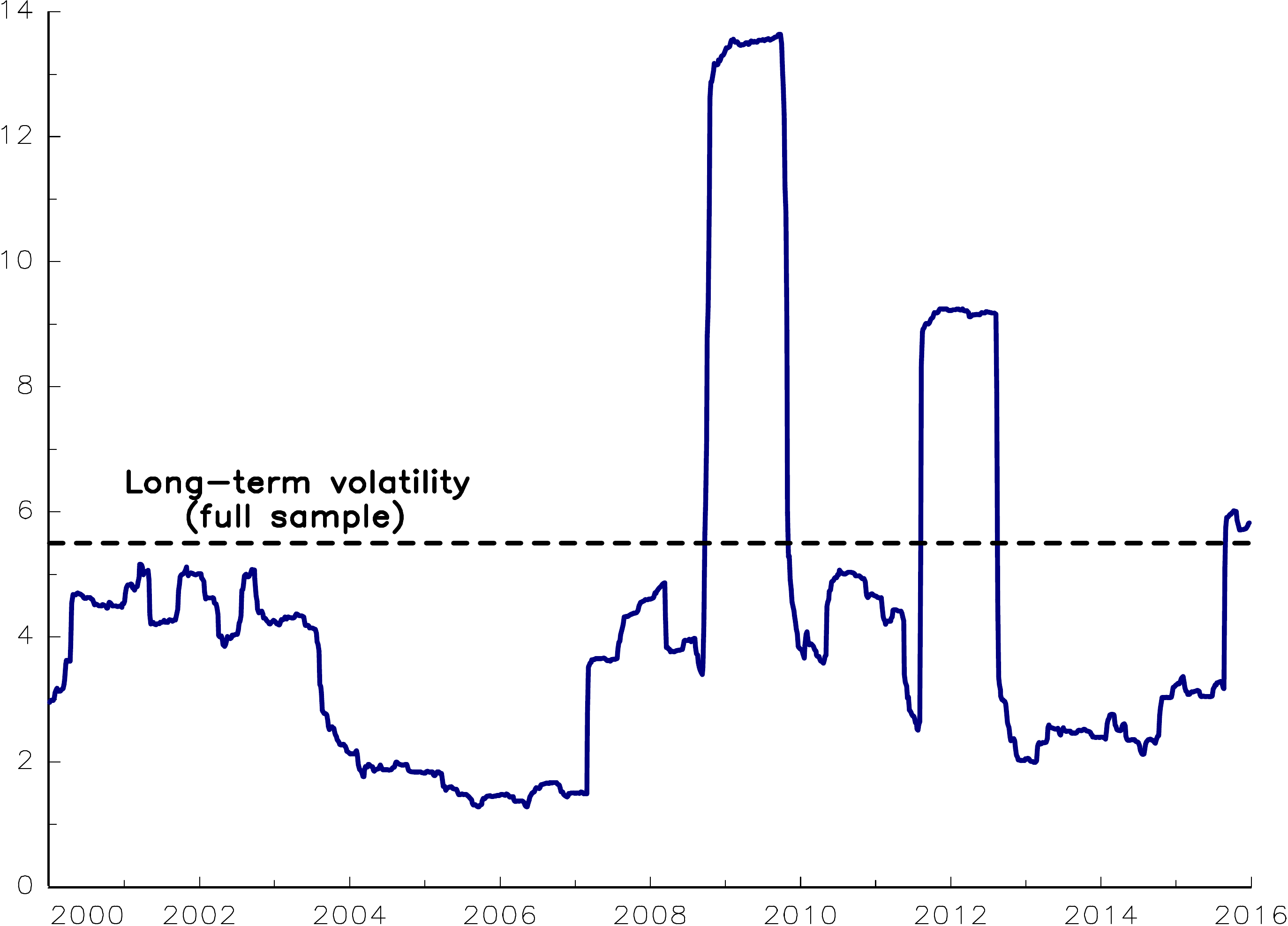}
\end{figure}

The previous results give the illusion that the Gaussian model is a
conservative model and can be compared to the mixture model. This is inaccurate
because the results presented are in-sample. For instance, the
first major drawdown of the carry strategy occurs in 2008. Before
2008, a Gaussian model will dramatically underestimate the skewness
risk of the carry strategy. By considering the full period (January
2000 -- December 2015), the covariance matrix of the Gaussian model
incorporates these jumps, meaning that the ex-post volatility risk
takes into account this skewness risk. To illustrate this bias, we
report the one-year rolling volatility of the short volatility
strategy in Figure \ref{fig:carry9-3}. We verify that our estimate
largely overestimates the volatility of the carry risk premium in
normal periods.

\subsection{Dynamics of out-of-sample ERC portfolios}

\begin{figure}[tbph]
\centering
\caption{Dynamics of the ERC weights (Gaussian model)}
\label{fig:carry9-4}
\figureskip
\includegraphics[width = \figurewidth, height = \figureheight]{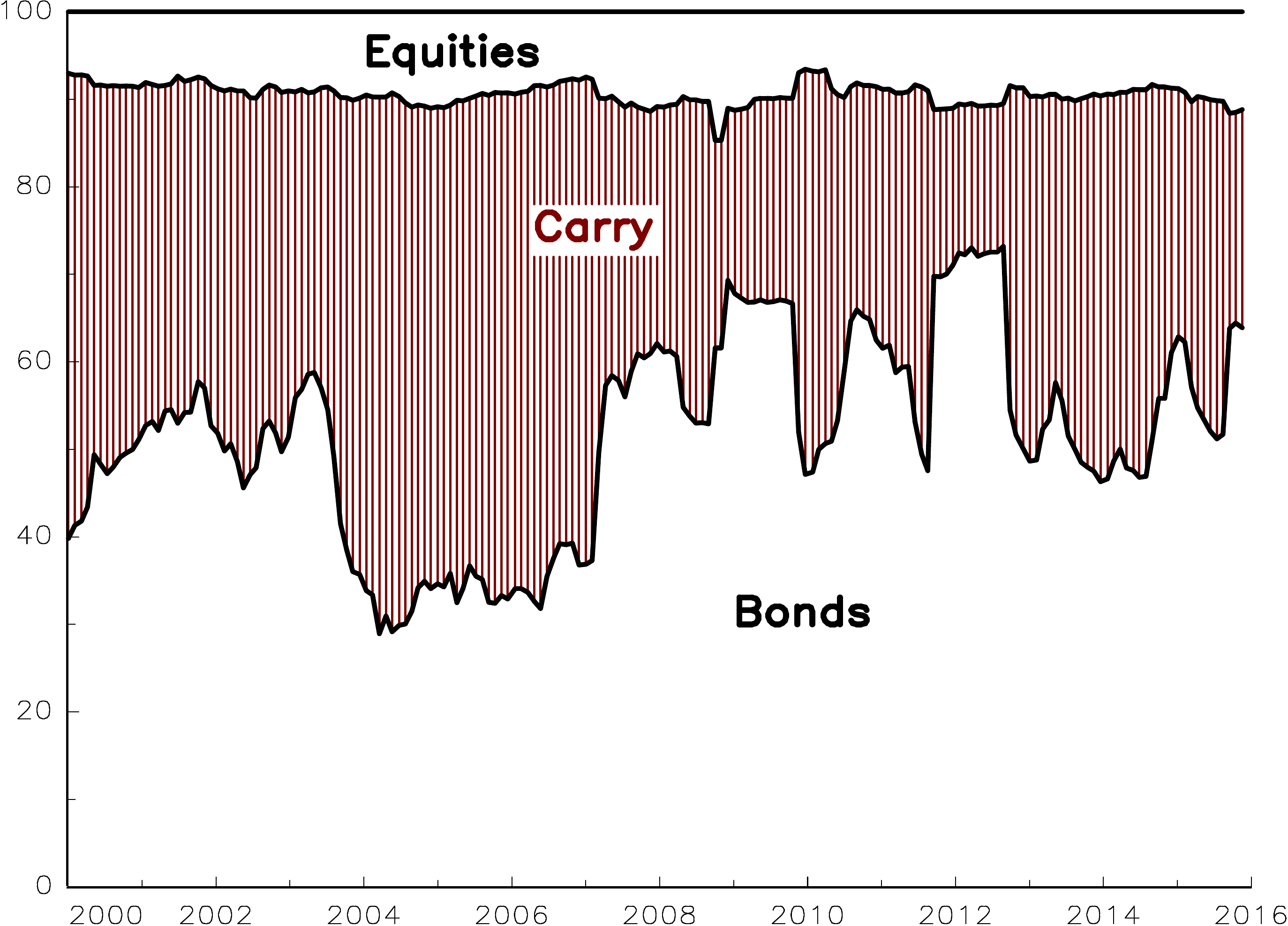}
\end{figure}

A more realistic backtest consists in rebalancing the portfolio at a
given frequency and to calculate the allocation by using only past
data. Therefore, the covariance matrix is no longer static, but
time-varying. $\Sigma_t$ is then calculated using information until
time $t-1$. In the case of the ERC portfolio, it is common to
consider a monthly rebalancing based on a one-year rolling
covariance matrix. The allocation of the ERC portfolio based on the
Gaussian model is reported in Figure \ref{fig:carry9-4}. We observe
that the allocation is not smooth. The weight of the carry risk
premium decreases sharply every time there is a jump in asset
returns. This negative jump is accompanied by a positive jump in the
allocation one year later when the jump in the asset returns exits
the rolling window. As a result, we obtain a high turnover and
changes in the allocation, which are mainly due to jumps in asset
returns.

When asset returns exhibit high skewness risk, the drawback of risk
parity portfolios is that they reduce the allocation after the
occurrence of a jump. However, it is generally too late because
jumps are not frequent and not correlated. Moreover, the occurrence
of a negative jump is generally followed by a positive performance
of the asset. Another drawback of such portfolios is that they will
overweight the asset that has no jump over the estimation period,
because the risk is underestimated. This implies that the weight
will be generally maximum just before the occurrence of a
jump.\bigskip

We now consider the dynamics of the ERC portfolio in the case of our
mixture model. We assume that the parameters $\pi$, $\tilde{\mu}$
and $\tilde{\Sigma}$ are given once and for all, and are set to the
previous estimates. At each rebalancing date, we have to estimate
the covariance matrix $\Sigma_t$ for the normal regime. For that, we
can apply the method of maximum likelihood to the mixture model with
the data of the rolling window:
\begin{equation*}
\left( \hat{\mu}_{t},\hat{\Sigma}_{t}\right) =\underset{\left( \mu
_{t},\Sigma _{t}\right) }{\arg \max }\sum_{s=1}^{n_{\mathrm{rw}}}\ln
\left( \left( 1-\pi \right) \phi _{3}\left( R_{t-s},\mu
_{t}\,\mathrm{d}t,\Sigma
_{t}\,\mathrm{d}t\right) +\pi \phi _{3}\left( R_{t-s},\mu _{t}\,\mathrm{d}t+%
\tilde{\mu},\Sigma _{t}\,\mathrm{d}t+\tilde{\Sigma}\right) \right)
\end{equation*}
where $n_{\mathrm{rw}}$ is the length of the rolling window.
Contrary to the previous analysis, the parameters $\pi$,
$\tilde{\mu}$ and $\tilde{\Sigma}$ are not estimated, but are
fixed. Only the parameters $\mu_{t}$ and $\Sigma_{t}$ are optimized.
The resulting allocation of the ERC portfolio%
\footnote{Recall that the risk measure corresponds to the $99\%$
expected shortfall. Moreover, the ERC portfolio is calculated by
considering that the expected returns in the normal regime are equal
to zero.} is reported on Page \pageref{fig:carry9-5}.\bigskip

However, estimating the covariance
matrix in the normal regime by maximum likelihood is not our preferred approach. Instead,
a popular approach consists in isolating continuous and jump
components (Ait-Sahalia and Jacod, 2012). Let $R_{t}$ be the return
of one asset. In the thresholding approach, we observe a jump at
time $t$ when the absolute return%
\footnote{The parameter $v$ is generally set to zero. This is equivalent to assume that
positive and negative jumps are symmetric.} is larger than a given level
$r^{\star }$:
\begin{equation*}
J_{t}=1\Leftrightarrow \left\vert R_{t} - v \right\vert \geq r^{\star }
\end{equation*}%
Using a sample of asset returns, we can then create two sub-samples:
\begin{itemize}
\item a sub-sample of asset returns without jump that satisfy $\left\vert
R_{t} - v\right\vert <r^{\star }$;

\item a sample of asset returns with jumps that satisfy $\left\vert
R_{t}- v\right\vert \geq r^{\star }$;
\end{itemize}
This method is simple and easy to understand. For instance, we
can easily estimate the normal volatility by considering the first
sub-sample. However, one of the issue is to choose the truncating
point $r^{\star}$. Another drawback is that the thresholding
approach is only valid in the one-dimensional case. Consequently we
prefer to use another approach called the filtering method. In this
case, we calculate the posterior probability of the jump regime:
\begin{equation*}
\hat{\pi}_{t}=\frac{\pi \phi _{n}\left( R_{t},\mu \,\mathrm{d}t+\tilde{\mu}%
,\Sigma \,\mathrm{d}t+\tilde{\Sigma}\right) }{\left( 1-\pi \right)
\phi _{n}\left( R_{t},\mu \,\mathrm{d}t,\Sigma \,\mathrm{d}t\right)
+\pi \phi
_{n}\left( R_{t},\mu \,\mathrm{d}t+\tilde{\mu},\Sigma \,\mathrm{d}t+\tilde{%
\Sigma}\right) }
\end{equation*}
We will say that we observe a jump at time $t$ when the posterior
probability is larger than a threshold $\pi^{\star }$:
\begin{equation*}
J_{t}=1\Leftrightarrow \hat{\pi}_{t} \geq \pi^{\star }
\end{equation*}%
In Appendix \ref{appendix:thresholding1}, we show that the filtering
procedure is equivalent to the thresholding procedure in the
one-dimensional case. However, the thresholding approach is no
longer valid in the multi-dimensional case, because the jumps across
assets are not necessarily synchronous and the jump regime depends
on asset correlations. Moreover, the thresholding rule becomes%
\footnote{See Appendix \ref{appendix:thresholding2} for the proof.}:
\begin{equation*}
J_{t}=1\Leftrightarrow \left( R_{t}-v\right) ^{\top } \left( \Sigma
\,\mathrm{d}t+\tilde{\Sigma}\right) ^{-1}\tilde{\Sigma}%
\left( \Sigma \,\mathrm{d}t\right) ^{-1} \left( R_{t}-v\right) \geq
r^{\star }
\end{equation*}%
where $v$ and $r^{\star}$ depend on the parameters $\mu $, $\Sigma $, $\tilde{%
\mu}$ and $\tilde{\Sigma}$. The solution of the
thresholding approach is therefore not unique. For this reason, we prefer the
filtering approach to detect jumps in the multi-dimensional
case.\bigskip

\begin{figure}[tbph]
\centering
\caption{Jump probabilities (in \%)}
\label{fig:carry11-4}
\figureskip
\includegraphics[width = \figurewidth, height = \figureheight]{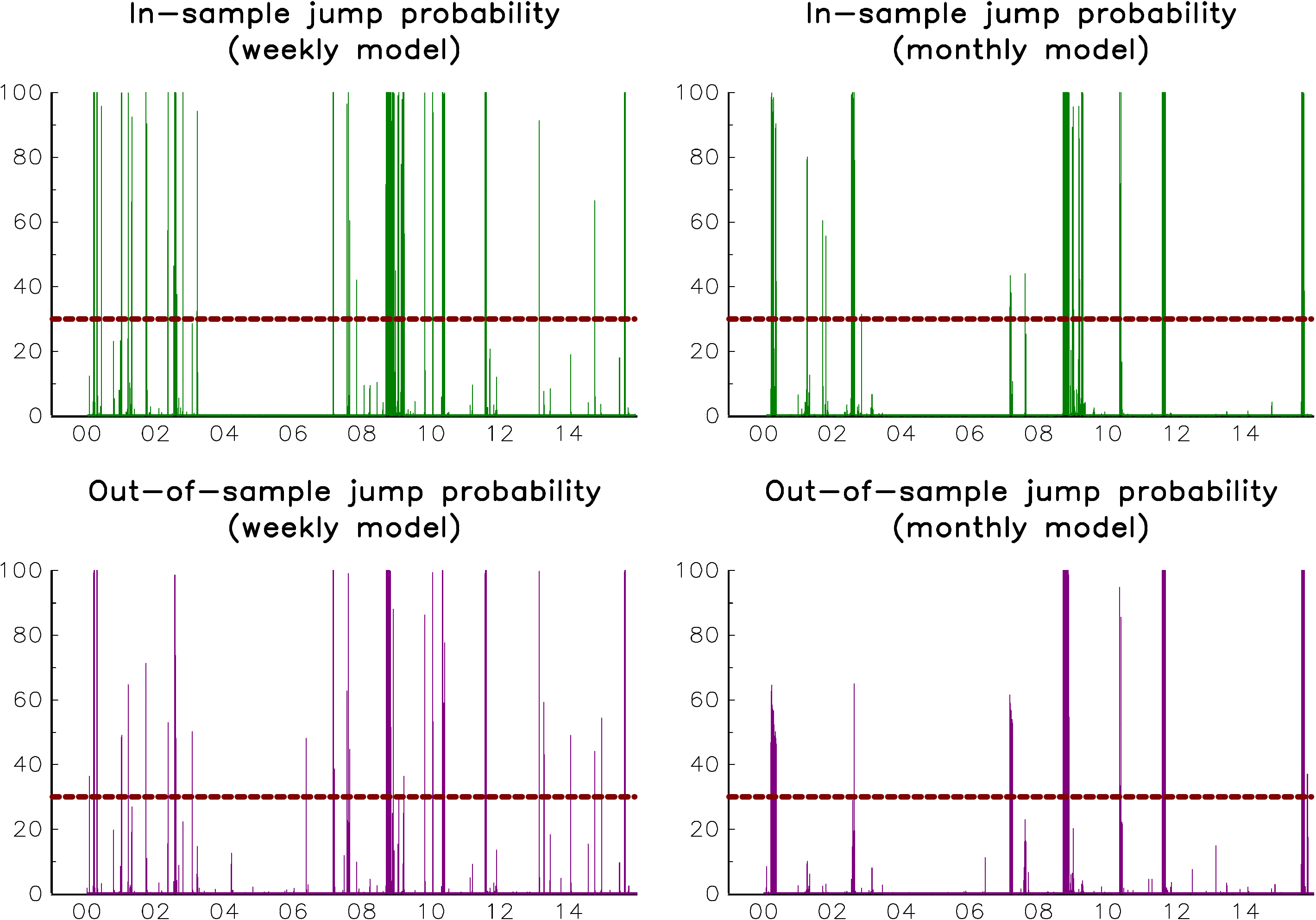}
\end{figure}

The filtering method can also be used to estimate the out-of-sample
parameters $\hat{\mu}_t$ and $\hat{\Sigma}_t$. The algorithm is
explained in Appendix \ref{appendix:filtering}. The underlying idea
is the following:
\begin{itemize}
\item Given $\hat{\mu}_{t-1}$ and $\hat{\Sigma}_{t-1}$, we
estimate the jump probability for each dates of the rolling window
that ends at time $t$;
\item Given the previous jump probabilities, we estimate the
parameters $\hat{\mu}_{t}$ and $\hat{\Sigma}_{t}$ by only considering the
dates of the rolling window, which do not correspond to a jump%
\footnote{We use the rule $\hat{\pi}_t \geq \pi^{\star}$.};
\item We iterate the algorithm until today.
\end{itemize}
The only difficulty is the initialization of the covariance matrix.
For that, we consider that there is no jump in the first rolling
window. To illustrate the filtering method, we report the jump
probability in Figure \ref{fig:carry11-4}. The in-sample
probabilities are calculated using the estimates given in Tables
\ref{tab:carry8-2-1} and \ref{tab:carry8-2-2}, which have been
obtained using the full sample. The out-of-sample probabilities
correspond to the average of posterior probabilities%
\footnote{At each date, we calculate the posterior probabilities for
all the dates in the rolling window. This implies that we calculate
$n_{\mathrm{rw}}$ jump probabilities for a given date, where
$n_{\mathrm{rw}}$ is the size of the rolling window.}, which are
calculated using the filtering algorithm, a one-year rolling window
and $\pi^{\star} = 30\%$.

\begin{remark}
In Markov regime-switching models, in-sample probabilities are
called smoothing probabilities, whereas out-of-sample probabilities
are called filtering probabilities.
\end{remark}

\begin{figure}[tbph]
\centering
\caption{Estimated volatility (in \%) of the carry risk premium (weekly model)}
\label{fig:carry12}
\figureskip
\includegraphics[width = \figurewidth, height = \figureheight]{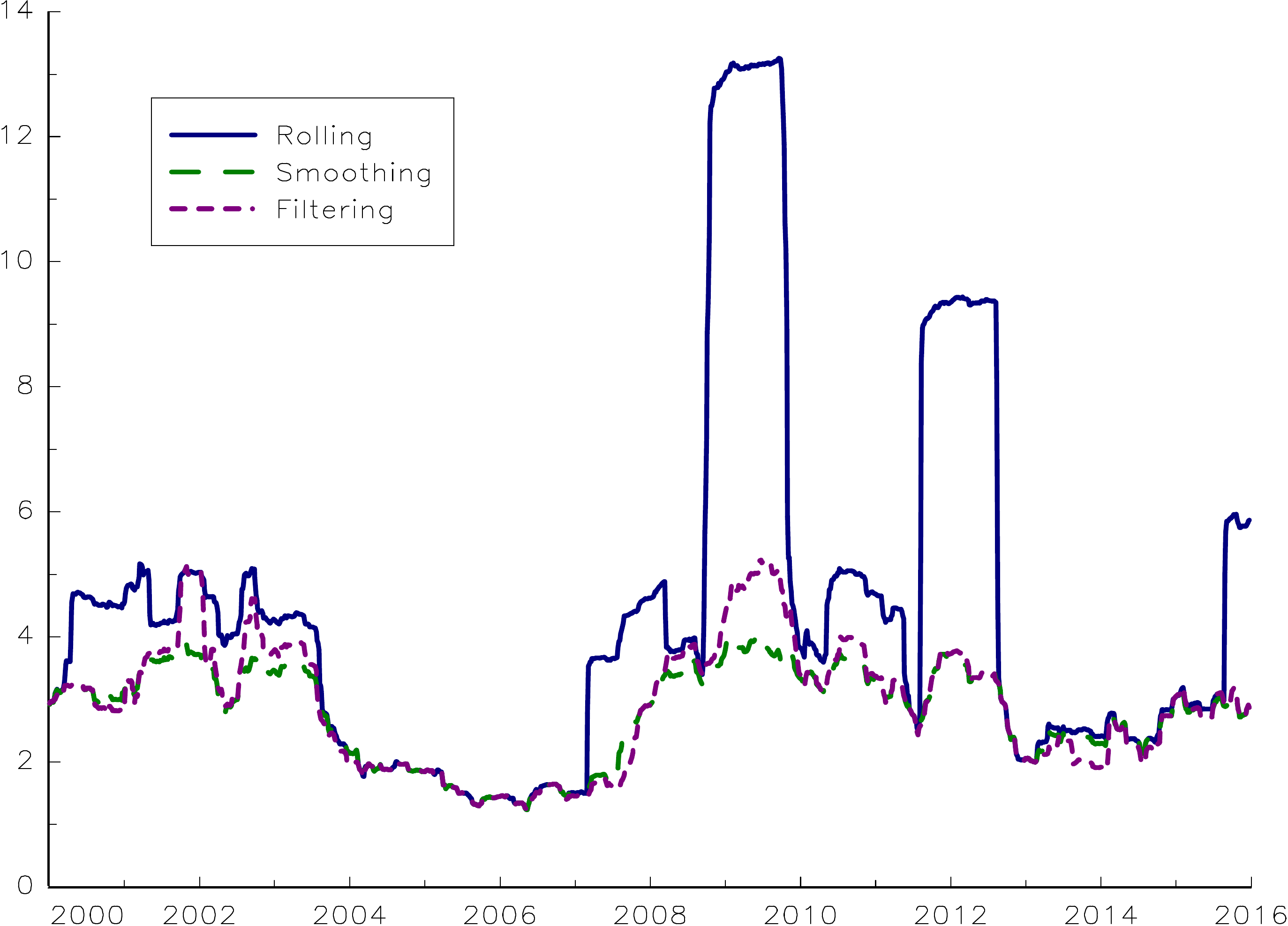}
\end{figure}

\begin{figure}[tbph]
\centering
\caption{Dynamics of the ERC weights (mixture model, filtering method)}
\label{fig:carry13}
\figureskip
\includegraphics[width = \figurewidth, height = \figureheight]{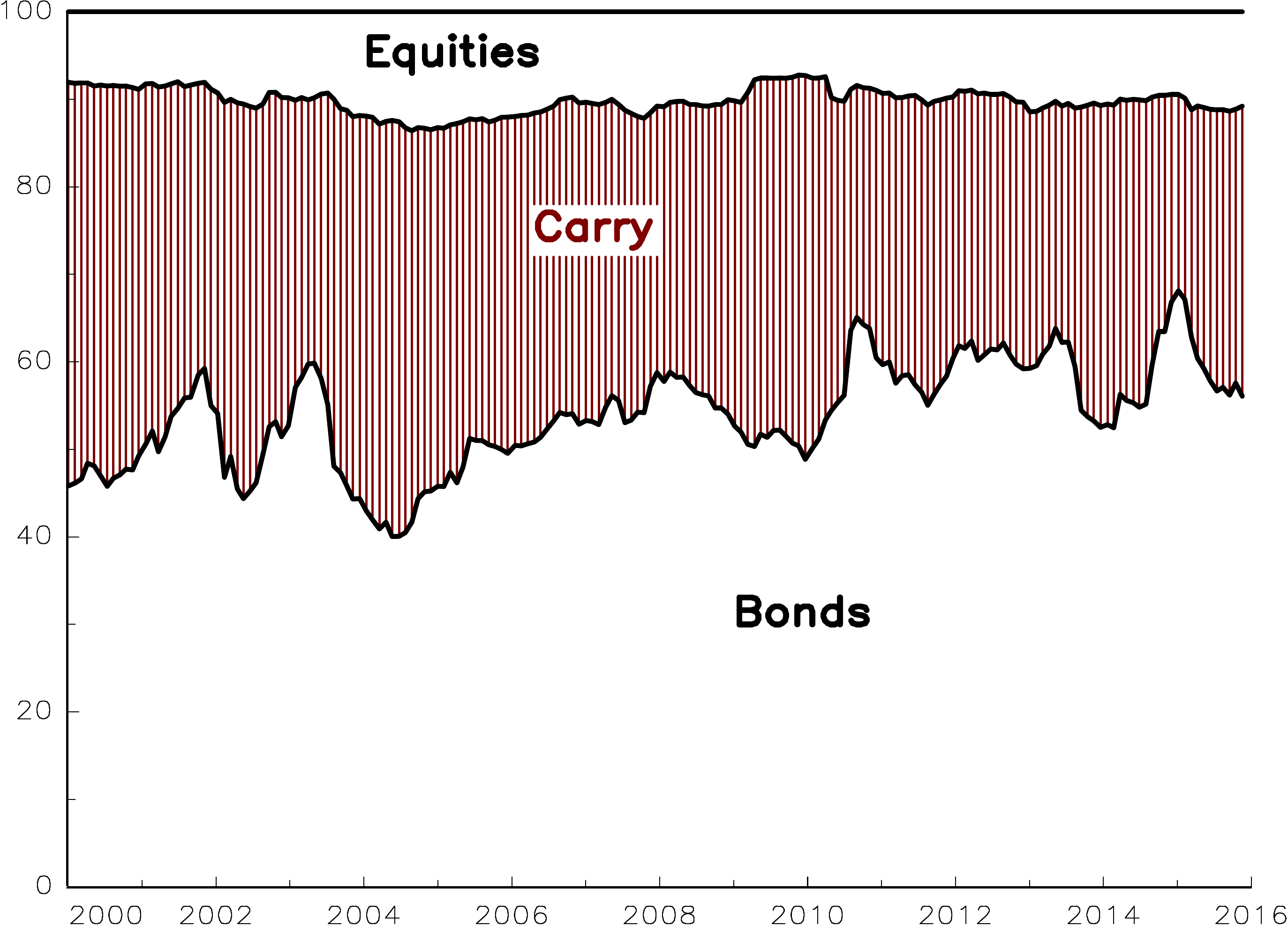}
\end{figure}

We notice some differences between smoothing and filtering
probabilities. One reason is that filtering probabilities
depend on the value taken by $\pi^{\star}$. If we use a small value,
we will delete a lot of observations when computing $\hat{\mu}_{t}$
and $\hat{\Sigma}_{t}$. Therefore, the covariance matrix will be
underestimated implying that the jump probabilities will be higher.
The opposite is also true. The choice of $\pi^{\star}$ is then a key
factor. In Figure \ref{fig:carry11-4}, we observe that some jumps
detected by the smoothing procedure are not identified by the
filtering algorithm. This explains that the filtering volatility of
the carry risk premium is slightly overestimated with respect to the
smoothing volatility (see Figure \ref{fig:carry12}). However, the
filtering procedure appears as a robust out-of-sample method if we
compare rolling and filtering volatilities. Moreover, it produces an
allocation (Figure \ref{fig:carry13}) that is close to the one
obtained by the maximum likelihood approach (see Figure
\ref{fig:carry9-5} on Page \pageref{fig:carry9-5}).

\subsection{Analysis of the results}

\begin{figure}[tbph]
\centering
\caption{Comparison of the carry allocation (weekly model)}
\label{fig:carry14}
\figureskip
\includegraphics[width = \figurewidth, height = \figureheight]{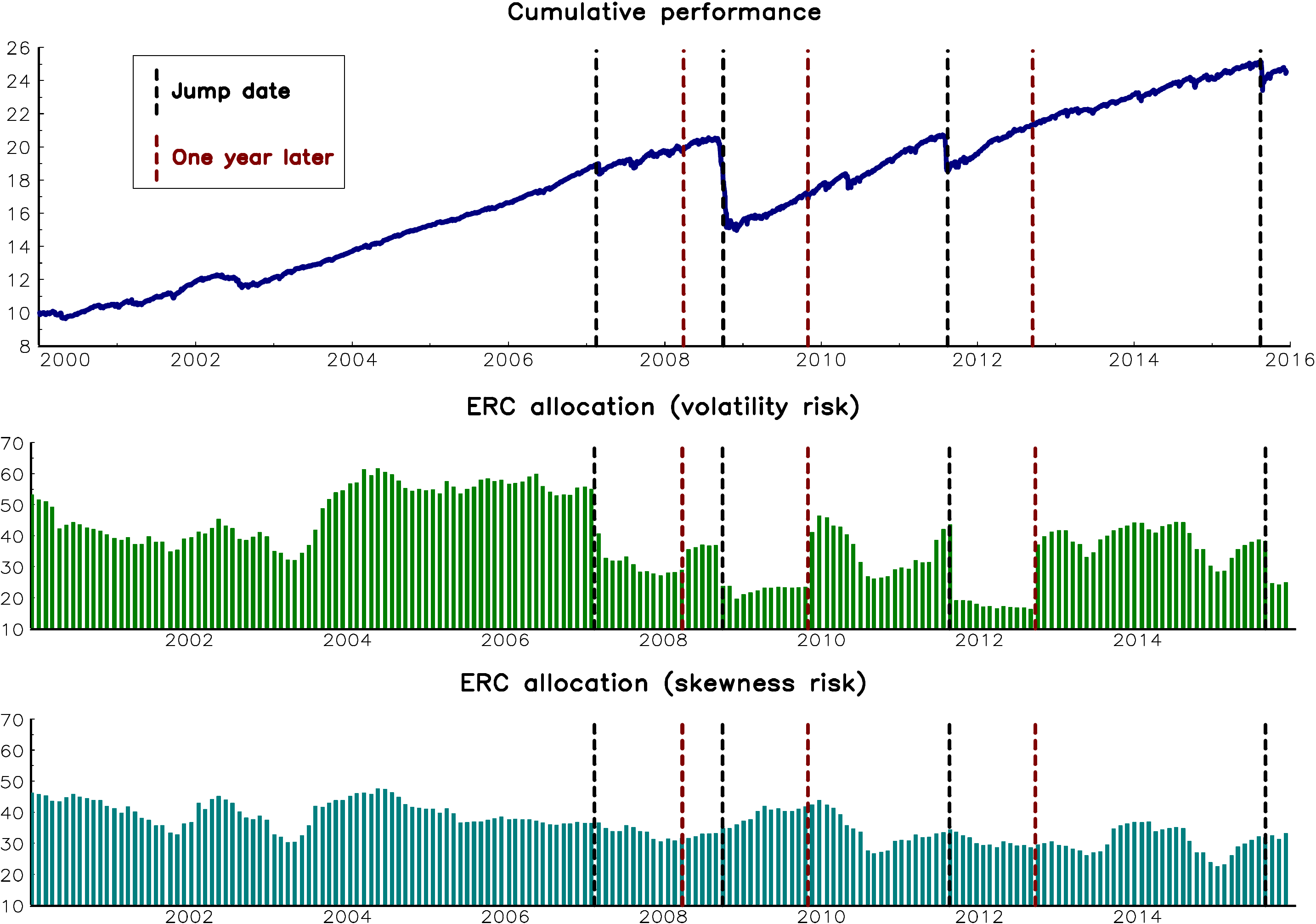}
\end{figure}

If we compare the previous out-of-sample ERC portfolios calculated using
the volatility risk measure and the skewness risk measure; we obtain the following results:
\begin{itemize}
\item The annual turnover of the skewness-based ERC allocation is $40\%$ lower than
the annual turnover of the volatility-based ERC allocation%
\footnote{The turnover is equal to $53.19\%$ for the
volatility-based ERC allocation and $31.45\%$ for the skewness-based
ERC allocation.}.
\item This additional turnover is mainly explained by the jumps of the short
volatility strategy. In Figure \ref{fig:carry14}, we report the
cumulative performance of the carry risk premium and the weights in
the ERC portfolios. In the case of the volatility-based ERC
allocation, we notice that each time we observe a jump, the
volatility-based weight decreases dramatically. One year later after
the jump, the weight highly increases because the jump falls outside
the rolling window.
\item In terms of historical performance, volatility and Sharpe ratio, the two portfolios are
equivalent. However, we observe two different periods. Before 2008,
the volatility-based ERC portfolio largely outperforms the
skewness-based ERC portfolio (28 bps more in terms of Sharpe ratio).
Since 2008, the opposite is true.
\end{itemize}

\section{Application to factor investing and alternative risk premia}

The extensive study on the equity/bond/volatility asset mix policy
in the previous section shows how the properties of risk parity
portfolios can be improved by taking into account skewness or jump
risk. This is particularly true if we consider transaction costs.
Indeed, the volatility-based risk parity portfolio reduces
dramatically the allocation of one asset immediately after the jump.
In this situation, portfolio rebalancing happens in a period of stress and
there is no guarantee that market liquidity will be enough to
ensure fair bid-ask spreads or even execute the sell order.\bigskip

Nowadays, the traditional%
\footnote{When it considers the volatility risk measure.} risk
parity portfolio (and especially the equal risk contribution
portfolio) has become the standard approach for allocating between
risk premia, far ahead mean-variance optimized portfolios (Maeso and
Martellini, 2016). In the case of standard and liquid assets, this
approach is robust. Indeed, the drawdowns of these assets are
generally in line with their volatility. This concerns equities,
investment grade bonds and commodities. However, as illustrated in
this article, traditional risk-based portfolios are not necessarily
optimal investments when assets present high skewness risk. This is
the case of the short volatility strategy, but also other
alternative risk premia like equity size and value risk factors.
This is also true with illiquid assets or low liquidity strategies like
the credit carry risk premium and some leveraged relative value
strategies.

\subsection{Factor investing}

Equity factor investing is a typical example of skewness risk. It
consists in building a diversified portfolio using the more relevant
long-only equity risk factors (size, value, momentum, low beta and
quality). This portfolio combines then two pure (or skewness) risk
premia -- size and value -- and three market anomalies -- momentum,
low beta and quality (Cazalet and Roncalli, 2014). The risk parity
portfolio allocates the same volatility risk to the five risk
factors. However, the risk parity portfolio ignores two main risks
which are difficult to measure using the volatility risk measure
(Hamdan \textsl{et al.}, 2016):
\begin{itemize}
\item The value strategy faces a distressed risk due to the default
risk of value stocks. An example was the impact of the Lehman
Brothers bankruptcy on the performance of value portfolios at the
end of 2008.
\item The size risk premium also faces a distressed risk due to the
liquidity risk of small cap stocks. For instance, the period of low
liquidity between June 2008 and March 2009 has weighted on the
performance of the size factor.
\end{itemize}
By using the framework developed here, we show that the size and
value allocation is generally overestimated in factor-based
investing, because the portfolio construction does not take into
account the skewness risk.

\subsection{Skewness risk premia}

By construction, our framework is particularly relevant for managing
a portfolio of assets that incorporates skewness risk premia. The
equity/bond/volatility asset mix policy is an emblematic
illustration, because the short volatility strategy is one of the
most famous example of skewness risk. We may also consider other
asset allocation problems, like the portfolio of cross-asset carry
risk premia or the introduction of less liquid strategies. Another
example concerns the equity cross-section momentum risk premium. In
a long/short format, this risk premium is subject to a high skewness
risk known as momentum crashes (Daniel and Moskowitz, 2016). In this
context, a volatility-based risk parity portfolio overestimates the
allocation of such risk premium. Moreover, our analysis shows that
this result depends on the implementation of the momentum risk
factor. In particular, the overestimation is more salient for the
winners-minus-losers strategy than for the winners-minus-market
strategy.

\subsection{Volatility hedging versus skewness hedging}

Hamdan \textsl{et al.} (2016) noticed that \textquotedblleft
\textsl{there is a floor to the hedging of the third moment, which
is not the case for the second moment. As a consequence, volatility
diversification or negative correlation leads to reduced volatility,
but increased tail risks in relative terms}\textquotedblright. Using
the equity/bond/carry example, we confirm that skewness hedging
is a difficult task. In table \ref{tab:carry17}, we report the composition
of some risk-based portfolios using the estimates of the weekly model. Using the Gaussian model,
the minimum variance (MV) portfolio is composed of $63.26\%$ of bonds, $2.23\%$ of equities and $34.51\%$
of the carry risk premium. The volatility and the skewness%
\footnote{These statistics are calculated under the assumption that
asset returns follow the Gaussian mixture model.} of the MV
portfolio are then equal to $2.62\%$ and $-2.75$. If we consider the
normal regime of the mixture model and compute the MV portfolio, the
weights become $36.06\%$ for bonds, $0\%$ for equities and $63.95\%$
for the carry risk premium. In both cases, the MV portfolio has a
low or zero allocation in equities, because this asset class is
highly volatile. The allocation in the short volatility strategy is
high when we consider the normal regime. This is the most frequent
situation because jumps are rare. However, the portfolio is also
highly risky because skewness is equal to $-19.81$!. We have
also computed the minimum expected shortfall (MES) portfolio at the
$99\%$ confidence level in the last column%
\footnote{We have also reported the skewness-based ERC portfolio
found previously in the fourth column. Its skewness is equal to
$-6.17$.}. In this case, the portfolio is fully invested in bonds.
Therefore, we notice a big difference between MV and MES portfolios.
Minimizing the volatility leads then to allocating a large part in the
carry risk premium, whereas minimizing the expected shortfall
implies replacing this asset by bonds. By focusing on
volatility, the investor may then neglect the skewness risk.
Moreover, this risk cannot be diversified contrary to the volatility
risk. These results explain that life performance of risk factor and
alternatives risk premia portfolios are not always in line with the
performance of their (in-sample) backtests.

\begin{table}[tbph]
\centering \caption{Volatility and skewness risks of risk-based
portfolios (weekly model)}
\label{tab:carry17}
\tableskip
\begin{tabular}{|c|c:ccc|}
\hline
Portfolio              &  MV              & MV               & ERC             & MES          \\ \hdashline
\multirow{2}{*}{Model} &  Gaussian        & \multicolumn{3}{c|}{Jump model}                   \\
                       &  (full sample)   & Normal           & Mixture         & Mixture      \\
\hline
Bonds                  &  ${\TsIII}63.26\%$ & ${\TsVIII}36.05\%$ & ${\TsIII}52.71\%$ &            $100.00\%$ \\
Equities               &  ${\TsVIII}2.23\%$ &  ${\TsXIII}0.00\%$ & ${\TsIII}10.36\%$ &        ${\TsX}0.00\%$ \\
Carry                  &  ${\TsIII}34.51\%$ & ${\TsVIII}63.95\%$ & ${\TsIII}36.93\%$ &        ${\TsX}0.00\%$ \\ \hdashline
$\sigma\left(x\right)$ &  ${\TsVIII}2.62\%$ &  ${\TsXIII}2.33\%$ & ${\TsVIII}2.75\%$ &        ${\TsX}4.17\%$ \\
$\gamma_1$             &   $-2.75{\TsVIII}$ &  $-19.81{\TsVIII}$ &  $-6.17{\TsVIII}$ & ${\TsX}0.00{\TsVIII}$ \\ \hline
\end{tabular}
\end{table}

\clearpage

\begin{remark}
Some investors think that CTA is a good strategy for hedging the
skewness risk. This belief comes from the 2008 financial crisis,
when CTA has posted positive returns while the drawdown of the
equity market was large. In fact, there is a confusion between two
risks: the CTA strategy has done a good job in hedging the
volatility risk, not the skewness risk. Indeed, the 2008 financial
crisis is more an event of high volatility risk than an episode of
pure skewness risk. This explains that the performance of CTA is
generally disappointing when skewness risks occur (e.g. 2011,
January 2015, etc.)
\end{remark}

\section{Conclusion}

This paper may be viewed as the end of a trilogy, which started with
the study of factor investing (Cazalet and Roncalli, 2014) and was
extended with the analysis of alternative risk premia (Hamdan
\textsl{et al.}, 2016). The two previous analysis have shown the
importance of skewness risk in the universe of risk premia and
confirmed the asymmetric tail risks of some strategies
(Lemp\'eri\`ere \textsl{et al.}, 2014). This is particularly true
for skewness risk premia, which may exhibit extreme jump
movements.\bigskip

As shown by Hamdan \textsl{et al.} (2016), it is an illusion to
think that we can fully hedge the skewness risk, because there is a
floor to skewness diversification. In this paper, we present a new
risk-based model, whose objective is to take into account these
limits. Our framework uses the risk budgeting approach by
considering the expected shortfall measure instead of the volatility
risk measure. However, contrary to Roncalli (2015) who assumed that
asset returns are Gaussian, we use a mixture distribution with two
regimes for modeling asset returns. The first regime is normal
whereas the second regime incorporates jumps. This model allows us to
obtain closed-form formulas for risk contributions, and we can then
compute risk-based portfolios.\bigskip

To illustrate the model, we consider the traditional equity/bond
asset mix policy by introducing the short volatility strategy, which
is a highly skewed asset. The comparison of volatility-based and
skewness-based risk parity portfolios showed that traditional
risk-based portfolios give too much weight to skewed assets in
periods of calm. Skewness-based risk parity portfolios are more robust and
remain invested in skewed assets once the jump risk has been
observed. This example shows that hedging the skewness is a
difficult task. Because the skewness is a statistical measure of
jumps, skewness hedging is equivalent to jump hedging. However,
contrary to the volatility, we show that jumps are difficult to
hedge. In particular, volatility diversification and optimization
can produce portfolios, which present low volatility risk but high
stress risk. This is why skewness diversification is an illusion.
The appropriate answer is then to size the position of the skewed asset
according to its real risk, and not only to its volatility risk.
This result is important for institutional investors and pension
funds managing portfolios of equity risk factors or alternative risk
premia. More generally, our results question the strategic asset
allocation when it includes skewed and illiquid assets. It is of
course the case of hedge fund strategies (Billio \textsl{et al.},
2012), but also other alternative assets like credit, real estate or
infrastructure.

\clearpage

\appendix

\section{Mathematical results}

\subsection{Characteristic function of the jump-diffusion process}

\label{appendix:characteristic-function}

We know that the characteristic function of the compound Poisson process $%
Z_{t}$ has the following expression:%
\begin{equation*}
\mathbb{E}\left[ e^{-iu.Z_{t}}\right] =e^{\lambda t\left( F\left(
u\right) -1\right) }
\end{equation*}%
where $F\left( u\right) =\exp \left( iu^{\top }\tilde{\mu}-\frac{1}{2}%
u^{\top }\tilde{\Sigma}u\right) $ is the characteristic function of
jump
amplitudes. It follows that the characteristic function of the Lévy process $%
\mathrm{d}L_{t}$ is equal to:%
\begin{equation*}
\mathbb{E}\left[ e^{-iu.\mathrm{d}L_{t}}\right] =e^{\left( iu^{\top }\mu -%
\frac{1}{2}u^{\top }\Sigma u\right) \,\mathrm{d}t+\lambda \left(
F\left( u\right) -1\right) \,\mathrm{d}t}
\end{equation*}%
If we assume $\lambda \,\mathrm{d}t$ small enough (or alternatively that
$\lambda $ is small), we can make the following approximation:%
\begin{align*}
\mathbb{E}\left[ e^{-iu.\mathrm{d}L_{t}}\right] & \approx \left( 1+\lambda \,%
\mathrm{d}t\left( F\left( u\right) -1\right) \right) e^{\left(
iu^{\top }\mu
-\frac{1}{2}u^{\top }\Sigma u\right) \,\mathrm{d}t} \\
& \approx \left( 1-\lambda \,\mathrm{d}t\right) \cdot e^{\left(
iu^{\top
}\mu -\frac{1}{2}u^{\top }\Sigma u\right) \,\mathrm{d}t}+\left( \lambda \,%
\mathrm{d}t\right) \cdot e^{\left( iu^{\top }\mu -\frac{1}{2}u^{\top
}\Sigma
u\right) \,\mathrm{d}t}e^{ iu^{\top }\tilde{\mu}-\frac{1}{2}u^{\top }%
\tilde{\Sigma}u }
\end{align*}%
Finally, we obtain:%
\begin{equation*}
\mathbb{E}\left[ e^{-iu.\mathrm{d}L_{t}}\right] \approx \left( 1-\lambda \,%
\mathrm{d}t\right) \cdot e^{\left( iu^{\top }\mu -\frac{1}{2}u^{\top
}\Sigma u\right) \,\mathrm{d}t}+\left( \lambda \,\mathrm{d}t\right)
\cdot e^{iu^{\top }\left( \mu \,\mathrm{d}t+\tilde{\mu}\right)
-\frac{1}{2}u^{\top }\left( \Sigma
\,\mathrm{d}t+\tilde{\Sigma}\right) u\ }
\end{equation*}

\subsection{Skewness of Gaussian mixture models}
\label{appendix:skewness}

We consider the mixture $Y$  of two normal random variables $Y_{1}$
and $Y_{2}$,
whose distribution function is:%
\begin{equation*}
f\left( y\right) =\pi _{1}f_{1}\left( y\right) +\pi _{2}f_{2}\left(
y\right)
\end{equation*}%
where $Y_{1}\sim \mathcal{N}\left( \mu _{1},\sigma _{1}^{2}\right) $, $%
Y_{2}\sim \mathcal{N}\left( \mu _{2},\sigma _{2}^{2}\right) $ and
$\pi
_{1}+\pi _{2}=1$. The $k$-th moment of $Y$ is given by:%
\begin{equation*}
\mathbb{E}\left[ Y^{k}\right] =\pi _{1}\mathbb{E}\left[
Y_{1}^{k}\right] +\pi _{2}\mathbb{E}\left[ Y_{2}^{k}\right]
\end{equation*}%
We know that $\mathbb{E}\left[ Y_{i}\right] =\mu _{i}$,
$\mathbb{E}\left[ Y_{i}^{2}\right] =\mu _{i}^{2}+\sigma _{i}^{2}$
and $\mathbb{E}\left[
Y_{i}^{3}\right] =\mu _{i}^{3}+3\mu _{i}\sigma _{i}^{2}$. We deduce that:%
\begin{equation*}
\mathbb{E}\left[ Y\right] =\pi _{1}\mu _{1}+\pi _{2}\mu _{2}
\end{equation*}%
and:%
\begin{eqnarray*}
\sigma ^{2}\left( Y\right)  &=&\pi _{1}\mathbb{E}\left[
Y_{1}^{2}\right]
+\pi _{2}\mathbb{E}\left[ Y_{2}^{2}\right] -\mathbb{E}^{2}\left[ Y\right]  \\
&=&\pi _{1}\left( \mu _{1}^{2}+\sigma _{1}^{2}\right) +\pi
_{2}\left( \mu _{2}^{2}+\sigma _{2}^{2}\right) -\left( \pi _{1}\mu
_{1}+\pi _{2}\mu
_{2}\right) ^{2} \\
&=&\pi _{1}\sigma _{1}^{2}+\pi _{2}\sigma _{2}^{2}+\pi _{1}\left(
1-\pi _{1}\right) \left( \mu _{1}^{2}+\mu _{2}^{2}\right) -2\pi
_{1}\pi _{2}\mu
_{1}\mu _{2} \\
&=&\pi _{1}\sigma _{1}^{2}+\pi _{2}\sigma _{2}^{2}+\pi _{1}\pi
_{2}\left( \mu _{1}-\mu _{2}\right) ^{2}
\end{eqnarray*}%
because $\pi _{2}=1-\pi _{1}$. Recall that the skewness
coefficient of $Y$
has the following expression:%
\begin{equation*}
\gamma _{1}\left( Y\right) =\mathbb{E}\left[ \left(
\frac{Y-\mathbb{E}\left[ Y\right] }{\sigma \left( Y\right) }\right)
^{3}\right]
\end{equation*}%
It follows:%
\begin{eqnarray*}
\mathbb{E}\left[ \left( Y-\mathbb{E}\left[ Y\right] \right) ^{3}\right]  &=&%
\mathbb{E}\left[ Y^{3}\right] -3\mathbb{E}\left[ Y\right] \sigma
^{2}\left(
Y\right) -\mathbb{E}^{3}\left[ Y\right]  \\
&=&\pi _{1}\left( \mu _{1}^{3}+3\mu _{1}\sigma _{1}^{2}\right) +\pi
_{2}\left( \mu _{2}^{3}+3\mu _{2}\sigma _{2}^{2}\right) - \\
&&3\left( \pi _{1}\mu _{1}+\pi _{2}\mu _{2}\right) \left( \pi
_{1}\sigma _{1}^{2}+\pi _{2}\sigma _{2}^{2}+\pi _{1}\pi _{2}\left(
\mu _{1}-\mu
_{2}\right) ^{2}\right) - \\
&&\left( \pi _{1}\mu _{1}+\pi _{2}\mu _{2}\right) ^{3} \\
&=&\pi _{1}\pi _{2}\left( \pi _{2}-\pi _{1}\right) \left( \mu
_{1}-\mu _{2}\right) ^{3}+3\pi _{1}\pi _{2}\left( \mu _{1}-\mu
_{2}\right) \left( \sigma _{1}^{2}-\sigma _{2}^{2}\right)
\end{eqnarray*}%
We deduce:%
\begin{equation}
\gamma _{1}\left( Y\right) =\frac{\pi _{1}\pi _{2}\left( \left( \pi
_{2}-\pi _{1}\right) \left( \mu _{1}-\mu _{2}\right) ^{3}+3\left(
\mu _{1}-\mu _{2}\right) \left( \sigma _{1}^{2}-\sigma
_{2}^{2}\right) \right) }{\left( \pi _{1}\sigma _{1}^{2}+\pi
_{2}\sigma _{2}^{2}+\pi _{1}\pi _{2}\left( \mu _{1}-\mu _{2}\right)
^{2}\right)^{\nicefrac{3}{2}}}  \label{eq:appendix-skewness1}
\end{equation}

\subsection{Derivation of the EM algorithm}
\label{appendix:em-algorithm}

The log-likelihood function is:%
\begin{equation*}
\ell \left( \theta \right) =\sum_{t=1}^{T}\ln \sum_{j=1}^{2}\pi
_{j}\phi _{n}\left( R_{t};\mu _{j},\Sigma _{j}\right)
\end{equation*}%
The derivative of $\ell \left( \theta \right) $ with respect to $\mu
_{j}$ is equal to:
\begin{equation*}
\frac{\partial \,\ell \left( \theta \right) }{\partial \,\mu _{j}}%
=\sum_{t=1}^{T}\frac{\pi _{j}\phi _{n}\left( R_{t};\mu _{j},\Sigma
_{j}\right) }{\sum_{s=1}^{2}\pi _{s}\phi _{n}\left( R_{t};\mu
_{s},\Sigma _{s}\right) }\Sigma _{j}^{-1}\left( R_{t}-\mu
_{j}\right)
\end{equation*}%
Therefore, the first-order condition is:%
\begin{equation*}
\sum_{t=1}^{T}\pi _{j,t}\Sigma _{j}^{-1}\left( R_{t}-\mu _{j}\right) =%
\mathbf{0}
\end{equation*}%
where:%
\begin{equation*}
\pi _{j,t}=\frac{\pi _{j}\phi _{n}\left( R_{t};\mu _{j},\Sigma _{j}\right) }{%
\sum_{s=1}^{2}\pi _{s}\phi _{n}\left( R_{t};\mu _{s},\Sigma
_{s}\right) }
\end{equation*}%
We deduce the expression of the estimator $\hat{\mu}_{j}$:%
\begin{equation}
\hat{\mu}_{j}=\frac{\sum_{t=1}^{T}\pi _{j,t}R_{t}}{\sum_{t=1}^{T}\pi
_{j,t}} \label{eq:em-appendix-3}
\end{equation}%
For the derivative with respect to $\Sigma _{j}$, we consider the function $%
g\left( \Sigma _{j}^{-1}\right) $ defined as follows:%
\begin{eqnarray*}
g\left( \Sigma _{j}^{-1}\right)  &=&\frac{1}{\left( 2\pi \right)
^{n/2}\left\vert \Sigma _{j}\right\vert ^{1/2}}e^{-\frac{1}{2}\left(
R_{t}-\mu _{j}\right)^{\top } \Sigma _{j}^{-1}\left( R_{t}-\mu
_{j}\right) }
\\
&=&\frac{\left\vert \Sigma _{j}^{-1}\right\vert ^{1/2}}{\left( 2\pi
\right) ^{n/2}}e^{-\frac{1}{2}\limfunc{trace}\left( \Sigma
_{j}^{-1}\left( R_{t}-\mu _{j}\right) \left( R_{t}-\mu _{j}\right)
^{\top }\right) }
\end{eqnarray*}%
It follows\footnote{%
We use the following results:%
\begin{eqnarray*}
\frac{\partial \,\left\vert A\right\vert }{\partial \,A}
&=&\left\vert
A\right\vert \left( A^{-1}\right) ^{\top } \\
\frac{\partial \,\limfunc{trace}\left( A^{\top }B\right) }{\partial
\,A} &=&B
\end{eqnarray*}%
}:%
\begin{eqnarray*}
\frac{\partial \,g\left( \Sigma _{j}^{-1}\right) }{\partial \,\Sigma
_{j}^{-1}} &=&\frac{1}{2}\frac{\left\vert \Sigma
_{j}^{-1}\right\vert ^{-1/2}\left\vert \Sigma _{j}^{-1}\right\vert
\Sigma _{j}}{\left( 2\pi \right)
^{n/2}}e^{-\dfrac{1}{2}\limfunc{trace}\left( \Sigma _{j}^{-1}\left(
R_{t}-\mu _{j}\right) \left( R_{t}-\mu _{j}\right) ^{\top }\right) } \\
&&-\frac{1}{2}\left( R_{t}-\mu _{j}\right) \left( R_{t}-\mu
_{j}\right) ^{\top }\frac{\left\vert \Sigma _{j}^{-1}\right\vert
^{1/2}}{\left( 2\pi \right)
^{n/2}}e^{-\dfrac{1}{2}\limfunc{trace}\left( \Sigma _{j}^{-1}\left(
R_{t}-\mu _{j}\right) \left( R_{t}-\mu _{j}\right) ^{\top }\right) } \\
&=&\frac{1}{\left( 2\pi \right) ^{n/2}\left\vert \Sigma
_{j}\right\vert ^{1/2}}e^{-\dfrac{1}{2}\left( R_{t}-\mu _{j}\right)
\Sigma _{j}^{-1}\left( R_{t}-\mu _{j}\right) ^{\top }}\frac{\left(
\Sigma _{j}-\left( R_{t}-\mu _{j}\right) \left( R_{t}-\mu
_{j}\right) ^{\top }\right) }{2}
\end{eqnarray*}%
We deduce:%
\begin{equation*}
\frac{\partial \,\ell \left( \theta \right) }{\partial \,\Sigma _{j}^{-1}}=%
\frac{1}{2}\sum_{t=1}^{T}\frac{\pi _{j}\phi _{n}\left( R_{t};\mu
_{j},\Sigma _{j}\right) }{\sum_{s=1}^{2}\pi _{s}\phi _{n}\left(
R_{t};\mu _{s},\Sigma _{s}\right) }\left( \Sigma _{j}-\left(
R_{t}-\mu _{j}\right) \left( R_{t}-\mu _{j}\right) ^{\top }\right)
\end{equation*}%
The first-order condition is then:%
\begin{equation*}
\sum_{t=1}^{T}\pi _{j,t}\left( \Sigma _{j}-\left( R_{t}-\mu
_{j}\right) \left( R_{t}-\mu _{j}\right) ^{\top }\right) =0
\end{equation*}%
It follows that the estimator $\hat{\Sigma}_{j}$ is equal to:%
\begin{equation}
\hat{\Sigma}_{j}=\frac{\sum_{t=1}^{T}\pi _{j,t}\left( R_{t}-\hat{\mu}%
_{j}\right) \left( R_{t}-\hat{\mu}_{j}\right) ^{\top
}}{\sum_{t=1}^{T}\pi _{j,t}}  \label{eq:em-appendix-4}
\end{equation}%
Regarding the mixture probabilities $\pi _{j}$, the first-order
condition implies:
\begin{equation*}
\sum_{t=1}^{T}\frac{\phi _{n}\left( R_{t};\mu _{j},\Sigma _{j}\right) }{%
\sum_{s=1}^{2}\pi _{s}\phi _{n}\left( R_{t};\mu _{s},\Sigma
_{s}\right) }=c
\end{equation*}%
where $c$ is the Lagrange multiplier associated to the constraint $%
\sum_{j=1}^{2}\pi _{j}=1$. We deduce that $c=T$. We conclude that it
is not possible to directly define the estimator $\hat{\pi}_{j}$.
This is why we have to use another route to obtain the ML
estimators.\bigskip

We introduce the estimator $\hat{\pi}_{j,t}$:%
\begin{equation}
\hat{\pi}_{j,t}=\frac{\pi _{j}\phi _{n}\left( R_{t};\mu _{j},\Sigma
_{j}\right) }{\sum_{s=1}^{2}\pi _{s}\phi _{n}\left( R_{t};\mu
_{s},\Sigma _{s}\right) }  \label{eq:em-appendix-1}
\end{equation}%
$\hat{\pi}_{j,t}$ is the posterior probability of the regime index
for the observation $t$. Knowing $\hat{\pi}_{j,t}$, the estimator
$\hat{\pi}_{j}$\
is given by:%
\begin{equation}
\hat{\pi}_{j}=\frac{\sum_{t=1}^{T}\hat{\pi}_{j,t}}{T}
\label{eq:em-appendix-2}
\end{equation}%
The EM algorithm consists in the following iterations:

\begin{enumerate}
\item Initialize the algorithm with starting values $\pi _{j}^{\left(
0\right) }$, $\mu _{j}^{\left( 0\right) }$ and $\Sigma _{j}^{\left(
0\right) }$; Set $k=0$.

\item Using Equation (\ref{eq:em-appendix-1}), we calculate the posterior
probabilities $\pi _{j,t}$:%
\begin{equation*}
\pi _{j,t}^{\left( k\right) }=\frac{\pi _{j}^{\left( k\right) }\phi
_{n}\left( R_{t};\mu _{j}^{\left( k\right) },\Sigma _{j}^{\left(
k\right) }\right) }{\sum_{s=1}^{2}\pi _{s}^{\left( k\right) }\phi
_{n}\left( R_{t};\mu _{s}^{\left( k\right) },\Sigma _{s}^{\left(
k\right) }\right) }
\end{equation*}

\item Using Equations (\ref{eq:em-appendix-2}), (\ref{eq:em-appendix-3}) and
(\ref{eq:em-appendix-4}), we update the estimators $\hat{\pi}_{j}$, $\hat{\mu%
}_{j}$ and $\hat{\Sigma}_{j}$:%
\begin{eqnarray*}
\pi _{j}^{\left( k+1\right) } &=&\frac{\sum_{t=1}^{T}\pi
_{j,t}^{\left(
k\right) }}{T} \\
\mu _{j}^{\left( k+1\right) } &=&\frac{\sum_{t=1}^{T}\pi
_{j,t}^{\left(
k\right) }R_{t}}{\sum_{t=1}^{T}\pi _{j,t}^{\left( k\right) }} \\
\Sigma _{j}^{\left( k+1\right) } &=&\frac{\sum_{t=1}^{T}\pi
_{j,t}^{\left( k\right) }\left( R_{t}-\mu _{j}^{\left( k+1\right)
}\right) \left( R_{t}-\mu _{j}^{\left( k+1\right) }\right) ^{\top
}}{\sum_{t=1}^{T}\pi _{j,t}^{\left( k\right) }}
\end{eqnarray*}

\item We iterate Steps 2 and 3 until convergence.

\item Finally, we have $\hat{\pi}_{j}=\pi _{j}^{\left( \infty \right) }$, $%
\hat{\mu}_{j}=\pi _{j}^{\left( \infty \right) }$ and $\hat{\Sigma}%
_{j}=\Sigma _{j}^{\left( \infty \right) }$.
\end{enumerate}

\subsection{Expression of the expected shortfall}
\label{appendix:expected-shortfall}

Let $Y\sim \mathcal{N}\left( \mu ,\sigma ^{2}\right) $ be a Gaussian
random
variable. We consider the following quantity:%
\begin{equation*}
\varphi =\mathbb{E}\left[ \mathds{1}\left\{ Y\geq a\right\} \cdot
Y\right]
\end{equation*}%
We have\footnote{%
We consider the change of variable $t=\sigma ^{-1}\left( x-\mu \right) $.}:%
\begin{eqnarray*}
\varphi  &=&\int_{a}^{\infty }\frac{y}{\sigma }\phi \left( \frac{y-\mu }{%
\sigma }\right) \,\mathrm{d}y \\
&=&\int_{\sigma ^{-1}\left( a-\mu \right) }^{\infty }\left( \mu
+\sigma
t\right) \phi \left( t\right) \,\mathrm{d}t \\
&=&\mu \left[ \Phi \left( t\right) \right] _{\sigma ^{-1}\left(
a-\mu \right) }^{\infty }+\frac{\sigma }{\sqrt{2\pi }}\int_{\sigma
^{-1}\left( a-\mu \right) }^{\infty }t\exp \left(
-\frac{1}{2}t^{2}\right) \,\mathrm{d}t
\\
&=&\mu \left( 1-\Phi \left( \frac{a-\mu }{\sigma }\right) \right) +\frac{%
\sigma }{\sqrt{2\pi }}\left[ -\exp \left( -\frac{1}{2}t^{2}\right)
\right]
_{\sigma ^{-1}\left( a-\mu \right) }^{\infty } \\
&=&\mu \Phi \left( -\frac{a-\mu }{\sigma }\right) +\sigma \phi \left( \frac{%
a-\mu }{\sigma }\right)
\end{eqnarray*}%
In the case of the model with jumps, the definition of the expected
shortfall is:%
\begin{equation*}
\limfunc{ES}\nolimits_{\alpha }\left( x\right) =\mathbb{E}\left[
L\left( x\right) \mid L\left( x\right) \geq
\limfunc{VaR}\nolimits_{\alpha }\left( x\right) \right]
\end{equation*}%
where $L(x)=-R(x)$ is the portfolio's loss. It follows that:%
\begin{eqnarray*}
\limfunc{ES}\nolimits_{\alpha }\left( x\right)  &=&\frac{1}{1-\alpha
}\cdot
\mathbb{E}\left[ \mathds{1}\left\{ L\left( x\right) \geq \limfunc{VaR}%
\nolimits_{\alpha }\left( x\right) \right\} \cdot L\left( x\right)
\right]
\\
&=&\frac{1}{1-\alpha }\int_{\limfunc{VaR}\nolimits_{\alpha }\left(
x\right) }^{\infty }yg\left( y\right) \,\mathrm{d}y
\end{eqnarray*}%
where $g\left( y\right) $ is the density function of $L\left(
x\right) $.
Using Equation (\ref{eq:jump1}), we have:%
\begin{equation*}
g\left( y\right) =\left( 1-\lambda \right) \frac{1}{\sigma
_{1}\left( x\right) }\phi \left( \frac{y+\mu _{1}\left( x\right)
}{\sigma _{1}\left( x\right) }\right) +\lambda \frac{1}{\sigma
_{2}\left( x\right) }\phi \left( \frac{y+\mu _{2}\left( x\right)
}{\sigma _{2}\left( x\right) }\right)
\end{equation*}%
We deduce:
\begin{eqnarray*}
\int_{\limfunc{VaR}\nolimits_{\alpha }\left( x\right) }^{\infty
}yg\left(
y\right) \,\mathrm{d}y &=&\left( 1-\lambda \right) \cdot \mathbb{E}\left[ %
\mathds{1}\left\{ L_{1}\left(x\right) \geq \limfunc{VaR}\nolimits_{\alpha }\left(
x\right)
\right\} \cdot L_{1}\left(x\right)\right] + \\
&&\lambda \cdot \mathbb{E}\left[ \mathds{1}\left\{ L_{2}\left(x\right) \geq \limfunc{VaR}%
\nolimits_{\alpha }\left( x\right) \right\} \cdot L_{2}\left(x\right) \right]
\end{eqnarray*}%
where $L_{1}\left(x\right)\sim \mathcal{N}\left( -\mu _{1}\left( x\right) ,\sigma
_{1}^{2}\left( x\right) \right) $ and $L_{2}\left(x\right)\sim \mathcal{N}\left(
-\mu _{2}\left( x\right) ,\sigma _{2}^{2}\left( x\right) \right) $.
Finally, we
obtain:%
\begin{eqnarray*}
\limfunc{ES}\nolimits_{\alpha }\left( x\right)  &=&\frac{1-\lambda }{%
1-\alpha }\left( \sigma _{1}\left( x\right) \phi \left( \frac{\limfunc{VaR}%
\nolimits_{\alpha }\left( x\right) +\mu _{1}\left( x\right) }{\sigma
_{1}\left( x\right) }\right) -\mu _{1}\left( x\right) \Phi \left( -\frac{%
\limfunc{VaR}\nolimits_{\alpha }\left( x\right) +\mu _{1}\left( x\right) }{%
\sigma _{1}\left( x\right) }\right) \right) + \\
&&\frac{\lambda }{1-\alpha }\left( \sigma _{2}\left( x\right) \phi
\left( \frac{\limfunc{VaR}\nolimits_{\alpha }\left( x\right) +\mu
_{2}\left( x\right) }{\sigma _{2}\left( x\right) }\right) -\mu
_{2}\left( x\right) \Phi \left(
-\frac{\limfunc{VaR}\nolimits_{\alpha }\left( x\right) +\mu
_{2}\left( x\right) }{\sigma _{2}\left( x\right) }\right) \right)
\end{eqnarray*}%
This expression depends on the value taken by the value-at-risk,
which is
defined by:%
\begin{equation*}
\Pr \left\{ L\left( x\right) \leq \limfunc{VaR}\nolimits_{\alpha
}\left( x\right) \right\} =\alpha
\end{equation*}%
It follows:%
\begin{equation*}
\int_{-\infty }^{\limfunc{VaR}\nolimits_{\alpha }\left( x\right)
}g\left( y\right) \,\mathrm{d}y=\alpha
\end{equation*}%
or:%
\begin{equation*}
\left( 1-\lambda \right) \cdot \Phi \left( \frac{\limfunc{VaR}%
\nolimits_{\alpha }\left( x\right) +\mu _{1}\left( x\right) }{\sigma
_{1}\left( x\right) }\right) +\lambda \cdot \Phi \left( \frac{\limfunc{VaR}%
\nolimits_{\alpha }\left( x\right) +\mu _{2}\left( x\right) }{\sigma
_{2}\left( x\right) }\right) =\alpha
\end{equation*}%
The calculation of the value-at-risk can then be done numerically
using a bisection algorithm.

\subsection{Calculation of the marginal expected shortfall}
\label{appendix:marginal-expected-shortfall}

We introduce the following notations:%
\begin{equation*}
h_{i}\left( x\right) =\frac{\limfunc{VaR}\nolimits_{\alpha }\left(
x\right) +\mu _{i}\left( x\right) }{\sigma _{i}\left( x\right) }
\end{equation*}%
Recall that the value-at-risk is defined by this implicit equation:%
\begin{equation*}
\left( 1-\lambda \right) \cdot \Phi \left( h_{1}\left( x\right)
\right) +\lambda \cdot \Phi \left( h_{2}\left( x\right) \right)
=\alpha
\end{equation*}%
We deduce:%
\begin{equation*}
\left( 1-\lambda \right) \cdot \phi \left( h_{1}\left( x\right)
\right)
\partial _{x}h_{1}\left( x\right) +\lambda \cdot \phi \left( h_{2}\left(
x\right) \right) \partial _{x}h_{2}\left( x\right) =0
\end{equation*}%
where:%
\begin{equation*}
\partial _{x}h_{1}\left( x\right) =\frac{\partial _{x}\limfunc{VaR}%
\nolimits_{\alpha }\left( x\right) +\mu }{\sigma _{1}\left( x\right) }-\frac{%
h_{1}\left( x\right) }{\sigma _{1}^{2}\left( x\right) }\Sigma x
\end{equation*}%
and:%
\begin{equation*}
\partial _{x}h_{2}\left( x\right) =\frac{\partial _{x}\limfunc{VaR}%
\nolimits_{\alpha }\left( x\right) +\mu +\tilde{\mu}}{\sigma
_{2}\left(
x\right) }-\frac{h_{2}\left( x\right) }{\sigma _{2}^{2}\left( x\right) }%
\left( \Sigma +\tilde{\Sigma}\right) x
\end{equation*}%
We finally obtain the following gradient of the value-at-risk:%
\begin{equation*}
\partial _{x}\limfunc{VaR}\nolimits_{\alpha }\left( x\right) =\frac{\varpi
_{1}\left( x\right) \left( \frac{h_{1}\left( x\right) }{\sigma
_{1}\left(
x\right) }\Sigma x-\mu \right) +\varpi _{2}\left( x\right) \left( \frac{%
h_{2}\left( x\right) }{\sigma _{2}\left( x\right) }\left( \Sigma +\tilde{%
\Sigma}\right) x-\left( \mu +\tilde{\mu}\right) \right) }{\varpi
_{1}\left( x\right) +\varpi _{2}\left( x\right) }
\end{equation*}%
where:%
\begin{equation*}
\varpi _{i}\left( x\right) =\frac{\pi _{i}\phi \left( h_{i}\left(
x\right) \right) }{\sigma _{i}\left( x\right) }
\end{equation*}%
Let us now consider the expected shortfall. We have:
\begin{eqnarray*}
\limfunc{ES}\nolimits_{\alpha }\left( x\right)  &=&\frac{1-\lambda }{%
1-\alpha }\sigma _{1}\left( x\right) \phi \left( h_{1}\left(
x\right) \right) -\frac{1-\lambda }{1-\alpha }\mu _{1}\left(
x\right) \Phi \left(
-h_{1}\left( x\right) \right) + \\
&&\frac{\lambda }{1-\alpha }\sigma _{2}\left( x\right) \phi \left(
h_{2}\left( x\right) \right) -\frac{\lambda }{1-\alpha }\mu
_{2}\left( x\right) \Phi \left( -h_{2}\left( x\right) \right)
\end{eqnarray*}%
We deduce\footnote{%
Because we have $\partial _{x}\phi \left( x\right) =-x\phi \left(
x\right) $
and $\phi \left( -x\right) =\phi \left( x\right) $.}:%
\begin{eqnarray*}
\partial _{x}\limfunc{ES}\nolimits_{\alpha }\left( x\right)  &=&\frac{%
1-\lambda }{1-\alpha }\left( \partial _{x}\sigma _{1}\left( x\right)
\phi \left( h_{1}\left( x\right) \right) -\sigma _{1}\left( x\right)
h_{1}\left( x\right) \phi \left( h_{1}\left( x\right) \right)
\partial _{x}h_{1}\left(
x\right) \right) - \\
&&\frac{1-\lambda }{1-\alpha }\left( \partial _{x}\mu _{1}\left(
x\right) \Phi \left( -h_{1}\left( x\right) \right) -\mu _{1}\left(
x\right) \phi \left( h_{1}\left( x\right) \right) \partial
_{x}h_{1}\left( x\right)
\right) + \\
&&\frac{\lambda }{1-\alpha }\left( \partial _{x}\sigma _{2}\left(
x\right) \phi \left( h_{2}\left( x\right) \right) -\sigma _{2}\left(
x\right) h_{2}\left( x\right) \phi \left( h_{2}\left( x\right)
\right) \partial
_{x}h_{2}\left( x\right) \right) - \\
&&\frac{\lambda }{1-\alpha }\left( \partial _{x}\mu _{2}\left(
x\right) \Phi \left( -h_{2}\left( x\right) \right) -\mu _{2}\left(
x\right) \phi \left( h_{2}\left( x\right) \right) \partial
_{x}h_{2}\left( x\right) \right)
\end{eqnarray*}%
We obtain:%
\begin{eqnarray*}
\partial _{x}\limfunc{ES}\nolimits_{\alpha }\left( x\right)  &=&\frac{\varpi
_{1}\left( x\right) }{1-\alpha }\delta _{1}\left( x\right)
+\frac{\varpi
_{2}\left( x\right) }{1-\alpha }\delta _{2}\left( x\right) - \\
&&\frac{1}{1-\alpha }\left( \left( 1-\lambda \right) \mu \Phi \left(
-h_{1}\left( x\right) \right) +\lambda \left( \mu
+\tilde{\mu}\right) \Phi \left( -h_{2}\left( x\right) \right)
\right)
\end{eqnarray*}%
where\footnote{%
We use the result: $\mu _{1}\left( x\right) -\sigma _{1}\left(
x\right) h_{1}\left( x\right) =\mu _{2}\left( x\right) -\sigma
_{2}\left( x\right)
h_{2}\left( x\right) =-\limfunc{VaR}\nolimits_{\alpha }\left( x\right) $.}:%
\begin{equation*}
\delta _{1}\left( x\right) =\left( 1+\frac{h_{1}\left( x\right)
}{\sigma _{1}\left( x\right) }\limfunc{VaR}\nolimits_{\alpha }\left(
x\right) \right)
\Sigma x-\limfunc{VaR}\nolimits_{\alpha }\left( x\right) \left( \partial _{x}%
\limfunc{VaR}\nolimits_{\alpha }\left( x\right) +\mu \right)
\end{equation*}%
and:%
\begin{eqnarray*}
\delta _{2}\left( x\right)  &=&\left( 1+\frac{h_{2}\left( x\right)
}{\sigma _{2}\left( x\right) }\limfunc{VaR}\nolimits_{\alpha }\left(
x\right) \right)
\left( \Sigma +\tilde{\Sigma}\right) x- \\
&&\limfunc{VaR}\nolimits_{\alpha }\left( x\right) \left( \partial _{x}%
\limfunc{VaR}\nolimits_{\alpha }\left( x\right) +\mu
+\tilde{\mu}\right)
\end{eqnarray*}

\subsection{Existence and uniqueness of the RB portfolio}
\label{appendix:uniqueness}

In our model, the portfolio's loss $L\left( x\right) $ can be decomposed in
two components:%
\begin{equation*}
L\left( x\right) =B\cdot L_{1}\left( x\right) +\left( 1-B\right) \cdot
L_{2}\left( x\right)
\end{equation*}%
where $B\sim \mathcal{B}\left( \pi _{1}\right) $, $L_{1}\left( x\right) $ is
the portfolio's loss under the first regime without jumps and $L_{2}\left(
x\right) $ is the portfolio's loss under the second regime with jumps. In
order to obtain some interesting results, we will bound the expected
shortfall of the jump model by another expected shortfall.

\subsubsection{VaR lower bound}

Let us consider the value-at-risk, which is defined as:%
\begin{equation*}
\int_{-\infty }^{\func{VaR}_{\alpha }\left( x\right) }g\left( y\right) \,%
\mathrm{d}y=\alpha
\end{equation*}%
where $g\left( y\right) =\pi _{1}g_{1}\left( y\right) +\pi _{2}g_{2}\left(
y\right) $, $g_{1}\left( y\right)$ is the density function of $L_{1}\left( x\right)$
and $g_{2}\left( y\right)$ is the density function of $L_{2}\left( x\right)$.
We also have:%
\begin{equation*}
\int_{\func{VaR}_{\alpha }\left( x\right) }^{+\infty } \left(\pi _{1}g_{1}\left(
y\right) +\pi _{2}g_{2}\left( y\right)\right) \,\mathrm{d}y=1-\alpha
\end{equation*}%
Since $\pi _{2}g_{2}\left( y\right) \geq 0$, we have:%
\begin{equation*}
\int_{\func{VaR}_{\alpha }\left( x\right) }^{+\infty }\pi _{1}g_{1}\left(
y\right) \,\mathrm{d}y\leq 1-\alpha
\end{equation*}%
and:%
\begin{equation*}
\int_{\func{VaR}_{\alpha }\left( x\right) }^{+\infty }g_{1}\left(
y\right) \,\mathrm{d}y\leq \frac{1-\alpha }{\pi _{1}}
\end{equation*}%
We set $1-\alpha ^{\prime }=\left( 1-\alpha \right) /\pi _{1}$. It follows:%
\begin{eqnarray*}
\alpha ^{\prime } &=&\frac{\pi _{1}-\left( 1-\alpha \right) }{\pi _{1}} \\
&=&\frac{\alpha -\lambda }{1-\lambda }
\end{eqnarray*}%
If we assume that\footnote{%
This is the normal case, because the confidence level $\alpha $ is generally
high ($\alpha \geq 50\%$), whereas the intensity parameter $\lambda $ is low
($\lambda \leq 50\%$).} $\alpha \geq \lambda $, we obtain:%
\begin{equation*}
\int_{\func{VaR}_{\alpha }\left( x\right) }^{+\infty }g_{1}\left(
y\right) \,\mathrm{d}y\leq 1-\alpha ^{\prime }=\int_{\func{VaR}_{\alpha
^{\prime }}^{1}\left( x\right) }^{+\infty }g_{1}\left( y\right) \,\mathrm{d}%
y
\end{equation*}%
where $\limfunc{VaR}_{\alpha ^{\prime }}^{1}(x)$ is the value-at-risk of the
portfolio's loss under the first regime at the confidence level $\alpha
^{\prime }$. As $\int_{a}^{b}g_{1}\left( y\right) \,\mathrm{d}y$ is a
decreasing function of $a$, we conclude:%
\begin{equation*}
\func{VaR}\nolimits_{\alpha }\left( x\right) \geq \func{VaR}%
\nolimits_{\alpha ^{\prime }}^{1}\left( x\right)
\end{equation*}%
We have found a lower bound $\func{VaR}^{-}$ of the value-at-risk, which
is equal to:%
\begin{equation*}
\limfunc{VaR}\nolimits^{-}=-\mu _{1}\left( x\right) +\Phi ^{-1}\left( \frac{%
\alpha -\lambda }{1-\lambda }\right) \sigma _{1}\left( x\right)
\end{equation*}

\subsubsection{ES lower bound}

The expected shortfall is equal to:%
\begin{eqnarray*}
\func{ES}\nolimits_{\alpha }\left( x\right)  &=&\frac{1}{1-\alpha }\int_{%
\func{VaR}_{\alpha }\left( x\right) }^{+\infty }yg\left( y\right) \,%
\mathrm{d}y \\
&=&\frac{\pi _{1}}{1-\alpha }\int_{\func{VaR}_{\alpha }\left( x\right)
}^{+\infty }yg_{1}\left( y\right) \,\mathrm{d}y+\frac{\pi _{2}}{1-\alpha }%
\int_{\func{VaR}_{\alpha }\left( x\right) }^{+\infty }yg_{2}\left(
y\right) \,\mathrm{d}y
\end{eqnarray*}%
We consider the worst case scenario when $g_{2}\left( y\right) $ is a Dirac
measure on $\func{VaR}_{\alpha }\left( x\right) $. Using the previous
lower bound of the value-at-risk, it follows:%
\begin{equation*}
\int_{\func{VaR}_{\alpha }\left( x\right) }^{+\infty }yg_{2}\left(
y\right) \,\mathrm{d}y\geq \left( 1-\alpha \right) \func{VaR}%
\nolimits_{\alpha }\left( x\right) \geq \left( 1-\alpha \right) \func{VaR}%
\nolimits_{\alpha ^{\prime }}^{1}\left( x\right)
\end{equation*}%
As the expected shortfall is an increasing function of the confidence level
and the value-at-risk, we obtain:%
\begin{equation*}
\int_{\func{VaR}_{\alpha }\left( x\right) }^{+\infty }yg_{1}\left(
y\right) \,\mathrm{d}y\geq \int_{\func{VaR}\nolimits_{\alpha ^{\prime
}}^{1}\left( x\right) }^{+\infty }yg_{1}\left( y\right) \,\mathrm{d}y=\left(
1-\alpha ^{\prime }\right) \func{ES}\nolimits_{\alpha ^{\prime
}}^{1}\left( x\right)
\end{equation*}%
where $\func{ES}\nolimits_{\alpha ^{\prime }}^{1}\left( x\right) $ is the
expected shortfall under the first regime at the confidence level $\alpha
^{\prime }$. We deduce:%
\begin{equation*}
\func{ES}\nolimits_{\alpha }\left( x\right) \geq \frac{\pi _{1}}{1-\alpha
}\left( 1-\alpha ^{\prime }\right) \func{ES}\nolimits_{\alpha ^{\prime
}}^{1}\left( x\right) +\frac{\pi _{2}}{1-\alpha }\left( 1-\alpha \right)
\func{VaR}\nolimits_{\alpha ^{\prime }}^{1}\left( x\right)
\end{equation*}%
or:%
\begin{equation*}
\limfunc{ES}\nolimits_{\alpha }\left( x\right) \geq \limfunc{ES}%
\nolimits_{\alpha ^{\prime }}^{1}\left( x\right) +\pi _{2}\limfunc{VaR}%
\nolimits_{\alpha ^{\prime }}^{1}\left( x\right)
\end{equation*}%
Using the analytical expression of $\limfunc{ES}\nolimits_{\alpha ^{\prime
}}^{1}\left( x\right) $ and $\limfunc{VaR}\nolimits_{\alpha ^{\prime
}}^{1}\left( x\right) $, the lower bound $\limfunc{ES}\nolimits^{-}$ of the
expected shortfall becomes:%
\begin{eqnarray*}
\limfunc{ES}\nolimits^{-} &=&  -\mu _{1}\left( x\right) +\frac{%
1-\lambda }{1-\alpha }\phi \left( \Phi ^{-1}\left( \frac{\alpha -\lambda }{%
1-\lambda }\right) \right) \sigma _{1}\left( x\right)  + \\
&&\lambda \left( -\mu _{1}\left( x\right) +\Phi ^{-1}\left( \frac{\alpha
-\lambda }{1-\lambda }\right) \sigma _{1}\left( x\right) \right)  \\
&=&-\left( 1+\lambda \right) \mu _{1}\left( x\right) +\left( \frac{1-\lambda
}{1-\alpha }\phi \left( \Phi ^{-1}\left( \frac{\alpha -\lambda }{1-\lambda }%
\right) \right) +\lambda \Phi ^{-1}\left( \frac{\alpha -\lambda }{1-\lambda }%
\right) \right) \sigma _{1}\left( x\right)
\end{eqnarray*}

\subsubsection{Main result}

Roncalli (2015) showed that when the risk measure $\mathcal{R}\left(
x\right) $ has the following form:%
\begin{equation*}
\mathcal{R}\left( x\right) =-\left( \mu \left( x\right) -r\right) +c\cdot
\sigma \left( x\right)
\end{equation*}%
then the RB portfolio exists and is unique if the following condition holds:%
\begin{equation*}
c\geq \limfunc{SR}\nolimits^{+}=\max \left( \sup_{x\in \left[ 0,1\right]
^{n}}\limfunc{SR}\left( x\mid r\right) ,0\right)
\end{equation*}%
In our case, it follows that the RB portfolio exists and is unique if we
have:
\begin{equation*}
\alpha \geq \max \left( \alpha ^{-},0\right)
\end{equation*}%
where:%
\begin{equation*}
\frac{1-\lambda }{1-\alpha ^{-}}\phi \left( \Phi ^{-1}\left( \frac{%
\alpha ^{-}-\lambda }{1-\lambda }\right) \right) +\lambda \Phi ^{-1}\left(
\frac{\alpha ^{-}-\lambda }{1-\lambda }\right) =\left( 1+\lambda
\right) \limfunc{SR}\nolimits_{1}^{+}
\end{equation*}%
and $\limfunc{SR}\nolimits_{1}^{+}$ is the maximum Sharpe ratio under the
first regime, which is calculated with a zero-interest rate.

\subsection{Equivalence between filtering and thresholding methods in the one-dimensional case}
\label{appendix:thresholding1}

We use the following parametrization of the probability density function:%
\begin{equation*}
f\left( y\right) =\left( 1-\pi \right) \phi _{1}\left( y_{t},\mu \,\mathrm{d}%
t,\sigma ^{2}\,\mathrm{d}t\right) +\pi \phi _{1}\left( y_{t},\mu \,\mathrm{d}%
t+\tilde{\mu},\sigma ^{2}\,\mathrm{d}t+\tilde{\sigma}^{2}\right)
\end{equation*}%
The posteriori probability to have a jump at time $t$ is:%
\begin{equation*}
\hat{\pi}_{t}=\frac{\pi \phi _{1}\left( y_{t},\mu \,\mathrm{d}t+\tilde{\mu}%
,\sigma ^{2}\,\mathrm{d}t+\tilde{\sigma}^{2}\right) }{\left( 1-\pi \right)
\phi _{1}\left( y_{t},\mu \,\mathrm{d}t,\sigma ^{2}\,\mathrm{d}t\right) +\pi
\phi _{1}\left( y_{t},\mu \,\mathrm{d}t+\tilde{\mu},\sigma ^{2}\,\mathrm{d}t+%
\tilde{\sigma}^{2}\right) }
\end{equation*}%
It follows that $\hat{\pi}_{t}\geq \pi ^{\star }$ is equivalent to:%
\begin{equation*}
\phi _{1}\left( y_{t},\mu \,\mathrm{d}t,\sigma ^{2}\,\mathrm{d}t\right) \leq
\frac{\pi \left( 1-\pi ^{\star }\right) }{\pi ^{\star }\left( 1-\pi \right) }%
\phi _{1}\left( y_{t},\mu \,\mathrm{d}t+\tilde{\mu},\sigma ^{2}\,\mathrm{d}t+%
\tilde{\sigma}^{2}\right)
\end{equation*}%
or:%
\begin{equation*}
\frac{\left( y_{t}-\mu \,\mathrm{d}t-\tilde{\mu}\right) ^{2}}{\sigma ^{2}\,%
\mathrm{d}t+\tilde{\sigma}^{2}}-\frac{\left( y_{t}-\mu \,\mathrm{d}t\right)
^{2}}{\sigma ^{2}\,\mathrm{d}t}\leq 2\ln \frac{\pi \left( 1-\pi ^{\star
}\right) }{\pi ^{\star }\left( 1-\pi \right) }+\ln \frac{\sigma ^{2}\,%
\mathrm{d}t}{\sigma ^{2}\,\mathrm{d}t+\tilde{\sigma}^{2}}
\end{equation*}%
We finally obtain:%
\begin{equation*}
\alpha y_{t}^{2}+2\beta y_{t}+\gamma \geq 0
\end{equation*}%
where:%
\begin{eqnarray*}
\alpha  &=&\tilde{\sigma}^{2} \\
\beta  &=&\left( \tilde{\mu}\sigma ^{2}-\mu \tilde{\sigma}^{2}\right) \,%
\mathrm{d}t \\
\gamma  &=&\left( \mu \,\mathrm{d}t\right) ^{2}\left( \sigma ^{2}\,\mathrm{d}%
t+\tilde{\sigma}^{2}\right) -\left( \mu \,\mathrm{d}t+\tilde{\mu}\right)
^{2}\sigma ^{2}\,\mathrm{d}t+ \\
&&\left( 2\ln \frac{\pi \left( 1-\pi ^{\star }\right) }{\pi ^{\star }\left(
1-\pi \right) }+\ln \frac{\sigma ^{2}\,\mathrm{d}t}{\sigma ^{2}\,\mathrm{d}t+%
\tilde{\sigma}^{2}}\right) \left( \sigma ^{2}\,\mathrm{d}t+\tilde{\sigma}%
^{2}\right) \sigma ^{2}\,\mathrm{d}t
\end{eqnarray*}%
Under some assumptions\footnote{%
We must have $\left\vert \tilde{\mu}\right\vert \geq \left\vert \mu
\right\vert \,\mathrm{d}t$ and $\tilde{\sigma}^{2}\gg \sigma ^{2}\,\mathrm{d}%
t$.}, we can show that $\Delta ^{\prime }=\beta ^{2}-\alpha \gamma \geq 0$
implying that there are two roots:%
\begin{equation*}
y^{-}=\frac{-\beta -\sqrt{\beta ^{2}-\alpha \gamma }}{\alpha }
\end{equation*}%
and:%
\begin{equation*}
y^{+}=\frac{-\beta +\sqrt{\beta ^{2}-\alpha \gamma }}{\alpha }
\end{equation*}%
We deduce:%
\begin{equation*}
\hat{\pi}_{t}\geq \pi ^{\star }\Leftrightarrow y_{t}\leq y^{-}\text{ or }%
y_{t}\geq y^{+}
\end{equation*}%
We conclude that we observe a jump if the following condition holds\footnote{%
We also have:%
\begin{equation*}
J_{t}=1\Leftrightarrow \left\vert y_{t}+\frac{\beta }{\alpha
}\right\vert
\geq \kappa _{\sigma }\cdot \left( \sigma \sqrt{\mathrm{d}t}\right)
\end{equation*}%
where:%
\begin{equation*}
\kappa _{\sigma }=\frac{\sqrt{\beta ^{2}-\alpha \gamma }}{\alpha
\sigma \sqrt{\mathrm{d}t}}
\end{equation*}%
According to Ait-Sahalia (2004), the multiplicative factor
$\kappa_{\sigma}$ takes its value between $3.5$ and $4.0$.}:%
\begin{equation*}
J_{t}=1\Leftrightarrow y_{t}\leq y^{-}\text{ or }y_{t}\geq y^{+}
\end{equation*}
\bigskip

\begin{figure}[tbph]
\centering
\caption{Detecting the jumps of the carry risk premium (weekly model)}
\label{fig:carry10-1}
\figureskip
\includegraphics[width = \figurewidth, height = \figureheight]{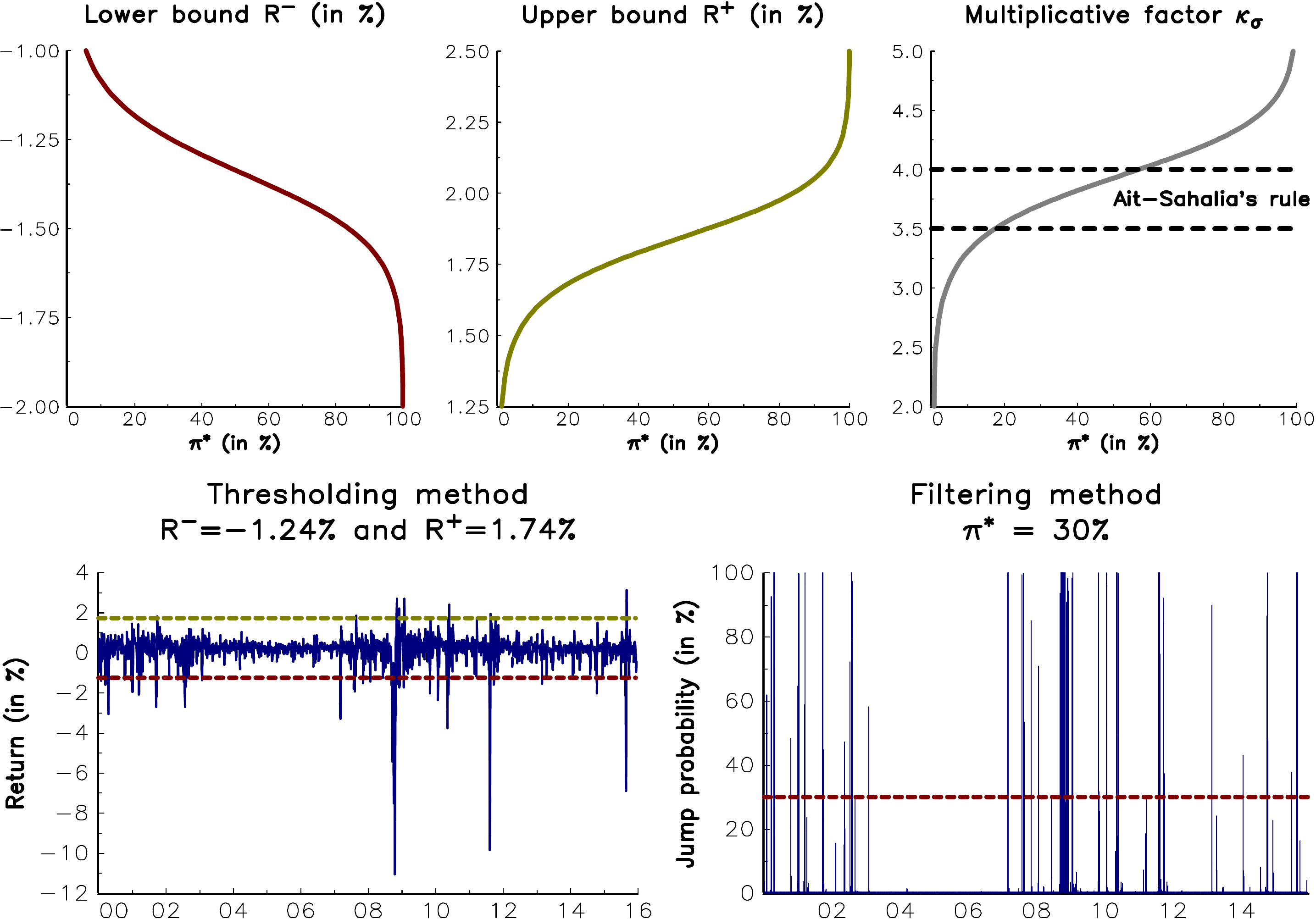}
\end{figure}

To show the equivalence of the two approaches, we consider the
returns $R_t$ of the carry risk premium. Using the weekly model, we
calculate the truncation points of the thresholding approach.
Results are given in top panels in Figure \ref{fig:carry10-1}. For
instance, when $\pi^{\star}$ is equal to $30\%$, we obtain $R^{-} =
-1.24\%$ and $R^{+} = 1.74\%$. Therefore, we believe that we have a
jump when the weekly return is lower than $-1.24\%$ or when it is
higher than $1.74\%$. In the bottom/left panel, we illustrate the
thresholding approach with $R^{-} = -1.24\%$ and $R^{+} = 1.74\%$
whereas the filtering approach is showed in the bottom/right panel
with $\pi^{\star}=30\%$. As expected, the two approaches detect the
jumps at the same times.

\subsection{Derivation of the thresholding approach in the multi-dimensional case}
\label{appendix:thresholding2}

We use the following parametrization of the probability density function:%
\begin{equation*}
f\left( y\right) =\left( 1-\pi \right) \phi _{n}\left( y_{t},\mu _{1},\Sigma
_{1}\right) +\pi \phi _{n}\left( y_{t},\mu _{2},\Sigma _{2}\right)
\end{equation*}%
The posteriori probability to have a jump at time $t$ is:%
\begin{equation*}
\hat{\pi}_{t}=\frac{\pi \phi _{n}\left( y_{t},\mu _{2},\Sigma _{2}\right) }{%
\left( 1-\pi \right) \phi _{n}\left( y_{t},\mu _{1},\Sigma _{1}\right) +\pi
\phi _{n}\left( y_{t},\mu _{2},\Sigma _{2}\right) }
\end{equation*}%
It follows that $\hat{\pi}_{t}\geq \pi ^{\star }$ is equivalent to:%
\begin{equation*}
\left( y_{t}-\mu _{1}\right) ^{\top }\Sigma _{1}^{-1}\left(
y_{t}-\mu _{1}\right) -\left( y_{t}-\mu _{2}\right) ^{\top }\Sigma
_{2}^{-1}\left( y_{t}-\mu _{2}\right) +2\ln \frac{\pi \left( 1-\pi
^{\ast }\right) }{\pi
^{\ast }\left( 1-\pi \right) }+\ln \frac{\left\vert \Sigma _{1}\right\vert }{%
\left\vert \Sigma _{2}\right\vert }\geq 0
\end{equation*}%
We deduce:%
\begin{eqnarray*}
&&y_{t}^{\top }\left( \Sigma _{1}^{-1}-\Sigma _{2}^{-1}\right)
y_{t}+2\left( \mu _{2}^{\top }\Sigma _{2}^{-1}-\mu _{1}^{\top
}\Sigma _{1}^{-1}\right)
y_{t}+ \\
&&\left( \mu _{1}^{\top }\Sigma _{1}^{-1}\mu _{1}-\mu _{2}^{\top }\Sigma
_{2}^{-1}\mu _{2}\right) +2\ln \frac{\pi \left( 1-\pi ^{\ast }\right) }{\pi
^{\ast }\left( 1-\pi \right) }+\ln \frac{\left\vert \Sigma _{1}\right\vert }{%
\left\vert \Sigma _{2}\right\vert }\geq 0
\end{eqnarray*}%
The quadratic form of this inequality is:%
\begin{equation*}
\left( y_{t}-v\right) ^{\top }Q\left( y_{t}-v\right) \geq r^{\star }
\end{equation*}%
where:%
\begin{eqnarray*}
Q &=&\Sigma _{1}^{-1}-\Sigma _{2}^{-1} \\
v &=&Q^{-1}\left( \Sigma _{1}^{-1}\mu _{1}-\Sigma _{2}^{-1}\mu _{2}\right)
\\
r^{\star } &=&v^{\top }Qv+\mu _{2}^{\top }\Sigma _{2}^{-1}\mu _{2}-\mu
_{1}^{\top }\Sigma _{1}^{-1}\mu _{1}-2\ln \frac{\pi \left( 1-\pi ^{\ast
}\right) }{\pi ^{\ast }\left( 1-\pi \right) }-\ln \frac{\left\vert \Sigma
_{1}\right\vert }{\left\vert \Sigma _{2}\right\vert }
\end{eqnarray*}%
With the parametrization $\mu _{1}=\mu \,\mathrm{d}t$, $\Sigma _{1}=\Sigma \,%
\mathrm{d}t$, $\mu _{2}=\mu \,\mathrm{d}t+\tilde{\mu}$ and $\Sigma
_{2}=\Sigma \,\mathrm{d}t+\tilde{\Sigma}$, we obtain\footnote{%
We use the following matrix formula: $Q=\Sigma _{2}^{-1}\left( \Sigma
_{2}-\Sigma _{1}\right) \Sigma _{1}^{-1}$.}:%
\begin{eqnarray*}
Q &=&\left( \Sigma \,\mathrm{d}t+\tilde{\Sigma}\right) ^{-1}\tilde{\Sigma}%
\left( \Sigma \,\mathrm{d}t\right) ^{-1} \\
v & = & \left( \mu - \Sigma \tilde{\Sigma}^{-1}\tilde{\mu}\right) \mathrm{d}t\\
r^{\star } &=&\tilde{\mu}^{\top }\tilde{\Sigma}^{-1}\tilde{\mu}-2\ln \frac{%
\pi \left( 1-\pi ^{\ast }\right) }{\pi ^{\ast }\left( 1-\pi \right) }-\ln
\frac{\left\vert \Sigma \,\mathrm{d}t\right\vert }{\left\vert \Sigma \,\mathrm{d}t+\tilde{\Sigma}\right\vert }
\end{eqnarray*}
The solution of the quadratic equation $\left( y_{t}-v\right) ^{\top
}Q\left( y_{t}-v\right) =r^{\star }$ is a $n$-dimensional ellipsoid
centered at $v$ and whose semi-axes are the square root of the
eigenvalues of the matrix $r^{\star }Q^{-1}$. We deduce that the
solution is not unique and can not be expressed in terms of
truncation points:
\begin{equation*}
J_{t}=1\nLeftrightarrow y_{i,t}\leq y_{i}^{-}\text{ and }y_{i,t}\geq
y_{i}^{+}
\end{equation*}
\bigskip

\begin{figure}[tbph]
\centering
\caption{Detecting the jumps in the multivariate case (weekly model)}
\label{fig:carry10-4}
\figureskip
\includegraphics[width = \figurewidth, height = \figureheight]{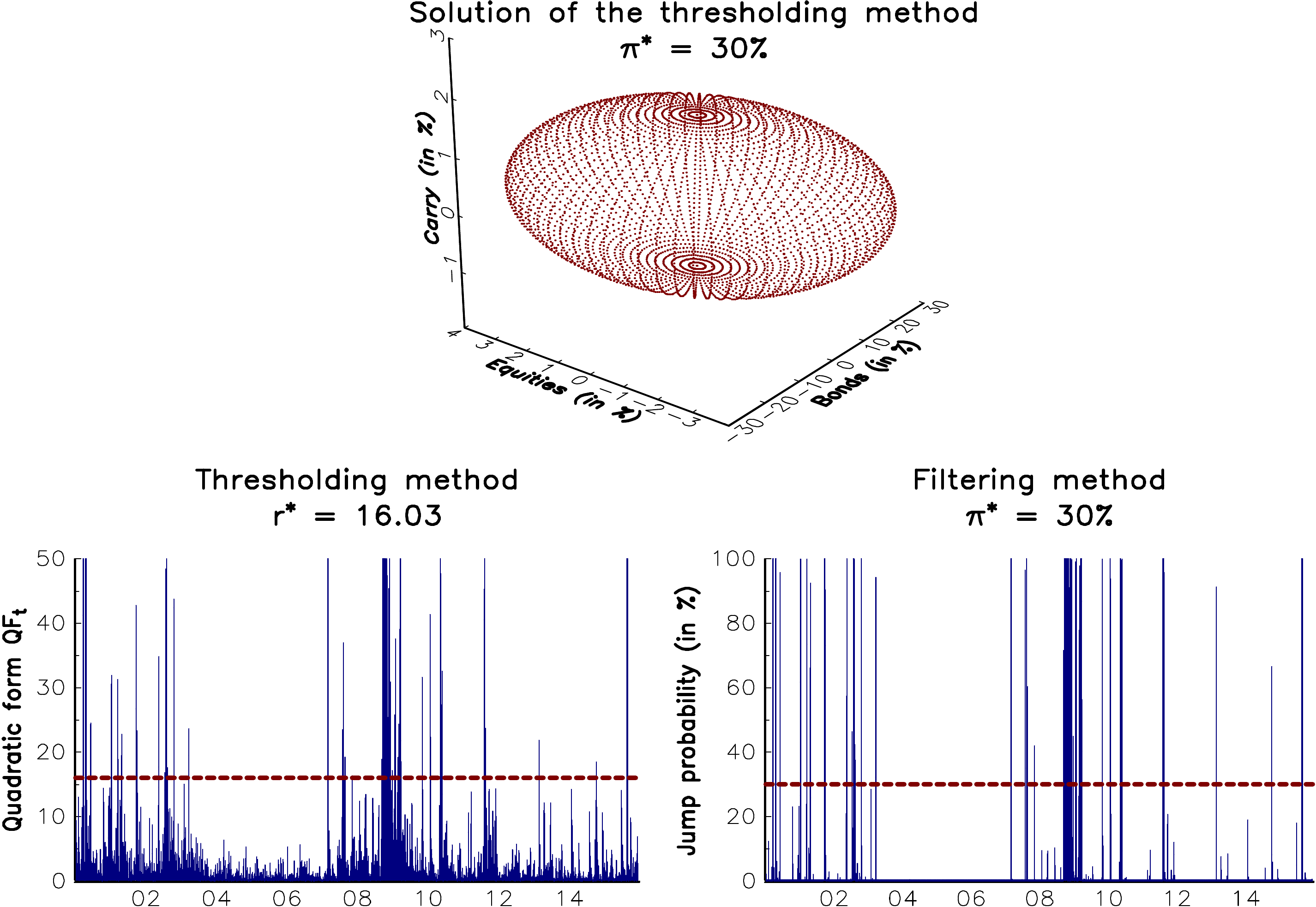}
\end{figure}

We consider again the equity/bond/volatility example. In the top
panel in Figure \ref{fig:carry10-4}, we report the solution of the
quadratic equation, which is an ellipsoid\footnote{The spherical
coordinates are:
\begin{equation*}
\left\{
\begin{array}{l}
R_{\mathrm{bond}}=\sqrt{\lambda _{1}}\cos \left( \theta \right) \cos
\left(
\phi \right)  \\
R_{\mathrm{equity}}=\sqrt{\lambda _{2}}\cos \left( \theta \right)
\sin
\left( \phi \right)  \\
R_{\mathrm{carry}}=\sqrt{\lambda _{3}}\sin \left( \theta \right)
\end{array}%
\right.
\end{equation*}%
where $-\pi /2\leq \theta \leq \pi /2$, $-\pi \leq \phi \leq \pi $
and $ \left\{ \lambda _{i}\right\} _{i\geq 1}$ are the eigenvalues
of $r^{\star }Q^{-1}$.}. For the thresholding method, we compare at
each date the quadratic form $\mathbb{QF}_{t}=\left( R_{t}-v\right)
^{\top }Q\left( R_{t}-v\right) $ with the threshold $r^{\star
}=16.03$ calculated using the numerical values of Table
\ref{tab:carry8-2-1}. We will say that we observe a jump if the
following conditions holds:
\begin{equation*}
J_{t}=1\Leftrightarrow \mathbb{QF}_{t}\geq r^{\star }
\end{equation*}%
The filtering method is given in the bottom/right panel. In this
case, we have:
\begin{equation*}
J_{t}=1\Leftrightarrow \pi _{t}\geq \pi ^{\star }
\end{equation*}

\subsection{Filtering algorithm for estimating $\hat{\mu}_{t}$ and $\hat{\Sigma}_{t}$}
\label{appendix:filtering}

We consider a sample $\left\{ y_{1},\ldots ,y_{T}\right\} $ of $T$
observations. We note $y_{i,t}$ the value of the $i^{\mathrm{th}}$
component at time $t$. Given $\pi $, $\tilde{\mu}$ and
$\tilde{\Sigma}$, we would like to estimate the parameters
$\hat{\mu}_{t}$ and $\hat{\Sigma}_{t}$ by
assuming that the probability density function of the mixture model is:%
\begin{equation*}
f\left( y\right) =\left( 1-\pi \right) \phi _{n}\left( y;\mu \,\mathrm{d}%
t,\Sigma \,\mathrm{d}t\right) +\pi \phi _{n}\left( y;\mu \,\mathrm{d}t+%
\tilde{\mu},\Sigma \,\mathrm{d}t+\tilde{\Sigma}\right)
\end{equation*}%
We assume that there is a jump at time $t$ if the filtering probability $%
\hat{\pi}_{t}$ is larger than a given threshold $\pi ^{\star }$:%
\begin{equation*}
J_{t}=1\Leftrightarrow \hat{\pi}_{t}\geq \pi ^{\star }
\end{equation*}%
The algorithm for estimating the parameters is then given by the
following steps:

\begin{enumerate}
\item Let $n_{rw}$ be the length of the rolling window. We initialize the
filtering procedure at time $n_{rw}+1$.

\item At time $t$, we calculate the posterior jump probabilities:%
\begin{equation*}
\hat{\pi}_{t-s}=\frac{\pi \phi _{n}\left( y_{t-s},\hat{\mu}_{t-1}\,\mathrm{d}%
t+\tilde{\mu},\hat{\Sigma}_{t-1}\,\mathrm{d}t+\tilde{\Sigma}\right)
}{\left(
1-\pi \right) \phi _{n}\left( y_{t-s},\hat{\mu}_{t-1}\,\mathrm{d}t,\hat{%
\Sigma}_{t-1}\,\mathrm{d}t\right) +\pi \phi _{n}\left( y_{t-s},\hat{\mu}%
_{t-1}\,\mathrm{d}t+\tilde{\mu},\hat{\Sigma}_{t-1}\,\mathrm{d}t+\tilde{\Sigma%
}\right) }
\end{equation*}%
for $s=0,\ldots ,n_{rw}-1$. The probabilities
$\hat{\pi}_{t-s} $ are based on the estimates $\hat{\mu}_{t-1}$ and
$\hat{\Sigma}_{t-1}$ calculated at time $t-1$.

\item We update the estimates $\hat{\mu}_{t-1}$ and $\hat{\Sigma}_{t-1}$
using the following formulas:%
\begin{equation*}
\hat{\mu}_{i,t}\,\mathrm{d}t=\frac{1}{n_{t}}\sum_{s=0}^{n_{rw}-1}\mathds{1}%
\left\{ \hat{\pi}_{t-s}\leq \bar{\pi}\right\} \cdot y_{i,t-s}
\end{equation*}%
\begin{equation*}
\hat{\Sigma}_{i,j,t}\,\mathrm{d}t=\frac{1}{n_{t}}\sum_{s=0}^{n_{rw}-1}%
\mathds{1}\left\{ \hat{\pi}_{t-s}\leq \bar{\pi}\right\} \cdot \left(
y_{i,t-s}-\hat{\mu}_{i,t}\right) \left(
y_{j,t-s}-\hat{\mu}_{j,t}\right)
\end{equation*}%
where $n_{t}$ indicates the number of observations that do not jump:
\begin{equation*}
n_{t}=\sum_{s=0}^{n_{rw}-1}\mathds{1}\left\{ \hat{\pi}_{t-s}\leq \bar{\pi}%
\right\}
\end{equation*}%
These new estimates  $\hat{\mu}_{t}$ and $\hat{\Sigma}_{t}$ will be
valid to calculate the posterior probabilities $\hat{\pi}_{t+1-s}$.

\item We go back to Step 2.
\end{enumerate}
To initialize the algorithm, we can use the traditional Gaussian
estimates.

\section{Additional results}
\label{appendix:additional-results}

\begin{table}[tbph]
\centering
\caption{Estimation of the Gaussian model (weekly model)}
\label{tab:carry1-2}
\tableskip
\begin{tabular}{|c|ccccc|}
\hline Asset    & $\mu_i$ & $\sigma_i$ & \multicolumn{3}{c|}{$\rho_{i,j}$} \\ \hline
Bonds    & $5.38$ & ${\TsV}4.17$ & ${\TsIII}100.00$ &      $      $ & $      $ \\
Equities & $6.09$ &      $18.38$ &         $-34.43$ &      $100.00$ & $      $ \\
Carry    & $6.00$ & ${\TsV}5.50$ &         $-18.22$ & ${\TsV}51.92$ & $100.00$ \\
\hline
\end{tabular}
\end{table}

\begin{table}[tbph]
\centering
\caption{Estimation of the Gaussian model (monthly model)}
\label{tab:carry1-3}
\tableskip
\begin{tabular}{|c|ccccc|}
\hline Asset    & $\mu_i$ & $\sigma_i$ & \multicolumn{3}{c|}{$\rho_{i,j}$} \\ \hline
Bonds    & $5.32$ & ${\TsV}4.31$ & ${\TsIII}100.00$ &      $      $ & $      $ \\
Equities & $5.91$ &      $16.97$ &         $-28.56$ &      $100.00$ & $      $ \\
Carry    & $6.26$ & ${\TsV}6.91$ &         $-14.50$ & ${\TsV}55.21$ & $100.00$ \\
\hline
\end{tabular}
\end{table}

\begin{table}[tbph]
\centering
\caption{Estimation of the Gaussian model (annually model)}
\label{tab:carry1-4}
\tableskip
\begin{tabular}{|c|ccccc|}
\hline Asset    & $\mu_i$ & $\sigma_i$ & \multicolumn{3}{c|}{$\rho_{i,j}$} \\ \hline
Bonds    & $5.29$ & ${\TsV}4.03$ & ${\TsIII}100.00$ &      $      $ & $      $ \\
Equities & $6.47$ &      $18.59$ &         $-69.26$ &      $100.00$ & $      $ \\
Carry    & $6.68$ & ${\TsV}8.16$ &         $-26.22$ & ${\TsV}63.25$ & $100.00$ \\
\hline
\end{tabular}
\end{table}

\begin{table}[tbph]
\centering
\caption{Estimation of the Gaussian mixture model when $\pi = 5\%$ (weekly model)}
\label{tab:carry7-1-1}
\tableskip
\begin{tabular}{|c|ccccc|}
\hline
Asset    & $\mu_i$          & $\sigma_i$   & \multicolumn{3}{c|}{$\rho_{i,j}$}           \\ \hdashline
Bonds    &  ${\TsVIII}4.65$ & ${\TsV}3.84$ & ${\TsIII}100.00$ &      $      $ & $      $ \\
Equities &  ${\TsIII}10.85$ &      $14.32$ &         $-34.78$ &      $100.00$ & $      $ \\
Carry    &  ${\TsIII}11.35$ & ${\TsV}2.53$ &         $-22.40$ & ${\TsV}57.72$ & $100.00$ \\ \hline
Asset    &  $\tilde{\mu}_i$ & $\tilde{\sigma}_i$ & \multicolumn{3}{c|}{$\tilde{\rho}_{i,j}$} \\ \hdashline
Bonds    &  ${\TsVIII}0.18$ & ${\TsV}0.78$ & ${\TsIII}100.00$ &      $      $ & $      $ \\
Equities &          $-1.26$ & ${\TsV}5.56$ &         $-35.80$ &      $100.00$ & $      $ \\
Carry    &          $-1.30$ & ${\TsV}2.04$ &         $-16.34$ & ${\TsV}53.86$ & $100.00$ \\
\hline
\end{tabular}
\end{table}

\begin{table}[tbph]
\centering
\caption{Estimation of the Gaussian mixture model when $\pi = 5\%$ (monthly model)}
\label{tab:carry7-1-2}
\tableskip
\begin{tabular}{|c|ccccc|}
\hline Asset    & $\mu_i$ & $\sigma_i$ & \multicolumn{3}{c|}{$\rho_{i,j}$} \\ \hdashline
Bonds    & ${\TsVIII}4.25$ & ${\TsV}4.01$ & ${\TsIII}100.00$ &      $      $ & $      $ \\
Equities & ${\TsIII}10.76$ &      $13.79$ &         $-33.04$ &      $100.00$ & $      $ \\
Carry    & ${\TsIII}10.14$ & ${\TsV}2.61$ &         $-19.03$ & ${\TsV}46.49$ & $100.00$ \\ \hline
Asset    & $\tilde{\mu}_i$ & $\tilde{\sigma}_i$ & \multicolumn{3}{c|}{$\tilde{\rho}_{i,j}$} \\ \hdashline
Bonds    & ${\TsVIII}1.18$ & ${\TsV}1.21$ & ${\TsIII}100.00$ &      $      $ & $      $ \\
Equities &         $-5.33$ & ${\TsV}8.97$ & ${\TsVIII}26.78$ &      $100.00$ & $      $ \\
Carry    &         $-4.14$ & ${\TsV}5.21$ & ${\TsVIII}32.51$ & ${\TsV}65.69$ & $100.00$ \\
\hline
\end{tabular}
\end{table}

\begin{table}[tbph]
\centering
\caption{Estimation of the Gaussian mixture model when $\pi = 1\%$ (weekly model)}
\label{tab:carry7-2-1}
\tableskip
\begin{tabular}{|c|ccccc|}
\hline Asset    & $\mu_i$ & $\sigma_i$ & \multicolumn{3}{c|}{$\rho_{i,j}$} \\ \hdashline
Bonds    & ${\TsVIII}4.75$ & ${\TsV}3.95$ & ${\TsIII}100.00$ &      $      $ & $      $ \\
Equities & ${\TsVIII}9.23$ &      $15.21$ &         $-34.60$ &      $100.00$ & $      $ \\
Carry    & ${\TsIII}10.51$ & ${\TsV}2.80$ &         $-22.91$ & ${\TsV}56.83$ & $100.00$ \\ \hline
Asset    & $\tilde{\mu}_i$ & $\tilde{\sigma}_i$ & \multicolumn{3}{c|}{$\tilde{\rho}_{i,j}$} \\ \hdashline
Bonds    & ${\TsVIII}0.29$ & ${\TsV}0.85$ & ${\TsIII}100.00$ &      $      $ & $      $ \\
Equities &         $-1.56$ & ${\TsV}6.66$ &         $-35.97$ &      $100.00$ & $      $ \\
Carry    &         $-1.98$ & ${\TsV}2.43$ &  ${\TsV}$$-7.65$ & ${\TsV}57.45$ & $100.00$ \\
\hline
\end{tabular}
\end{table}

\begin{table}[tbph]
\centering
\caption{Estimation of the Gaussian mixture model when $\pi = 1\%$ (monthly model)}
\label{tab:carry7-2-2}
\tableskip
\begin{tabular}{|c|ccccc|}
\hline Asset    & $\mu_i$ & $\sigma_i$ & \multicolumn{3}{c|}{$\rho_{i,j}$} \\ \hdashline
Bonds    & ${\TsVIII}4.57$ & ${\TsV}4.07$ & ${\TsIII}100.00$ &      $      $ & $      $ \\
Equities & ${\TsVIII}9.16$ &      $14.54$ &         $-32.14$ &      $100.00$ & $      $ \\
Carry    & ${\TsVIII}9.70$ & ${\TsV}2.90$ &         $-19.56$ & ${\TsV}47.68$ & $100.00$ \\ \hline
Asset    & $\tilde{\mu}_i$ & $\tilde{\sigma}_i$ & \multicolumn{3}{c|}{$\tilde{\rho}_{i,j}$} \\ \hdashline
Bonds    & ${\TsVIII}1.36$ & ${\TsV}1.33$ & ${\TsIII}100.00$ &      $      $ & $      $ \\
Equities &         $-5.84$ &      $10.17$ &  ${\TsIII}30.97$ &      $100.00$ & $      $ \\
Carry    &         $-5.94$ & ${\TsV}5.77$ &  ${\TsIII}55.18$ & ${\TsV}74.83$ & $100.00$ \\
\hline
\end{tabular}
\end{table}

\begin{table}[tbph]
\centering
\caption{Estimation of the Gaussian mixture model when $\lambda = 10\%$ (weekly model)}
\label{tab:carry7-3-1}
\tableskip
\begin{tabular}{|c|ccccc|}
\hline Asset    & $\mu_i$ & $\sigma_i$ & \multicolumn{3}{c|}{$\rho_{i,j}$} \\ \hdashline
Bonds    & ${\TsVIII}4.85$ & ${\TsV}3.99$ & ${\TsIII}100.00$ &      $      $ & $      $ \\
Equities & ${\TsVIII}8.53$ &      $15.70$ &         $-35.21$ &      $100.00$ & $      $ \\
Carry    & ${\TsIII}10.10$ & ${\TsV}2.94$ &         $-23.44$ & ${\TsV}56.95$ & $100.00$ \\ \hline
Asset    & $\tilde{\mu}_i$ & $\tilde{\sigma}_i$ & \multicolumn{3}{c|}{$\tilde{\rho}_{i,j}$} \\ \hdashline
Bonds    & ${\TsVIII}0.33$ & ${\TsV}0.89$ & ${\TsIII}100.00$ &      $      $ & $      $ \\
Equities &         $-1.71$ & ${\TsV}7.22$ &         $-32.16$ &      $100.00$ & $      $ \\
Carry    &         $-2.46$ & ${\TsV}2.62$ &  ${\TsV}$$-1.96$ & ${\TsV}60.59$ & $100.00$ \\
\hline
\end{tabular}
\end{table}

\begin{table}[tbph]
\centering
\caption{Estimation of the Gaussian mixture model when $\lambda = 10\%$ (monthly model)}
\label{tab:carry7-3-2}
\tableskip
\begin{tabular}{|c|ccccc|}
\hline Asset    & $\mu_i$ & $\sigma_i$ & \multicolumn{3}{c|}{$\rho_{i,j}$} \\ \hdashline
Bonds    & ${\TsVIII}4.59$ & ${\TsV}4.08$ & ${\TsIII}100.00$ &      $      $ & $      $ \\
Equities & ${\TsVIII}8.96$ &      $14.64$ &         $-31.97$ &      $100.00$ & $      $ \\
Carry    & ${\TsVIII}9.64$ & ${\TsV}2.93$ &         $-19.58$ & ${\TsV}48.02$ & $100.00$ \\ \hline
Asset    & $\tilde{\mu}_i$ & $\tilde{\sigma}_i$ & \multicolumn{3}{c|}{$\tilde{\rho}_{i,j}$} \\ \hdashline
Bonds    & ${\TsVIII}1.38$ & ${\TsV}1.34$ & ${\TsIII}100.00$ &      $      $ & $      $ \\
Equities &         $-5.79$ &      $10.28$ &  ${\TsIII}30.71$ &      $100.00$ & $      $ \\
Carry    &         $-6.15$ & ${\TsV}5.83$ &  ${\TsIII}58.45$ & ${\TsV}76.92$ & $100.00$ \\
\hline
\end{tabular}
\end{table}

\begin{table}[tbph]
\centering
\caption{Estimation of the constrained mixture model when $\pi = 1\%$ (monthly model)}
\label{tab:carry8-2-2}
\tableskip
\begin{tabular}{|c|ccccc|}
\hline Asset    & $\mu_i$ & $\sigma_i$ & \multicolumn{3}{c|}{$\rho_{i,j}$} \\ \hdashline
Bonds    & ${\TsVIII}5.32$ & ${\TsV}4.31$ & ${\TsIII}100.00$ &      $      $ & $      $ \\
Equities & ${\TsVIII}8.69$ &      $14.53$ &         $-32.57$ &      $100.00$ & $      $ \\
Carry    & ${\TsVIII}9.54$ & ${\TsV}2.94$ &         $-22.21$ & ${\TsV}48.50$ & $100.00$ \\ \hline
Asset    & $\tilde{\mu}_i$ & $\tilde{\sigma}_i$ & \multicolumn{3}{c|}{$\tilde{\rho}_{i,j}$} \\ \hdashline
Bonds    & ${\TsVIII}0.00$ & ${\TsV}0.00$ &      $100.00$   &      $      $ & $      $ \\
Equities &         $-5.46$ &      $10.86$ &  ${\TsX}0.00$   &      $100.00$ & $      $ \\
Carry    &         $-6.16$ & ${\TsV}5.93$ &  ${\TsX}0.00$   & ${\TsV}73.28$ & $100.00$ \\
\hline
\end{tabular}
\end{table}

\begin{figure}[tbph]
\centering
\caption{Expected monthly drawdown (in \%) of the carry risk premium}
\label{fig:carry8-1-2}
\figureskip
\includegraphics[width = \figurewidth, height = \figureheight]{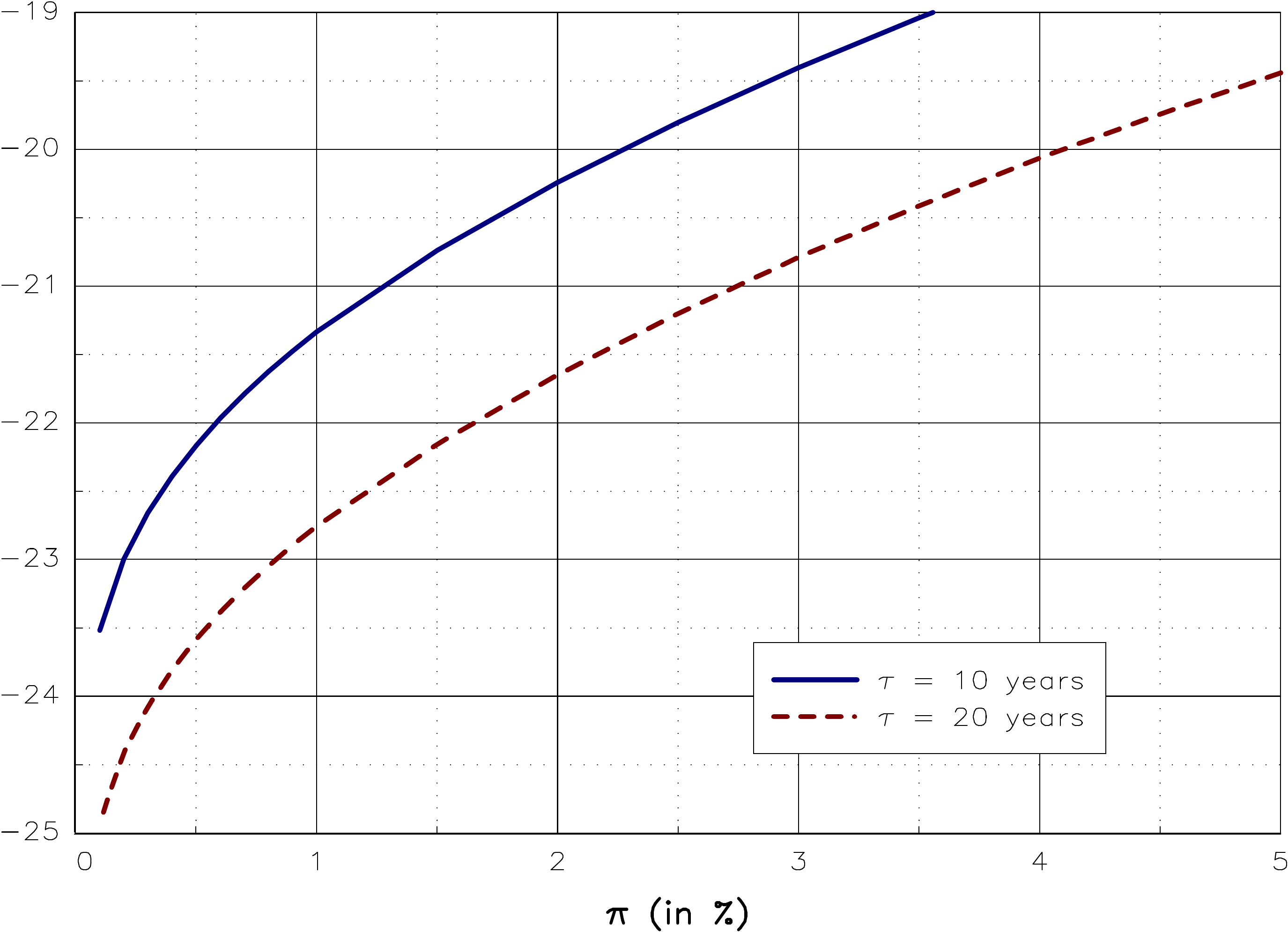}
\end{figure}

\begin{figure}[tbph]
\centering
\caption{Probability density function of asset returns in the normal regime (monthly model)}
\label{fig:carry8-4}
\figureskip
\includegraphics[width = \figurewidth, height = \figureheight]{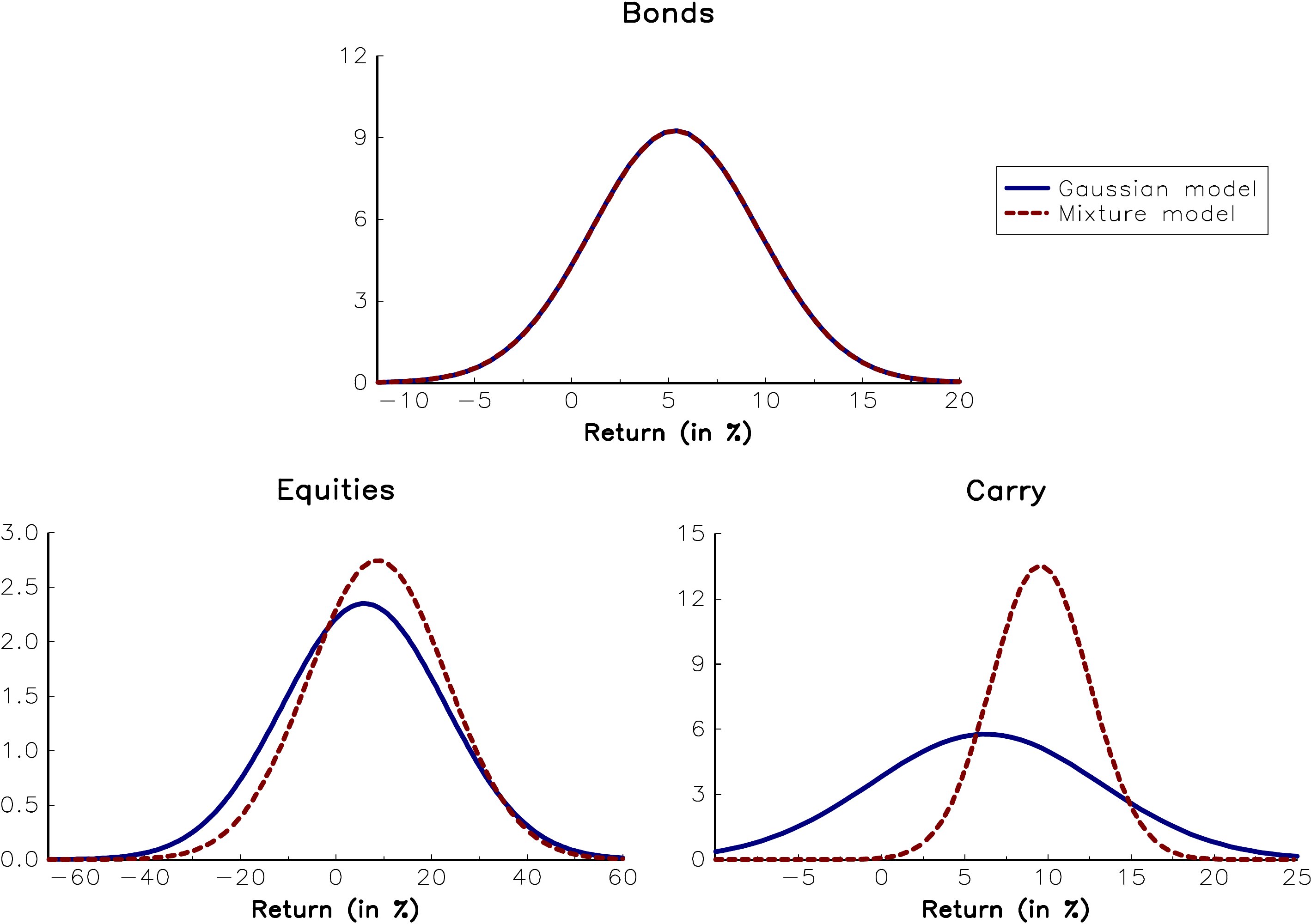}
\end{figure}

\begin{figure}[tbph]
\centering
\caption{Dynamics of the ERC weights (mixture model, ML method)}
\label{fig:carry9-5}
\figureskip
\includegraphics[width = \figurewidth, height = \figureheight]{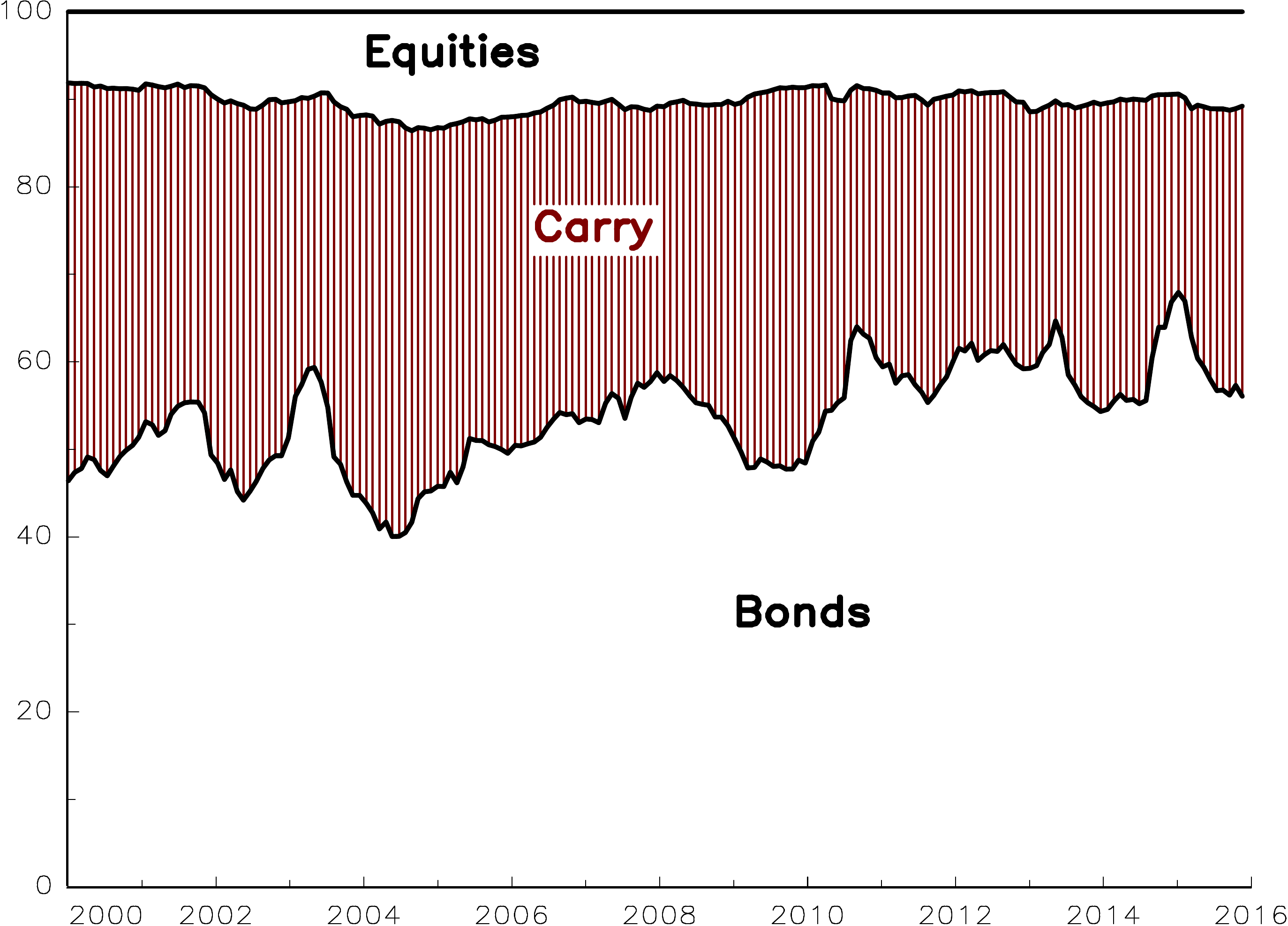}
\end{figure}

\end{document}